\documentclass[journal]{IEEEtran}
\usepackage{bm}
\usepackage{cite}
\usepackage{amssymb}
\usepackage{setspace}

\ifCLASSINFOpdf
\usepackage[pdftex]{graphicx}
\else
\fi

\usepackage{caption}
\usepackage{amsmath,amsthm,bm,amssymb}

\newtheorem{prop}{Proposition}

\newtheorem{thm}{Theorem}

\usepackage{amsfonts} 
\usepackage{algorithm}  
\usepackage{algorithmic}
\usepackage{color}
\usepackage[dvipsnames]{xcolor}

\usepackage{array}
\usepackage{subfigure}
\usepackage{url}
\begin{document}
\title{Time-of-use Pricing for Energy Storage Investment}
\author{Dongwei~Zhao,~\IEEEmembership{Student~Member,~IEEE,}
		Hao~Wang,~\IEEEmembership{Member,~IEEE,}\\
		Jianwei~Huang,~\IEEEmembership{Fellow,~IEEE,}
		and~Xiaojun~Lin,~\IEEEmembership{Fellow,~IEEE}
		\thanks{This work is supported by the Shenzhen Science and Technology Program (JCYJ20210324120011032), Shenzhen Institute of Artificial Intelligence and Robotics for Society, and the Presidential Fund from the Chinese University of Hong Kong, Shenzhen. It is also supported by National Science Foundation by grant ECCS-2129631.  Part of the results have appeared in IEEE SmartGridComm 2020 \cite{smartgrid2020}.}
		\thanks{Dongwei Zhao is with MIT Energy Initiative, Massachusetts Institute of Technology, Cambridge, MA 02139, USA (e-mail: zhaodw@mit.edu). 
		
		Hao Wang is with the Department of Data Science and Artificial Intelligence, Faculty of Information Technology, Monash University, Melbourne, VIC 3800, Australia (e-mail: hao.wang2@monash.edu). 
		
		Jianwei Huang is with the School of Science and Engineering, The Chinese University of Hong Kong, Shenzhen, and the Shenzhen Institute of Artificial Intelligence and Robotics for Society (corresponding author, e-mail: jianweihuang@cuhk.edu.cn).
		
		Xiaojun Lin is with the School of Electrical and Computer Engineering, Purdue University, West Lafayette, IN 47907, USA (e-mail: linx@ecn.purdue.edu).}
	}
	\maketitle
	
\begin{abstract} 
Time-of-use (ToU) pricing is widely used by the electricity utility to shave peak load. Such a pricing scheme provides users with incentives to invest in behind-the-meter energy storage and to shift peak load towards low-price intervals.  
However, without considering the implication on energy storage investment, an improperly designed ToU pricing scheme may lead to significant welfare loss, especially when users over-invest the storage, which leads to new energy consumption peaks. 
In this paper,  we will study how to design a social-optimum ToU pricing scheme by explicitly considering its impact on storage investment.  We model the interactions  between the utility and users  as a two-stage optimization problem. To resolve the challenge of asymmetric information due to users' private storage cost, we propose a ToU pricing scheme
based on different storage types and the aggregate demand per type. Each user does not need to reveal his private cost information. We can further compute the optimal ToU pricing with only a linear complexity.  
Simulations based on real-world data show that the suboptimality gap of our proposed ToU pricing, compared with the social optimum achieved under complete information,  is less than 5\%.

\end{abstract}


\IEEEpeerreviewmaketitle

\section{Introduction}

\subsection{Background and motivation}
Time-of-use (ToU) pricing is a  electricity tariff that is widely used by the electricity utility. It can help shave the system peak load and reduce the system overall cost \cite{tous}. In ToU pricing, the utility usually divides one day into two or three periods with different electricity prices. In a typical two-period ToU pricing \cite{timeofusey}, the utility defines a peak period (e.g., 4 PM to 9 PM) and an off-peak period (e.g., 10 PM to 3 PM). The  price for the peak period is higher than that of the  off-peak period.  The ToU pricing can incentivize users to  shift elastic loads from the peak period to the off-peak period to reduce their energy costs.
 
Besides changing the energy consumption pattern, users may further consider investing in energy storage to take advantage of the price difference in a ToU pricing \cite{pimm2018time}. Specially, during off-peak hours  with a lower electricity price, users with storage can purchase more electricity (than the actual needed consumption)  and charge it into storage for later use. During peak hours with a high electricity price, users can discharge the storage to partially fulfill their energy demands.  In the ideal case, such operations of storage not only reduce users' electricity bills but also help shave the system peak load and reduce the social cost. Note that although some part of user's demand may be elastic, there always exists a substantial part of the demand that is inelastic, the latter of which can only be shifted by storage.  The ToU pricing itself cannot  shift users' inelastic demand and reduce the system peak load unless with the help of users' storage.

However, the increasing deployment of energy storage on the end-user side poses new challenges for the ToU pricing design. If the ToU pricing design does not consider the impact of storage, it may lead to new and even higher system peaks.  To understand this, note that the storage investment decision depends on both the peak/off-peak price difference and the storage cost.  A small price difference (compared with the storage cost) cannot incentivize sufficient storage investment from users. A higher price difference, however,  may incentivize too much storage investment. Consider the extreme case where  all the users invest in storage and shift the demand from the peak period to the off-peak period, such that the original peak period will have zero demand and the original off-peak period will become a new peak. Both the new peak and the large storage  investment cost may increase the social cost. Although the utility may reduce the future price difference in the ToU pricing to flatten the new peak, the sunken cost of storage investment can not be recovered. This increases the social cost and leads to social welfare loss, which also harms users' interests. Therefore, a proper design of the  ToU pricing considering users'  storage investment and operation  is critical to the performance of the  electricity system.

The above discussions motivate us to answer the key question in this paper: 
\begin{itemize}
    \item 
\textit{How to design a ToU pricing to induce proper users'  storage investment in order to  achieve the social optimum?}
\end{itemize}

 The challenge for designing such ToU pricing   is the private information of individual users' storage costs, which makes it challenging to
    incentivize low-cost users to invest in storage while discouraging high-cost users from investing. To address the  challenge, we  define a set of storage types based on the possible storage costs on the market,  and classify users based on such types. We propose a ToU pricing scheme based on each type's storage cost and aggregate demand, instead of individual users' private storage cost and demand. {Such a ToU pricing scheme does not require users' private information.}
 
 We compare our proposed pricing scheme (without individual users' private information) against two other cases:
 \begin{itemize}
 \item A ToU pricing scheme assuming knowledge of
individual users’ private information. 

\item The social-optimum benchmark where a social planner decides the storage investment for all users with complete system information. 
\end{itemize}

\subsection{Main results and contributions}
To the best of our knowledge, our paper is the first work that studies the ToU pricing design considering the impact of the end-users' storage investment. Our results can guide users' storage investment and operation to minimize the social cost.

To decide the optimal ToU pricing, we formulate a two-stage optimization problem between the utility and users over two timescales. In Stage I, before the investment horizon, the utility determines the peak and off-peak prices for the ToU pricing. In Stage II, at the beginning of the investment horizon, each user decides the optimal investment capacity of storage. Then, in each operational horizon (one day), each user determines the charging and discharging of the storage given the storage capacity and realized load profiles.

In our proposed ToU pricing,  the utility only needs to know the storage cost of each storage type, and the aggregate demand of users in each type. It does not require knowledge of individual users' private cost or demand  information. We prove that the social cost under our proposed type-based ToU pricing is higher than that under individual-based ToU pricing, which is further higher than that under a social-optimum benchmark. However, extensive simulations based on real-world data show that  our proposed type-based ToU pricing can induce a social cost very close to the social-optimum benchmark.

The main contributions of this paper are listed as follows.
	\begin{itemize}
		\item \textit{Storage-aware ToU pricing}:
		As far as we know, this is the first work that studies the ToU pricing design considering the impact of users' storage investment decisions, with the purpose of achieving social optimum. Such a  storage-aware ToU pricing can significantly improve the performance  of the electricity system. 
		\item \textit{Pricing scheme without private information}: The key challenge for designing the ToU pricing scheme is users' private storage investment costs.
		We propose a simple yet effective pricing scheme for the utility based on the storage types, which does not require each user's private information but only each type's storage cost and aggregate demand. Such aggregation incurs no information loss if users demands' are perfectly positively correlated.

		\item \textit{{Threshold-based algorithm}}: We formulate a two-stage optimization problem that is non-convex and challenging to solve. Despite such difficulty, we characterize a step-wise structure for the social cost with respect to the price, based on which we design an efficient algorithm to determine the optimal pricing by searching finite threshold points. The number of  threshold points is linear in the numbers of demand outcomes and storage types.
		
		\item \textit{Performance of  the proposed pricing scheme}: 
	Extensive simulations based on real-world data  validate the  near-optimal  performance of the proposed pricing scheme, where the suboptimality gap comparing with the social optimum is less than 5\%.  A surprising result is that  an increased level of user demand uncertainty (within a certain range) can improve the performance of the pricing scheme by smoothing users' storage investment decisions. 

	\end{itemize}

\section{Related works}

There have been a series of active studies on the design of ToU pricing (e.g., \cite{contract3,touren,charwand2018optimal}). Chen \textit{et al.} \cite{contract3} designed the optimal ToU pricing for households, which minimizes the system peak load and maximizes the utility's profit. K{\"o}k \textit{et al.} \cite{touren} designed the optimal ToU pricing considering the impact of renewable energy investment. Charwand \textit{et al.} \cite{charwand2018optimal} proposed a robust midterm framework to optimize ToU pricing strategies. However, these studies did not consider the impact of end-users' storage investment, which can significantly affect the system load and the ToU pricing strategy.

Some recent literature considered the optimal storage operation and investment under the ToU pricing (e.g., \cite{timestorage1,carpinelli2016probabilistic,Kalathil2019Storage}). Nguyen \textit{et al.}\cite{timestorage1} optimized the operation of energy storage to minimize users' energy costs under the ToU pricing. Carpinelli \textit{et al.} \cite{carpinelli2016probabilistic} proposed a probabilistic method to size the energy storage under the ToU pricing. Kalathil \textit{et al.}  \cite{Kalathil2019Storage} studied the game-theoretic model for storage sharing under the ToU pricing. However, the ToU prices in these prior literature  are exogenously given, without considering the impact of users' proactive decisions in storage investment and operation on the system. To our best knowledge, there has been no literature studying the design of ToU pricing that directly takes into account  the end-users' storage investment decisions.   

Multi-stage optimization models have been widely adopted in energy systems (e.g.,\cite{bistoragesharing,retailer,Zhao2020Storagesharing}). Chen \textit{et al.} \cite{bistoragesharing} formulated a two-stage model for the central storage sharing between a distribution company and customers. Wei \textit{et al.} \cite{retailer} optimized the energy pricing and dispatch for electricity retailers considering users' demand response. Both \cite{bistoragesharing} and \cite{retailer} solved the two-stage optimization problem by constructing an equivalent single optimization problem, e.g., a mixed-integer linear programming problem, which requires all the users' private information and is often solved with  high computational complexity. Zhao \textit{et al.} \cite{Zhao2020Storagesharing} proposed a distributed algorithm based on the information exchange between Stage I and Stage II, still assuming a truthful report of private information from Stage II. In our work, we design and solve the pricing scheme based on storage types' information, which does not require any individual users' private information. We also develop an efficient algorithm 
by searching a finite number of threshold points, which corresponds to  low linear complexity in key system parameters.

\section{System Model}

We consider one electricity utility serving a group of users. The utility sets a  two-period  ToU pricing for users, with a higher electricity price for the peak period and a lower price for the off-peak period.\footnote{Both two-period pricing and three-period pricing exist in practice, and both of them can incentivize the storage investment of end users. Our work focuses on the two-period pricing because it is simple and can always help us demonstrate the impact of ToU pricing on storage investment. We will consider the three-period pricing in the future work.}

\begin{figure}[t]
	\centering
	\includegraphics[width=3.1in]{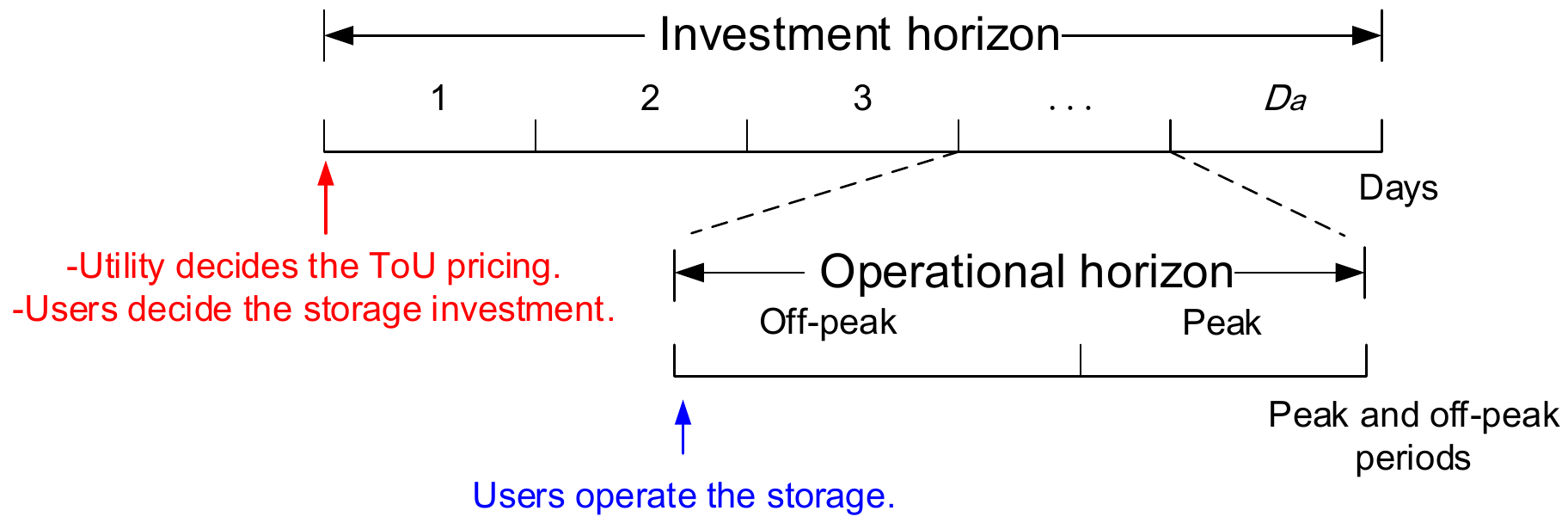}
	\vspace{-1mm}
	\caption{\small Two timescales.}
	\label{fig:time0}
	\vspace{-5mm}
\end{figure}

 We illustrate two timescales of decision-making between the utility and users in Figure \ref{fig:time0}.  Before an investment horizon of $D_a$ days (e.g., $D_a$ can correspond to many years), the utility announces the ToU pricing to users. Then, at the beginning of the investment horizon, users decide how much storage to invest in.\footnote{Note that, in order to show the impact of the utility's ToU pricing on the users' storage investment, we focus on a fixed investment horizon and assume that the utility's ToU pricing shares the same time horizon as the investment horizon of users' storage. The utility can make the  ToU pricing decision sometime before the investment horizon but the ToU price should be effective over the whole investment horizon.}  The investment horizon is divided into operational horizons. Each operational horizon corresponds to one day, which is further divided into two periods $\mathcal{T}\hspace{-1mm}=\hspace{-1mm}\{p,o\}$: the peak period $p$ and the off-peak period $o$. Each peak period and off-peak period can contain multiple hours. During each day, each user utilizes  storage to minimize his energy cost  through proper charging and discharging decisions.  Next we will introduce the detailed models for users and utility.

\subsection{Users}

We consider a group  $\mathcal{I}=\{1,2,\ldots ,I\}$ of users that face the ToU pricing from the utility.  Based on ToU pricing, users can invest and operate the storage  to shift the demand and reduce the electricity bill. Next, we introduce the model of users' demands and storage costs.  

\subsubsection{Demands}  Users' electricity bills only depend on the total demand at peak and off-peak periods. For each user,  his peak and off-peak demands vary across days, so we model each user's peak and off-peak demands for one day as  random variables.  We let $\bm{\mathcal{D}}_i=(\mathcal{D}_i^o,\mathcal{D}_i^p)$ denote the random demand of user $i$ in one day,  where $\mathcal{D}_i^p$ and $\mathcal{D}_i^o$ denote  his peak and off-peak demands, respectively.
We  denote the vector of all the users'  peak and off-peak demand  as  $\boldsymbol{\mathcal{D}} = (\boldsymbol{\mathcal{D}}_i, \forall i \in \mathcal{I})$. We assume that the random variable  $\mathcal{D}_i^x$  has CDF $F_i^x$ with a range $[\underline{\mathcal{D}}_i^x,\overline{\mathcal{D}}_i^x]$,  $x\in\{o,p\}$. Across all the users, we assume a general joint CDF $F$  for the random vector  $\bm{\mathcal{D}}$, where users' demands can be independent or dependent. To examine the impact of storage, we focus on the users' inelastic demands \cite{fundamentals} in the main text.\footnote{In Appendix.H, we generalize our model to incorporate the elastic demand, and provide additional simulation results about the impact of elastic demand. Our high-level finding is that additional elastic demand with a low shift cost  will reduce users' demand for storage but benefit the social welfare.}
Note that each user's load depends on both his inelastic demand and storage operation. When a user charges the storage, his load is higher than the demand. When a user discharges, his load is smaller than his demand.

Demand distribution can be estimated using users' historical load data \cite{haojoint}. In the simulation of Section \ref{section:simulation}, we use one-year load data of users to build the discrete distribution.

\subsubsection{Storage cost}
Users can have heterogeneous  storage costs, as they can choose different storage technologies, e.g., Lithium-ion storage or Lead-acid storage \cite{fisher2019can}. 
We denote the unit capacity investment cost of storage for user $i$  as  $\theta_i'$.

The main cost of storage is  the one-time investment cost. To facilitate the optimization problem formulation, we convert  the one-time unit investment cost  $\theta_i'$ into a daily cost $\theta_i$ according to  $\theta_i=r^f \theta_i'$ based on a scaling factor $r^f$.  To derive $r^f$, we first calculate the present value of an annuity (a sequence of equal annual cash flows) with the  annual interest rate $r$, and then we divide the annuity equally to each day. This leads to the following formulation of the factor $r^f$
	\begin{align}
	r^f=\frac{r(1+r)^y}{(1+r)^y-1}\cdot \frac{1}{Y_d},\label{eq:factor}
	\end{align}
where  $y$ is the number of years over the total time horizon, and $Y_d$ is the number of days (e.g., 365) in one year.
For example,  Tesla Powerwall's price is 6500\$ for 13.5 kWh with the warranty of 10 years \cite{Teslap}. Here, if we set the annual interest rate $r=5\%$, we can calculate $r^f=3.55\times 10^{-4}$. Then, $\theta_i=r^f\cdot 6500/13.5=0.171 \$/$kWh.

\subsection{Electricity utility}

The utility sets the ToU pricing for users and bears the energy supply cost of meeting users'   demand. We assume that  the utility is regulated \cite{fundamentals}, which aims to maximize the social welfare, i.e., minimize the social cost. Next, we  introduce the model of ToU pricing and supply cost for the utility.

\subsubsection{ToU pricing}
The  ToU pricing  is announced once and is valid for the entire investment horizon. We assume that peak hours and off-peak hours are given as parameters, with $H^{p}$ hours for the peak period and $H^{o}$ hours for the off-peak period, where $H^{p}+H^{o}=24$.   For example, the peak period can be set  from
4PM to 9PM and the off-peak period can be from  10PM to 3PM\cite{timeofusey}, hence $H^p=6$ and $H^o =18$.  Such division is based on the historical observations of the energy loads in the network.  The utility decides the electricity price $p^p$  for the  peak period and the price $p^o$ for the  off-peak period for all users, with $p^p\geq p^o$.  

\subsubsection{Energy supply cost}We consider a quadratic supply cost, which is commonly used for thermal power plants \cite{touren}. 

Note that the power consumption here is the aggregated load in the system. The supply cost for power $p_t$ in hour $t$ is given by $g(p_t)=\alpha p_t^2+\beta p_t+ \gamma$, where the coefficients $\alpha>0$, $\beta\geq 0$ and   $\gamma\geq 0$ are based on practical measurements  given in the literature, such as in  \cite{wu2011generationcost}.  

Our model focuses on the two-period ToU pricing in practice, which charges users based on their total demands in the peak period and off-peak period, respectively. The two-period pricing does not directly regulate users' demand in each hour. To calculate the supply cost based on the total demand in the peak and off-peak periods, we adopt an approximation of constant load in each period. Specifically,  we approximate the power of the peak period  and off-peak  period (with multiple hours) by the average power (in MWh per hour) in these periods,  respectively. For example,  for  the  peak period of 12 hours  with total load 12 MWh, we use an average load of 1MWh per hour. The main purpose of such an approximation is to capture the load difference between the peak  period and off-peak period for the two-period pricing structure.\footnote{Based on realistic data, we can show that such an assumption of two-period constant power can still provide a good approximation for the more elaborate model of 24-hour variable load in terms of  the supply cost.  The supply cost under the 2-period constant-load approximation has a small gap of  6.2\% comparing with the supply cost computed based on the 24-hour variable load. This shows that the 2-period constant-power approximation is quite accurate in terms of predicting the total supply cost. We show more details about this approximation in Appendix.G.}
Then, for the peak period, if the total load is $L^{p}$ in the system, the average load is approximated by $L^{p}/H^{p}$. The total peak period's supply cost $g^{p}$ is given by
\begin{align}
g^{p}(L^{p})=H^p g\left(\frac{L^p}{H^p}\right) =\frac{ \alpha }{H^{p}}(L^{p})^2+\beta L^{p}+ \gamma H^{p}.
\end{align}
Similarly, the  total supply cost for the load  $L^{o}$ in the off-peak period is 
\begin{align}
g^{o}\left(L^{o}\right)= \frac{\alpha}{H^{o}} (L^{o})^2+\beta L^{o}+ \gamma H^{o}.
\end{align} 

\section{Two-stage optimization formulation} \label{section:twostage}

\begin{figure}[t]
	\centering
	\includegraphics[width=2.9in]{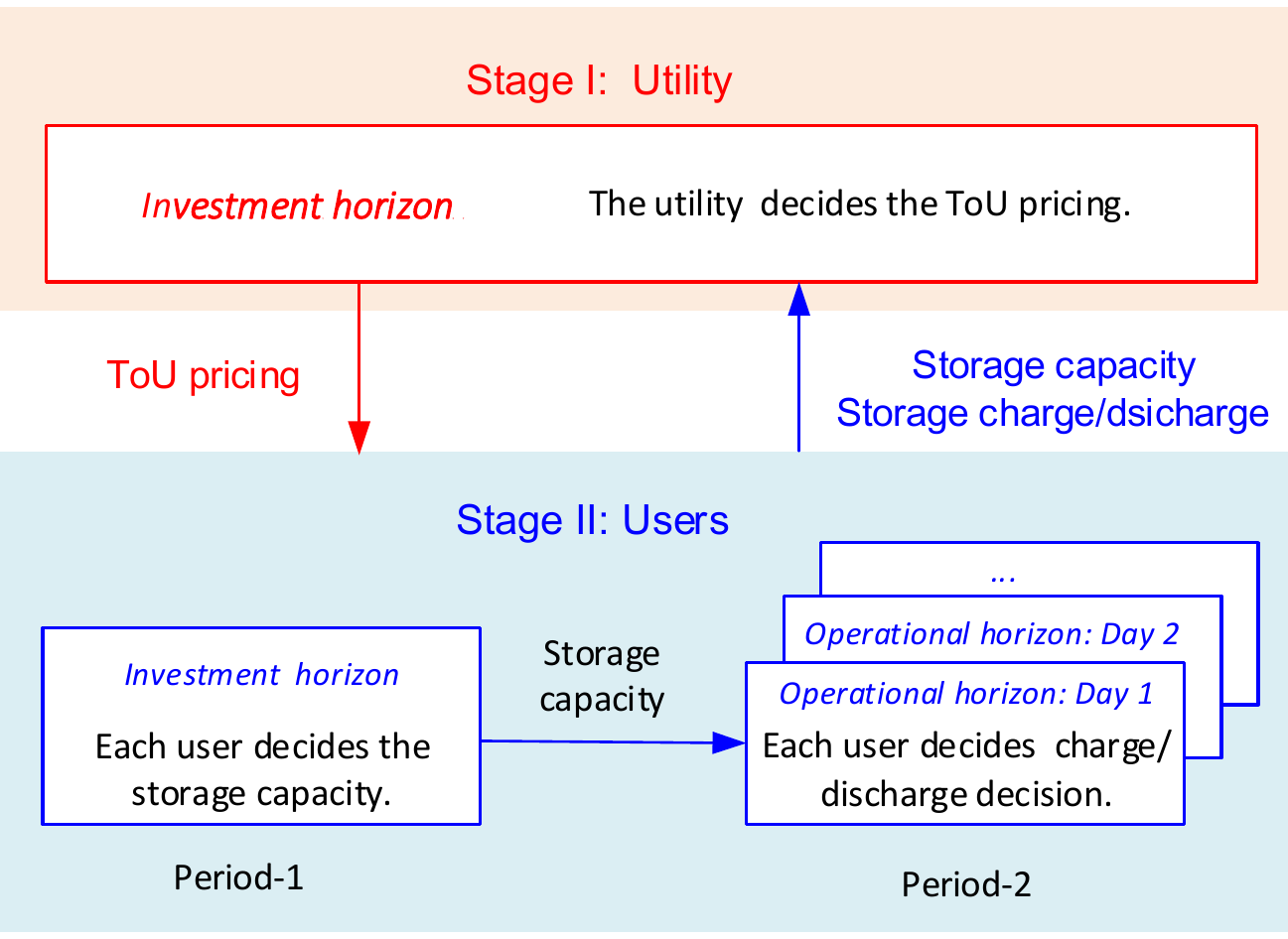}
	\caption{\small Two-stage optimization.}
	\label{fig:stage}
\end{figure}

To decide the optimal ToU pricing, we formulate a two-stage optimization problem between the utility and users, as illustrated in Figure \ref{fig:stage}. Recall that we consider two timescales of decision-making: investment horizon and operational horizon. In Stage I, before the investment horizon, the utility decides  the peak and  off-peak prices of ToU pricing  to minimize the social cost.  In Stage II, at the beginning of the investment horizon (Period-1),   each user decides the storage capacity to invest in.  Then, for each operational horizon (Period-2), each user decides the storage charging and discharging decision. Each user aims to minimize his expected energy cost over the investment horizon.

We can model such a two-stage optimization problem as a dynamic game with incomplete information. The challenges of analyzing such a game  are twofold. First, the social cost includes individual users' storage investment costs, which can be users' private information not known by the utility. Second,  the utility's optimization problem is non-convex even if the utility knows individual users' private information.

To solve the private information problem, we will first define the storage types based on statistical information of storage costs. Then, we formulate a  pricing problem for the utility based on the type information, which does not require users' private information. To solve the two-stage optimization problem, later in Section \ref{section:solution}, we will first characterize the structure of the social cost for the utility based on backward induction and then propose an efficient algorithm by searching a finite set of threshold points.

\subsection{Storage type} 

We consider  a set  $\mathcal{K}=\{1,2,\ldots,K\}$ of storage types, corresponding to different storage costs available in the market. The unit daily cost of storage capacity for type $k$ is  $\theta^k$. 
We rank the storage types in an increasing order of the storage costs, i.e., $\theta_1<\theta_2<\dots<\theta_K$. Each user's type is determined by the storage type that he can obtain. Multiple users can belong to the same type. 

Similar to the individual user's demand, we denote random daily aggregate peak and off-peak demand for a (user or storage) type $k$ as  $\mathcal{D}_k^{p}$ and $\mathcal{D}_k^{o}$, respectively. We let $\bm{\mathcal{D}}_k=(\mathcal{D}_k^{p},\mathcal{D}_k^{o})$ be the vector of the random daily demand for type $k$, and let  $\tilde{\bm{\mathcal{D}}}=(\bm{\mathcal{D}}_k,\forall k \in \mathcal{K})$ be the vector of all types'  peak and off-peak demand.	We assume that the random variable  $\mathcal{D}_k^{x}$  has a CDF  $F_k^{x}$ over the support of $[\underline{\mathcal{D}}_k^{x},\overline{\mathcal{D}}_k^{x}]$, where $x\in\{o,p\}$. Across all the  types, we assume a joint CDF $\tilde{F}$  for the random vector  $\tilde{\bm{\mathcal{D}}}$.

We consider two different information structures  for the utility. In the first case, the utility knows each individual user's storage cost  and  demand distribution. In the second more realistic case, the utility only knows each type's storage cost and aggregate demand distribution, without knowing each individual user's information. Such aggregated information can be obtained  through surveys,  historical data of storage incentive programs\cite{storageprogram2019utility}, or market share of different storage products \cite{storagemarket2019residential}.

Under the two information structures, we propose the following two  pricing schemes for the utility. The first one (\textbf{PI}) is based on  each \textbf{I}ndividual user's information and the second one (\textbf{PT}) is based on each \textbf{T}ype's information.

\begin{itemize}
	\item  \textit{Pricing scheme based on individual's information (\textbf{PI})}: In Stage I, the utility decides the ToU pricing based on   each individual user's  storage  cost and joint demand distribution among users. In Stage II, each user decides the optimal storage capacity and operation based on the ToU pricing and individual user's demand information.
	\item  \textit{Pricing scheme based on type's information (\textbf{PT})}:  In Stage I, the utility decides the ToU pricing based on   each type's  storage  cost and joint demand distribution among types. In Stage II, each type decides the optimal storage capacity and operation based on the ToU pricing and the type's demand information.
\end{itemize}

Note that for both pricing schemes   \textbf{PI}  and  \textbf{PT}, after receiving the ToU pricing, each individual user will invest and operate the storage based on his own storage cost and demand in practice. In  \textbf{PT},  when the utility designs the ToU pricing in Stage I, it considers each type's information and predicts its storage investment decisions as an aggregate in Stage II. However, once the ToU pricing is announced, each individual user still  makes his own storage investment decision based on his own information based on the ToU pricing. Therefore, compared with \textbf{PT}, the pricing scheme \textbf{PI} is more accurate in designing the ToU pricing. However, it requires  each user's private information, which can be difficult to implement in practice.  In \textbf{PT}, the utility only needs to know each type's aggregated demand and storage cost.  In this sense, the pricing scheme \textbf{PT}
 is more flexible than \textbf{PI} and requires less information, which we refer to as ``information loss". As a result, the pricing scheme \textbf{PT} can only achieve a sub-optimal performance in  designing the ToU pricing compared with \textbf{PI}.

The modeling and solution method are similar for the pricing schemes \textbf{PI} and \textbf{PT}, so we will focus on the discussion of  \textbf{PT}. To derive the modeling and solution method of  \textbf{PI}, we just treat each user as one type, by  replacing each  type $k$'s information $(\theta_k,\mathcal{D}_k^p,\mathcal{D}_k^o)$ and decisions  variable $(c_k,s_k)$ in \textbf{PT}  by each user $i$'s information  $(\theta_i,\mathcal{D}_i^{p},\mathcal{D}_i^{o})$ and decision variable $(c_i,s_i)$, respectively. 
In Sections \ref{section:analysis} and \ref{section:simulation}, 
we will also compare the performance of \textbf{PI} and \textbf{PT} with the social optimum.

\subsection{Stage II: Type $k$'s cost minimization}
In Stage II, each type needs to make decisions in two periods. In period-1,  i.e., at the beginning of the investment horizon, each type decides the optimal storage capacity. In period-2,  i.e., for each operational horizon (day), based on the invested capacity, each type decides the optimal storage charging and discharging decision for each demand realization.
	 
The overall objective of each type $k$ is to minimize  its energy cost (scaled into one day),  which includes the electricity bill  and the cost of storage investment (scaled into one day). We first introduce types'  storage investment cost and  electricity bill, and then formulate types' optimization problem.

\subsubsection{Storage investment cost} At the beginning of the investment horizon,  type $k$ decides the invested storage capacity $c_k$. Recall  the  unit daily capacity cost of storage for type $k$  denoted by  $\theta_k$ per day. Thus,  type $k$'s daily storage cost is $\theta_k c_k$. 

\subsubsection{Electricity bill}
We will first discuss the  electricity consumption of types with storage, and then calculate the electricity bill. For each realization $\bm{D}_k$ of random demand $\bm{\mathcal{D}}_k$,    in the off-peak period,  if type $k$ purchases $s_k$ amount of  energy from the utility only for the purpose of charging the storage,\footnote{The payment in ToU pricing is  based on the total energy consumption in peak and off-peak periods, which does not consider demand variation across hours. Thus, we use only  $s_k$ to denote the total charge and discharge energy. We  assume  types' charge and discharge of storage across hours can be regulated by the utility \cite{storageprogram2019utility}, so as to smooth the system load.} the total electricity consumption from the utility will be $D_k^o+s_k$. Here, the charge decisions $s_k\geq 0 $ is constrained  by storage capacity, i.e., $s_k\leq c_k$. As a result, in the peak period, the total consumption from the utility will be $D_k^p-s_k\geq 0$.\footnote{We do not consider the negative demand in the current model, i.e., we do not allow types to sell back energy from the storage to the utility \cite{storageprogram2019utility}.}   All the energy charged into the   storage during the off-peak period will be discharged to serve demand in the peak period.\footnote{In the main text, we consider the perfect charge and discharge efficiency, and no degradation cost of the storage. We generalize our model in Appendix.I, which further incorporates the imperfect charge and discharge efficiency as well as linear degradation cost (with respect to charge and discharge amount). }
Then, type $k$'s  electricity bill is  $ p^p (D_k^p-s_k)+p^o (D_k^o+s_k)$ for a demand realization $\bm{D}_k$. Therefore, given the storage capacity $c_k$, type $k$ minimizes the electricity bill in Period-2  for each demand realization $\bm{D}_k$ as follows.
\vspace{-1mm}
\begin{align}
\text{(Period-2)}~{Q}(c_k,\bm{D}_k):= \min~& p^p (D_k^p\hspace{-0.8mm}-\hspace{-0.8mm}s_k)\hspace{-0.8mm}+\hspace{-0.8mm} p^o (D_k^o\hspace{-0.8mm}+\hspace{-0.8mm}s_k)\\
~\text{s.t.~} &0\leq s_k\leq c_k, \\
&s_k\leq D_k^{p}, \\
\text{var}:&~ s_k.\notag
\end{align}\par{\vspace{-1mm}}
\noindent Given the pricing $\bm{p}=(p^p,p^o)$,  we denote type $k$'s optimal charging decision as $s_k^*(\bm{p},\bm{D}_k)$ for the demand realization $\bm{D}_k$.

Combining the storage investment cost and electricity bill, we formulate Problem \textbf{PT-Stage-II} for type $k$, which minimizes its total energy cost (scaled to one day).

\noindent \textbf{Problem {PT-Stage-II}:  Type $k$'s Cost Minimization}
\vspace{-1mm}
\begin{align}
\text{(Period-1)}~ \min ~&\theta_k c_k+\mathbb{E}_{\bm{\mathcal{D}}_k}[{Q}(c_k,\bm{\mathcal{D}}_k)]\\
~\text{s.t.} ~&c_k\geq 0,\\
\text{var:} ~&c_k.\notag
\end{align}

Problem \textbf{PT-Stage-II} is a two-period  stochastic programming problem, which will be solved  in Section \ref{section:solution}. Given the ToU pricing $\bm{p}$, we denote the optimal solution of type $k$'s storage capacity as $c_k^*(\bm{p})$. 

\subsection{Stage I: Utility's  pricing problem} 
Before the investment horizon, the utility decides the optimal ToU pricing $p^p$ and  $p^o$ for  all the types, which aims to minimize the social cost (scaled into one day).

The social cost includes the total storage investment cost and the supply cost for satisfying types' demands. The storage investment cost over the investment horizon is $\sum_{k\in\mathcal{K}}\theta_k c_k$, where $c_k$ is type $k$'s storage capacity  in Stage II. The supply cost is based on all the types' aggregated load profiles as well as the storage charging and discharging decisions over the operational horizon.  For each demand realization $ \bm{{D}}$, the supply cost is  $G(\bm{s},\bm{{D}}):= g^p\left(\sum_{k\in\mathcal{K}} (D_k^{p}-s_k(\bm{D}_k))\right)+g^o\left(\sum_{k\in\mathcal{K}} (D_k^{o}+s_k(\bm{D}_k))\right)$.
 
We formulate the utility's optimization problem \textbf{PT-Stage-I} under the pricing scheme \textbf{PT} as follows. 

\noindent \textbf{Problem PT-Stage-I: Type-based Pricing for  Social Cost Minimization }
\vspace{-1mm}
\begin{align}
\min~&  \sum_{k\in\mathcal{K}}\theta_k c_k(\bm{p})+\mathbb{E}_{\bm{\mathcal{D}}}~  G(\bm{s}(\bm{p},\bm{\mathcal{D}}),\bm{\mathcal{D}})\label{eq:stage1}\\
\text{s.t.} ~&{p^p\geq p^o\geq 0}\\
\text{var:}~ &{p^p,p^o}\notag,
\end{align}
where the invested capacity ${c}_k(\bm{p})$, and charging and discharging decision  $s_k(\bm{p},\bm{D}_k)$ are  type $k$'s decisions in Stage II.

In the next section, we solve the two-stage problem through backward induction.  We first characterize the solution in Stage II, and then solve the utility's  pricing problem in Stage I.

\section{Solution method for  utility's pricing problem}\label{section:solution}

The utility's pricing problem  is non-convex with the two-stage hierarchical  structure and  challenging to solve \cite{colson2007overview}. We adopt backward induction and characterize the solution structure  to solve the problem. We will first characterize  each type's optimal solution  (in Stage II) under an arbitrary fixed ToU pricing. Then, we  incorporate types' decisions into Stage I to characterize the properties of the social cost,  and propose an algorithm to determine the  optimal ToU pricing. We present  the proofs of all mathematical results  in Appendix.A-D.

For the solutions in both Stage II and Stage I, we will first consider a general distribution of type's demand, and then focus on a  discrete distribution of type's demand. The discrete distribution is  much more common in the decision-making of electricity planning based on the realistic data of load and renewable energy \cite{dai2017optimum}.   The discrete distribution can also make the computation tractable, as we will show that the utility only needs to search a set of threshold points, the size of which is linear in the number of demand outcomes and types. Furthermore, even given a continuous distribution, we can approximate it using the discrete distribution \cite{kennan2006note}\cite{Kazempour2018Stochastic}.

\subsection{Storage deployment solution of Stage II}

We will first solve the Stage-II problem under a general distribution of type's demand. Then, we focus on the solution under a  discrete distribution of type's demand. 
  
\subsubsection{Storage deployment under a general demand distribution}
We define the price difference between peak and off-peak price as $p^\Delta\triangleq p^p-p^o$. We characterize the optimal storage capacity $c_k^*$ and charging/discharging decision $s_k^*$ of type $k$ in Stage II as a  functions of $p^\Delta$  in Proposition \ref{prop:usercontinuous}.

\begin{prop}[type $k$'s optimal solution with a general demand distribution] Under a given $p^\Delta$,  the optimal solution of Stage II is as follows.\footnote{Here we adopt the generalized inverse distribution function: $F_k^{p^{-1}}(z)=\inf \{x\in \mathbb{R} :F_k^{p}\geq z \}$, which can be applied to the case when the CDF is not strictly increasing, e.g.,  for discrete random variables.}
\begin{itemize}
		\item \textit{Period-1} for $c_k$: 
	\begin{itemize}
		\item If $p^\Delta<\theta_k$, $c_k^*(p^\Delta)=0$.
		\item If $p^\Delta>\theta_k$, $c_k^*(p^\Delta)=F_k^{p^{-1}}\left(\frac{p^\Delta -\theta_k }{p^\Delta }\right)$.
		\item If $p^\Delta=\theta_k$, $c_k^*(p^\Delta)$ can be any value in $ [0, \underline{\mathcal{D}}_k^p].$
	\end{itemize}
	\item \textit{Period-2} for $s_k$ at any demand realization $\bm{D}_k\in\bm{\mathcal{D}_k}$: $s_k^*(p^\Delta,\bm{D}_k)=\min \left(c_k^*(p^\Delta),D_k^p\right)$.

\end{itemize}\label{prop:usercontinuous}
\end{prop}

Proposition \ref{prop:usercontinuous} 
shows that when the  price difference $p^\Delta$ is lower than the storage cost $\theta_k$, the type will not invest in any storage.  When the price difference   $p^\Delta$ is higher than the storage cost $\theta_k$, the optimal storage capacity is increasing with the price difference $p^\Delta$, and is bounded by the type's maximum peak demand. Figure \ref{fig:ill}(a) illustrates the optimal capacity $c_k^*$ as a function of $p^\Delta$, when  the CDF $F_k^p$ of peak demand is strictly increasing and continuous.

\subsubsection{Storage deployment under discrete demand distribution} 
We define the discrete  random variable  $\mathcal{D}_k^x$ of type $k$ over a sample space  $\Omega_k^x$, where $x\in \{p,o\}$. Each outcome ${D}_k^{x,\omega}$, for $\omega\in \{1,2,3,\ldots, \mid\Omega_k^x\mid\}$, occurs with probability  $\rho_k^{x, \omega}$. We denote the  sample space of the joint peak and off-peak demands across all the types as $\Omega$.

To characterize the solution of  type $k$, we first sort the outcomes of its peak demand in an increasing order,  i.e.,    $D_k^{p,1}\leq D_k^{p,2}\leq \ldots\leq D_k^{p,\mid \Omega_k^p\mid}.$ We  characterize the type's optimal solution in Proposition \ref{prop:userdiscrete}. For ease of exposition, we define $D_k^{p,0}=0$, and $\sum_{\omega=x}^{y} \rho^{\omega}=0$ if $x>y$. We later also use $\rho^{\omega}$ for $\rho_k^{p, \omega}$ for simplicity.

\begin{prop}[Type $k$'s optimal solution with discrete demand distribution] Given a fixed $p^\Delta$,  the optimal solution of Stage II is as follows.
\begin{itemize}
	\item \textit{Period-1} for $c_k$: 
			\begin{itemize}
	\item If $p^\Delta<\theta_k$, $c_k^*(p^\Delta)=0$.
	\item If $p^\Delta\geq \theta_k$, for $1\leq m \leq \mid \Omega_k^p\mid$:
	\begin{itemize}
		\item 	 $c_k^*(p^\Delta)={D}_k^{p,m}$, if there exists $m$ such that $\sum_{\omega=m}^{\mid \Omega_k^p\mid} \rho^{\omega} p^\Delta> \theta_k$ and  $\sum_{\omega=m+1}^{\mid \Omega_k^p\mid} \rho^{\omega} p^\Delta<{\theta_k}$. 
		\item  	 $c_k^*(p^\Delta)$ can be any value in $  [{D}_k^{p,m-1},{D}_k^{p,m}]$, if there exists $m$ such that $ p^\Delta= \frac{\theta_k}{\sum_{\omega=m}^{\mid \Omega_k^p\mid} \rho^{\omega}}$. 
	\end{itemize}
	Note that the optimal $m$ always exists.
\end{itemize}
	\item \textit{Period-2} for $s_k$: $s_k^{\omega*}(p^\Delta)=\min \left(c_k^*(p^\Delta),D_k^{p,\omega}\right)$.
	\end{itemize}
 \label{prop:userdiscrete} 
\end{prop}

Proposition  \ref{prop:userdiscrete} shows that the optimal storage capacity $c_k^*(p^\Delta)$ is a step-wise function of the price difference $p^\Delta$. For type $k$'s step-wise function $c_k^*(p^\Delta)$,  we construct the set  $\mathcal{P}_k$  of thresholds points for $p^\Delta$  as follows.
\begin{align}
 \mathcal{P}_k=\{0\}\bigcup \left\{ \frac{\theta_k}{\sum_{\omega=m}^{\mid \Omega_k^p\mid} \rho^{\omega}} ,\forall m=1,2,\ldots,  \mid \Omega_k^p\mid \right\}.\label{eq:threshold_price}   
\end{align}
We let $P_k^m=\theta_k/\sum_{\omega=m}^{\mid \Omega_k^p\mid} \rho^{\omega},\forall m\geq 1$. Note that $P_k^1=\theta_k$.

\begin{figure}[t]
		\centering
		\hspace{-1ex}
		\subfigure[]{
			\raisebox{-2mm}{\includegraphics[width=1.53in]{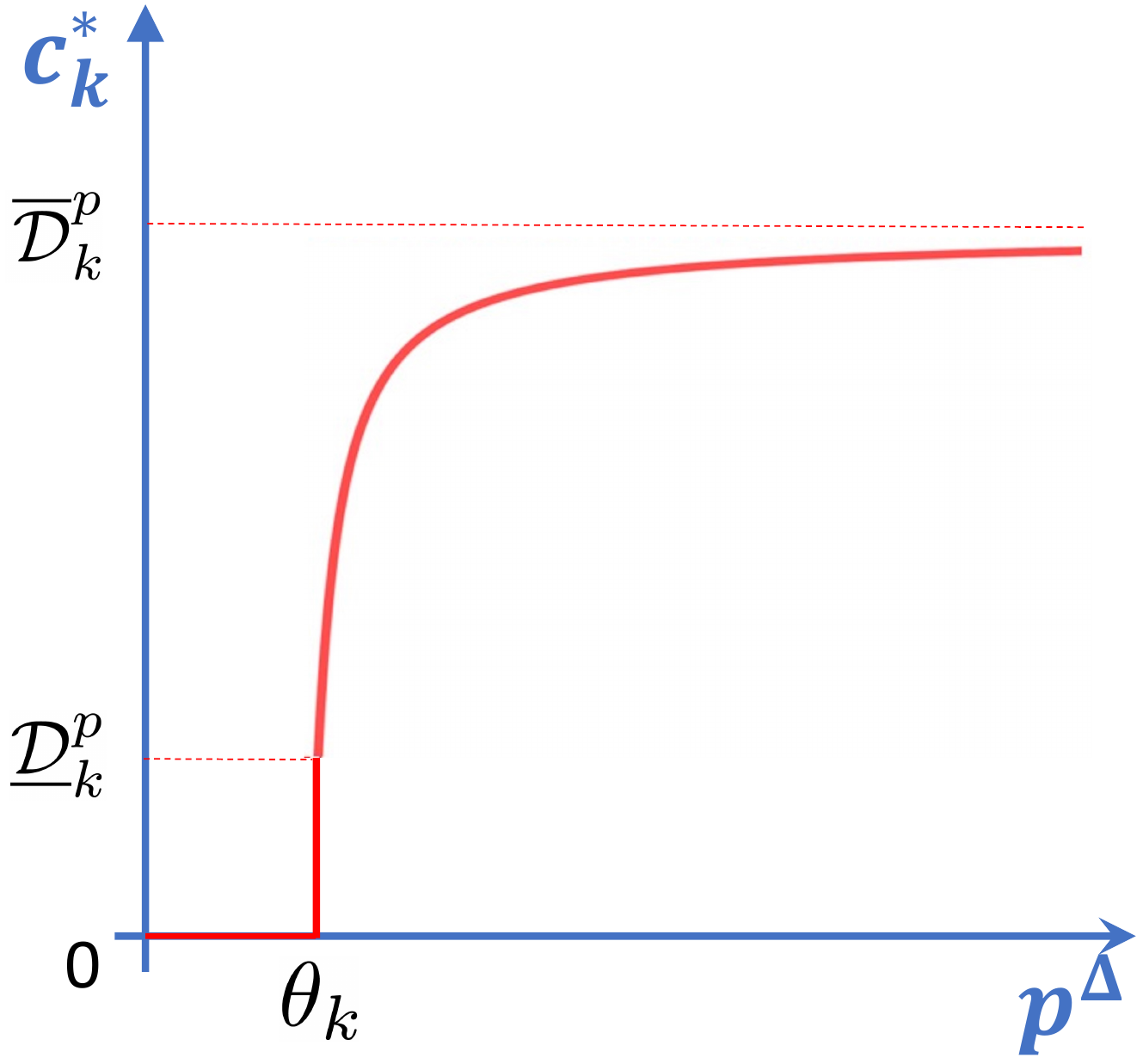}}}
		\hspace{-2ex}
		\subfigure[]{
			\raisebox{-2mm}{\includegraphics[width=1.72in]{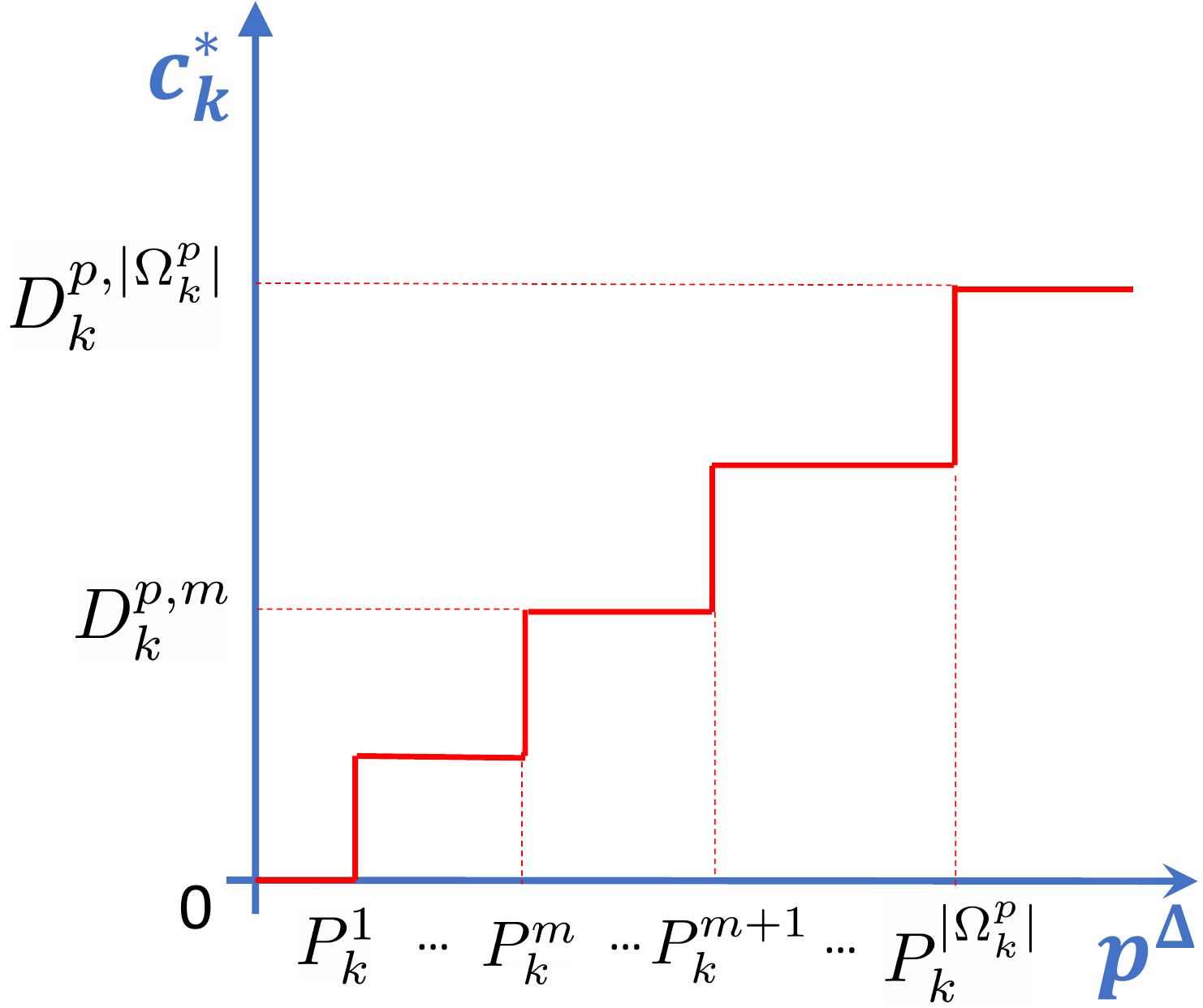}}}
		\vspace{-2mm}
		\caption{(a) \small Type $k$'s optimal capacity $c_k^*$ with $p^\Delta$ under a strictly increasing and continuous CDF $F_k^p$. (b) type $k$'s optimal capacity $c_k^*$ with $p^\Delta$ under a discrete peak-demand distribution.}
		\label{fig:ill}
		\vspace{-3mm}
	   \end{figure}

We illustrate such a step-wise property in Figure \ref{fig:ill}(b). 
The optimal storage capacity $c_k^*$ increases in a step-wise fashion as the price difference  $p^\Delta$ increases. When the price difference  $p^\Delta$  is higher than the threshold  $ \theta_k/\rho^{\mid \Omega_k^p\mid}$, the invested capacity is the maximum value ${D}_k^{p,\mid \Omega_k^p\mid} $ of the peak demands in the sample space.  Note that at each positive threshold-price point, the optimal invested capacity is not unique.

\subsection{Solution method of Stage I}
According to the solution of Stage II, only the price difference $p^\Delta$ will affect the types' decisions. Thus, in Stage I, the utility only needs to decide the optimal price difference $p^{\Delta*}$, while the specific peak price $p^p$ and off-peak price $p^o$ can be flexibly adjusted for regulating the utility's profit.
\subsubsection{Pricing under a general demand distribution}
It is highly  challenging to solve the utility's problem $\textbf{PT-Stage-I}$ based on the general distribution of demands. Since we have closed-form solutions  of Stage II and reduce two pricing variables of (peak and off-peak) into one variable of price difference $p^\Delta$, we can perform a heuristic  exhaustive search by discretizing $p^\Delta$ to find a close-optimal value of  $p^\Delta$. 

\subsubsection{Pricing under discrete demand  distribution}  

Based on the step-wise structure of the types' decisions in Stage II as shown in Proposition \ref{prop:userdiscrete},  we further analyze the structure of the social cost as a function of  the price difference  $p^\Delta$. We then propose an efficient algorithm to achieve the social optimum.

First, we show a step-wise structure of the social cost with respect to the price difference in Theorem \ref{prop:utility}, which is due to the step-wise solution structure in Stage II.
\begin{thm}[Step-wise structure of social cost]
Under types' optimal decisions in Stage II, the social cost is step-wise in the price difference $p^{\Delta}$, with the threshold set $\bigcup_k \mathcal{P}_k$.

\label{prop:utility}
\end{thm}

Based on Theorem \ref{prop:utility}, we propose  Algorithm \ref{alg:A}  that searches all the threshold prices to find the social optimum. Specifically, the utility first  calculates the set $\mathcal{P}_k$  of  threshold prices from each type $k$, which can be executed in a distributed fashion at the type side based on equation \eqref{eq:threshold_price} (Steps 2-5). Then, the utility searches all the threshold prices from the threshold-price set $\bigcup_{k}\mathcal{P}_k$ to obtain the optimal one (Steps 6-18).  Note that there can be  multiple solutions for $c_k$ at  each threshold price. To eliminate the ambiguity, we choose a sufficiently small parameter $\epsilon>0$.  The utility will search over $p +\epsilon$ for each $p \in \bigcup_{k}\mathcal{P}_k$. Among those search steps  6-18, given the announced price difference $p +\epsilon$ (Step 8), each type computes and report the optimal storage deployment decisions in a distributed fashion (Steps 9-12). Finally, the utility computes the optimal  ${p}^{\Delta*}$   that  minimizes the social cost (Step 19). 
 
When we consider $K$ types and $|\Omega|$ outcomes in the sample space of the joint demand  distribution across types, the utility needs to search at most $K |\Omega|+1$ threshold prices to find the optimal value. Therefore, the computational complexity is  $O(K |\Omega|)$. 
Note that Algorithm \ref{alg:A} is for the pricing scheme \textbf{PT}. The algorithm for \textbf{PI} is similar by  replacing each type’s information and decisions with each user.

	\begin{algorithm}  
	\setstretch{0.95}
	\caption{Computing the optimal price difference  ${p}^{\Delta*}$}
	\label{alg:A}  
	\begin{algorithmic}[1] 
		\STATE \textbf{initialization}: set a sufficiently small error $\epsilon>0$; set a  sufficiently high social cost $SCM^*>0$; set $j=0$ and $p^{\Delta*}=\epsilon$;
		\FOR {type $k\in \mathcal{K}$ \textbf{in parallel}} 
		\STATE  Computes the price threshold set $\mathcal{P}_k$ as in equation \eqref{eq:threshold_price};
		\STATE  Report  $\mathcal{P}_k$ to the utility;	
		\ENDFOR
		\FOR {each $p +\epsilon$, with $p \in \mathcal{P}_a=\bigcup_{k}\mathcal{P}_k$}
		\STATE $j=j+1$;
		\STATE The utility announces  $p^{\Delta(j)}=p +\epsilon$ to all the types.
			\FOR {type $k\in \mathcal{K}$ \textbf{in parallel}}
			\STATE Compute the invested storage capacity $c_k^*\left(p^{\Delta(j)}\right)$ and the charge/discharge decision $s_k^{\omega*}\left(p^{\Delta(j)}\right)$ for each outcome $\omega$ as in Proposition \ref{prop:userdiscrete};
			\STATE Report the results to the utility;	
			\ENDFOR
		\STATE 	The utility calculates the  social cost $SCM^{(j)}=SC$ as in equation \eqref{eq:stage1};
		\IF {$SCM^{(j)}<SCM^*$}
		\STATE $SCM^*=SCM^{(j)}$;
		\STATE $p^{\Delta*}=p^{\Delta(j)}$;
		\ENDIF
		\ENDFOR
		\STATE {\textbf{output}}:  ${p}^{\Delta*}$.
	\end{algorithmic}
	\vspace{-1mm}
\end{algorithm}

\section{Performance analysis of the pricing scheme} \label{section:analysis}

To examine the performance of the pricing schemes \textbf{PT} and \textbf{PI}, we first formulate a benchmark problem \textbf{SO}, where  a social planner centrally decides and controls the optimal storage investment and operation decisions for each user.  Then, we present theoretical comparisons among \textbf{PI}, \textbf{PT},  and   \textbf{SO}. Finally,  we characterize the upper bound for the ratio between the social cost under  \textbf{PT} and the social optimum under \textbf{SO}, when the storage costs approach zero or are sufficiently high.

\subsection{Benchmark: Social optimum}

For the benchmark problem \textbf{SO}, we consider two periods as  in Section \ref{section:twostage}.   In Period-1, i.e., before the investment horizon, the social planner decides the optimal invested storage capacity for each user. In  Period-2, i.e., for each operational horizon, the social planner decides the optimal charging and discharging decisions for each user.

\noindent \textbf{Benchmark SO: Social Optimum by Social Planner}
\begin{align}
(\text{Period-1})~~~	\min ~&\sum_{i\in\mathcal{I}}\theta_i c_i+ \mathbb{E}_{\bm{\mathcal{D}}}~  G(\bm{c},\bm{\mathcal{D}})\\
\text{s.t.~} 
& 0\leq c_i, ~\forall i\in \mathcal{I},\\
\text{var}: ~&\bm{c}.\notag
\end{align}\par{\vspace{-1mm}}
For each demand realization $\bm{D}$ of  $\bm{\mathcal{D}}$, \text{we have}
\begin{align}
(\text{Period-2})~~~G(\bm{c},\bm{{D}}):=	\min~  & g^p\Big(\sum_{i=1}^I (D_i^{p}-s_i)\Big)\notag\\&+g^o\Big(\sum_{i=1}^I (D_i^{o}+s_i)\Big)\\
\text{s.t.~} 
& 0\leq s_i\leq c_i, ~\forall i\in \mathcal{I},\\
& s_i\leq D_i^{p}, ~\forall i\in \mathcal{I},\\			\text{var:}~& s_i, \forall i \in \mathcal{I}.\notag
\end{align}

It is challenging to solve Benchmark \textbf{SO} based on the general  demand distribution. Fortunately, based on the discrete demand distribution, Benchmark \textbf{SO} is a quadratic programming problem whose optimal solution can be efficiently solved\cite{quad}. For the continuous demand distribution, we can adopt discrete approximations for computation \cite{kennan2006note}\cite{Kazempour2018Stochastic}.

We will compare   social costs under the pricing schemes $\textbf{PI}$ and $\textbf{PT}$ with that of the  benchmark \textbf{SO}. We denote the social costs  induced by   \textbf{PT},  \textbf{PI}, and  \textbf{SO} as $SC^{\text{PT}}$, $SC^{\text{PI}}$, and $SC^\text{SO}$, respectively.  Although in the pricing scheme \textbf{PT}, the utility decides the pricing  based on types' information, the actual social cost is calculated based on each individual user's storage decision in response to the announced ToU pricing. We show the comparison of social costs  in Proposition \ref{prop:sc}.

\begin{prop}[Social costs comparison] 
The social costs of the three schemes satisfy
$SC^{\text{PT}}\geq 	SC^{\text{PI}}\geq 	SC^{\text{SO}}.$
	\label{prop:sc}
\end{prop}

\noindent\textbf{Proof}: The optimal storage investment and operation decision induced in the pricing scheme \textbf{PI} is a feasible solution to  Benchmark \textbf{SO}. Thus, we must have $SC^{\text{PI}}\geq SC^{\text{SO}}$.

Note that the social cost is always calculated based on the individual users' information. 
In the pricing scheme \textbf{PI}, the utility designs the ToU pricing based on individual users' information. Thus,  the social cost $SC^{\text{PI}}$ is optimal under the ToU pricing.
 Thus, we always have  $SC^{\text{PT}}\geq SC^{\text{PI}}$. Overall, we obtain  $SC^{\text{PT}}\geq SC^{\text{PI}}\geq SC^{\text{SO}}$.\qed

The gap between   {\textbf{PT}} and  \textbf{PI} is due to the information loss during the aggregation of user's demands of each type.  The gap between the pricing scheme  {\textbf{PI}} and the benchmark \textbf{SO}  is because the ToU pricing may not achieve the social optimum. Next, we show some theoretical results for the gaps between  {\textbf{PT}}, \textbf{PI}, and  \textbf{SO}.

\subsection{Gap analysis among \textbf{PT},  \textbf{PI}, and \textbf{SO}. }

\subsubsection{Comparison between \textbf{PT} and \textbf{PI}} The difference between \textbf{PT} and \textbf{PI} is affected by the correlation of users' demand in each type. In each type, if users' peak demands have perfect positive correlations, the pricing scheme \textbf{PT} will be equivalent to \textbf{PI},  since no information is lost in the aggregation. In Appendix.F, we will further present numerical results when users' peak demands are negatively correlated or weakly positively correlated, which shows that a stronger positive correlation will reduce the gap between the pricing schemes \textbf{PT} and \textbf{PI}.

\subsubsection{Comparison between  \textbf{PI} and \textbf{SO}}  In our simulation results in Section \ref{section:simulation}, we find that the pricing schemes \textbf{PT} and \textbf{PI} often achieve social costs very close to the benchmark \textbf{SO}. One reason behind such results is that \textbf{PI}  and \textbf{PT} can lead to a similar storage investment structure among  storage types, as in the benchmark \textbf{SO}.

To illustrate the insights, we first present the  storage investment structure of different storage types under the benchmark \textbf{SO} in Proposition \ref{prop:benstorage}.

\begin{prop}[Investment structure of \textbf{SO}]
	In the benchmark \textbf{SO} with $\underline{\mathcal{D}}_i^{p}>0$ for each user $i$, we denote the number of users who invest in storage at the optimal solution as $M$. These users belong to $M'(\leq M)$ storage types.  Then,  the storage costs of those users must be  the lowest $M'$ costs $\theta^1< \theta^2<\ldots <\theta^{M'} $ in the system. Users can also be classified into three classes.
	\begin{itemize}
		\item for any user $i$ with  $\theta^{1}\leq \theta_i \leq \theta^{M'-1}$, the optimal  capacity $c_i^{*} $ satisfies  $\underline{\mathcal{D}}_i^{p}\leq  c_i^{*}  \leq \overline{\mathcal{D}}_i^{p}$;
		\item for  any user $i$ with $\theta_i =\theta^{M'}$,   $0\leq c_i^{*} \leq \overline{\mathcal{D}}_i^{p}$;
		\item for users $k$ with $\theta_i \geq\theta^{M'+1}$, $c_i^{*}=0$.
	\end{itemize} 
\label{prop:benstorage}
\end{prop}

We denote the users with the storage cost $\theta^{M'}$ as the \textit{boundary} users, who are the highest-cost users that invest in positive storage capacity. The boundary users' optimal investment capacity can be any value 
between $[0, \overline{\mathcal{D}}_i^{p}]$. Next, we present the storage investment structure of storage types for the pricing scheme \textbf{PI}  in Proposition \ref{prop:pricestorage}.

\begin{prop}[Investment structure of \textbf{PI}]
	In \textbf{PI}, we denote the number of users who invest in storage at the optimal solution as $N$. These users belong to $N'(\leq N)$ storage types.  Then,  the storage costs of those users must be  the lowest $N'$ costs $\theta^1< \theta^2<,\ldots, <\theta^{N'} $ in the system. Further, users can be classified into two classes.
	\begin{itemize}
		\item for any user $i$ with  $\theta^{1}\leq \theta_i \leq \theta^{N'}$, the optimal  capacity $c_i^{*} $ satisfies  $\underline{\mathcal{D}}_i^{p}\leq  c_i^{*}  \leq \overline{\mathcal{D}}_i^{p}$;
		\item for any user $i$ with  $\theta_{i}\geq \theta^{N'+1}$, $ c_i^{*}=0$.
	\end{itemize}
\label{prop:pricestorage} 
\end{prop}

The pricing scheme  \textbf{PT} induces the same structure as \textbf{PI}. Comparing Propositions \ref{prop:benstorage} and  \ref{prop:pricestorage}, we can see that the pricing scheme \textbf{PI} (or \textbf{PT}) can induce a structure of storage investment for users very similar  to  the benchmark \textbf{SO}. In both cases, the low-cost users will invest in a capacity within their peak-demand range, while the high-cost users will invest in no storage. 

However, there are two differences  between  \textbf{PI} (or \textbf{PT})  and  \textbf{SO}.   First,  compared with \textbf{SO} (Proposition \ref{prop:benstorage}), there is no so-called boundary users in \textbf{PT} or \textbf{PI} (Proposition \ref{prop:pricestorage}) due to the limitation of ToU pricing in inducing social optimum. Second, when demands vary across days, the invested capacity can be different between \textbf{PT}, \textbf{PI} and \textbf{SO}, even though they all follow similar structures. In the special case when the peak  demand is fixed across days, i.e., $\underline{\mathcal{D}}_i^{p}= \overline{\mathcal{D}}_i^{p}$ for each user $i$, each user's storage investment in \textbf{PI} (or \textbf{PT}) will be the same as \textbf{SO}, except the possible boundary users \cite{smartgrid2020}. In this case, in \textbf{PI} (or \textbf{PT}), each user will either invest in zero capacity or the amount of peak demand (all-or-nothing).\footnote{For the deterministic demand, we design a contract in the conference paper \cite{smartgrid2020} for users to minimize the social cost  considering the boundary-user impact.}  Later in Section \ref{section:simulation}, we will present simulation results that  show the impact of the demand variance on 
the performance of \textbf{PI} and \textbf{PT}. 
	
\subsection{Performance bound}

	As the pricing scheme \textbf{PT} is the easiest to implement, we are interested in characterizing its relative performance to the benchmark. We define  $\kappa^{\text{PT}}=	SC^{\text{PT}}/	SC^{\text{SO}}$ as the ratio between the social costs under the pricing scheme {\textbf{PT}} and under the social-optimum benchmark {\textbf{SO}}. We characterize  upper bounds of  $\kappa^{\text{PT}}$  for two special cases:   (i) the storage costs approach zero and (ii) the storage costs are sufficiently high.   Later in Section \ref{section:simulation}, we will show more simulation results for the ratio $\kappa^{\text{PT}}$ under different storage costs using realistic data.
	
 Proposition \ref{prop:zerocost}  considers the case where users' storage costs approach zero. 

	\begin{prop}[Zero storage cost] \mbox{}
		When each user's  storage cost approaches zero,  $\kappa^{\text{PT}}$  is upper-bounded as follows.
		$$\kappa^{\text{PT}} \leq \min\left(\frac{H^{p}+H^{o}}{H^{o}},\frac{H^{p}+H^{o}}{H^{p}}\right).$$\par{\vspace{-1mm}}
		\label{prop:zerocost}
The upper bound is tight. 
	\end{prop}
\noindent	Proposition \ref{prop:zerocost}  shows that the worst upper-bound of the ratio is 2, when $H^p=H^o$, i.e., the number of peak hours equal to the number of off-peak hours. If the gap between $H^p$ and $H^o$ is large, the upper-bound is close to 1 and the scheme \textbf{PT} is close to the social optimum. We can construct an extreme example where only one type has positive peak demand in each demand realization, such that $\kappa^{\text{PT}} $ can reach the upper bound. We show the example in detail in Appendix.D.
	
Then, when each user's storage cost is sufficiently high, the ratio $\kappa^{\text{PT}}$  will be 1 since no users will invest in any storage in both the pricing scheme \textbf{PT} and the benchmark \textbf{SO}. 

	\begin{prop}[High storage cost] 
	When each user's storage cost is higher than a certain threshold,  $\kappa^{\text{PT}}$  will be 1.
		\label{prop:highcost}
	\end{prop}

Later in Section \ref{section:simulation}, we will show more simulation results of $\kappa^{\text{PT}}$ under different storage costs using realistic data.

\section{Numerical Study}\label{section:simulation}
We use the realistic data of users' demand in Austin and New York, US \cite{diverseload} to perform the simulation. We will first show the importance of designing the ToU pricing considering the storage impact. Then, we show that the pricing scheme \textbf{PT} achieves good performance with the ratio $\kappa^{\text{PT}}$ always lower than 1.05. Finally,  we investigate the impact of demand variance on the performance of  \textbf{PT} and \textbf{PI}, where a higher demand variance may improve the performance.

\subsection{Simulation setup}

\subsubsection{Load  profile}
\begin{figure}[t]
	\centering
	\hspace{-2ex}
	\subfigure[]{
		\raisebox{-3mm}{\includegraphics[width=1.77in]{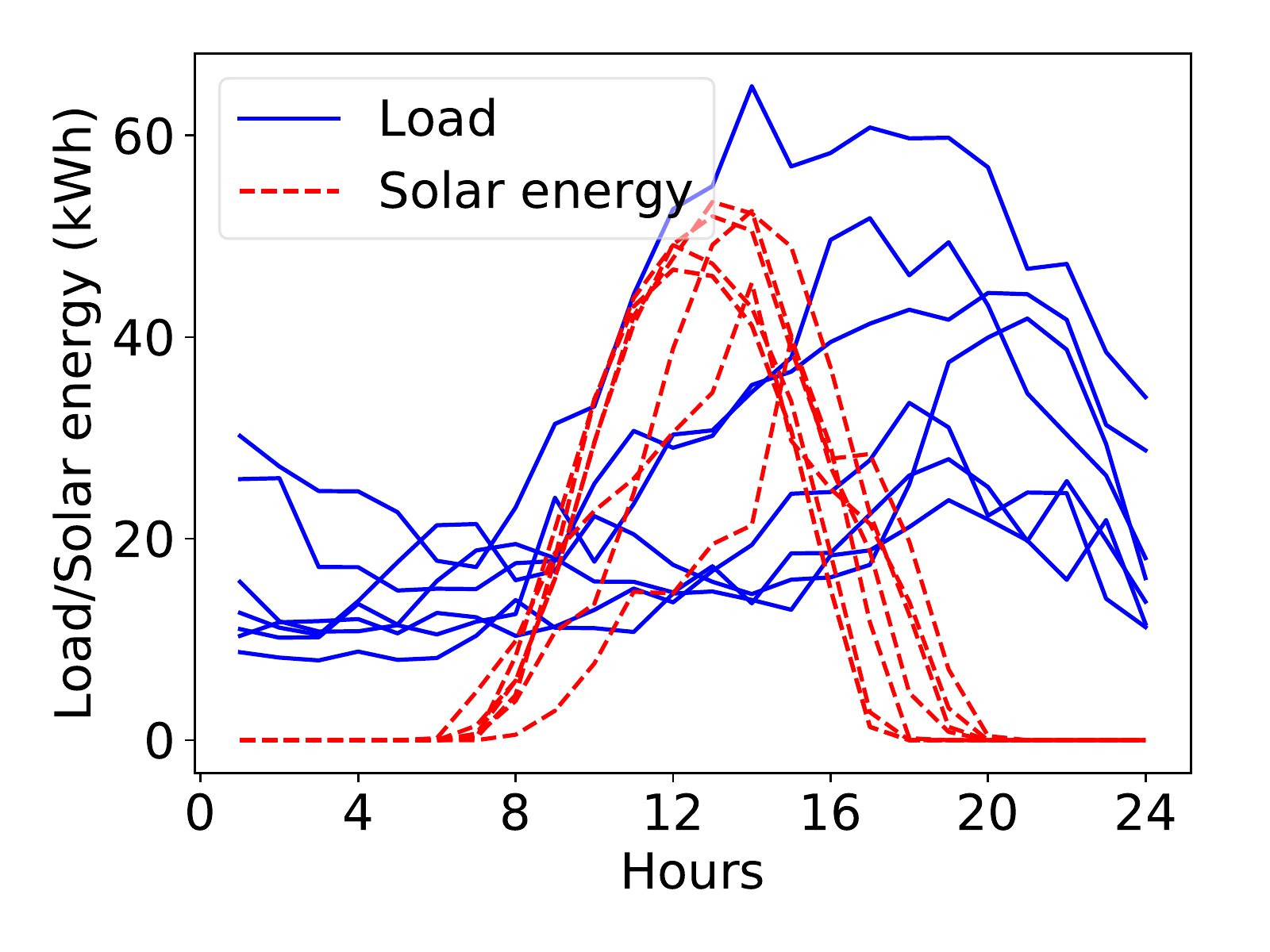}}}
	\hspace{-3ex}
	\subfigure[]{
		\raisebox{-3mm}{\includegraphics[width=1.77in]{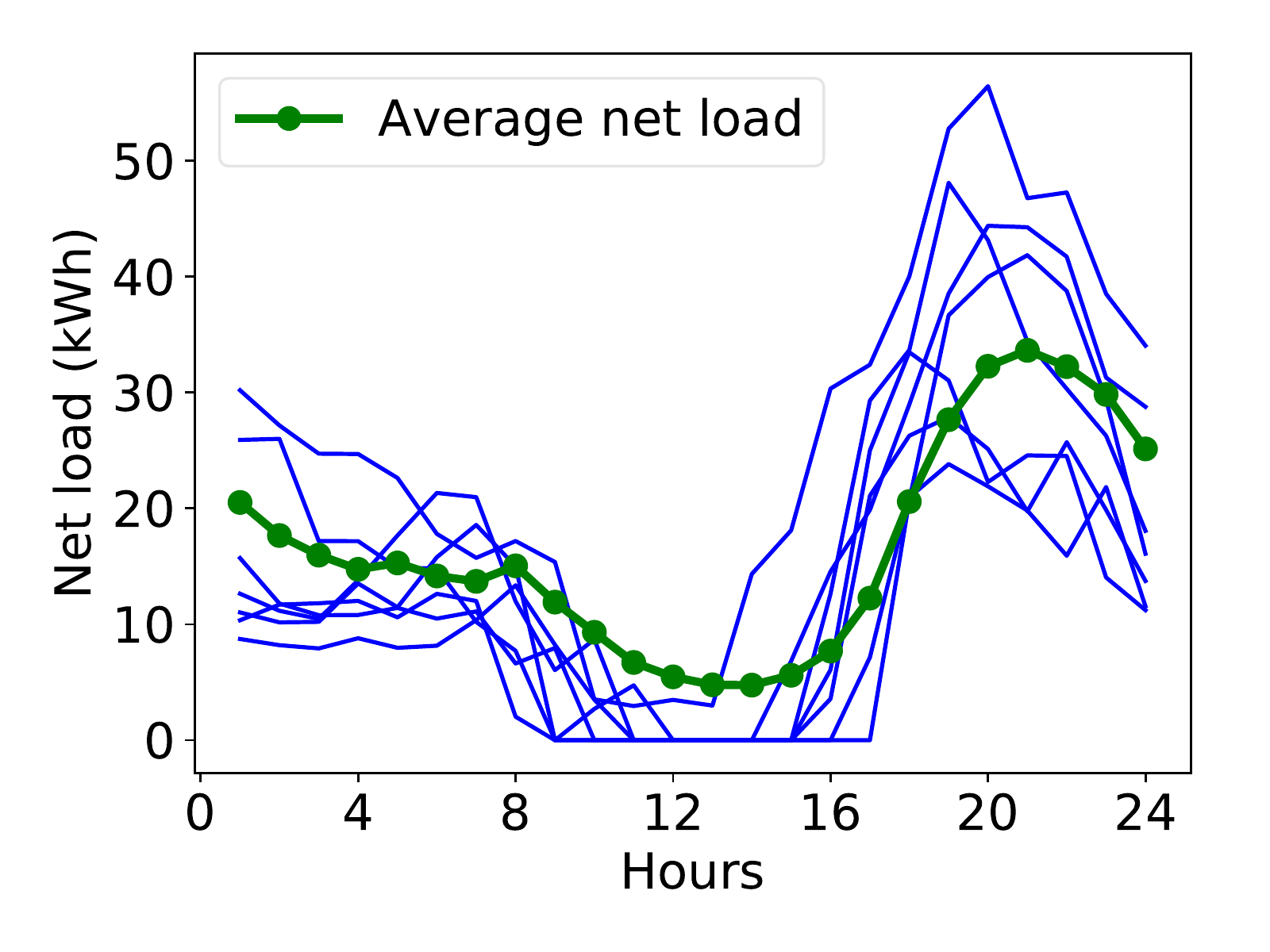}}}
	\vspace{-4mm}
	\caption{\small	(a) Aggregate load /solar energy; (b) Aggregate net load.}
	\label{fig:load}
	\vspace{-5mm}
\end{figure}

Based on the Pecan Street load dataset \cite{diverseload}, we consider  hourly  load and solar energy generations  of 16 (households) users in one year (with valid data for 361 days) from Austin (USA).  In Figure \ref{fig:load}(a), we show the aggregate energy profile with seven randomly picked days in one year, where the blue curves and red curves represent the aggregate loads and solar energy generations, respectively. In Figure  \ref{fig:load}(b), we show the aggregate net load (load minus solar energy)\footnote{We let users curtail the surplus renewable energy in simulations.} of seven randomly-picked days  (blue curves). We also show the mean value of the entire year data in the  green curve. We construct users' demand distribution based on their net load data of the entire year, e.g., 361 joint demand outcomes with a probability 1/361 for each.
\subsubsection{Peak  and off-peak periods of ToU pricing} Based on the average net load of all users in Figure \ref{fig:load}(b), we empirically set the  peak period  from 18:00 to 00:00 (7 hours), and the off-peak period  from  01:00 to 17:00 (17 hours). 

\subsubsection{Storage cost} We consider  4 storage types with the corresponding (daily) investment costs of  $[\theta_1,\theta_2,\theta_3,\theta_4]$ $=\big[\bar{\theta}(1-1.5\delta^s)$ $,\bar{\theta}(1-0.5\delta^s),\bar{\theta}(1+0.5\delta^s),\bar{\theta}(1+1.5\delta^s)\big]$.  The mean value of  the storage costs is $\bar{\theta}$. The coefficient $\delta^s$ indicates the level of storage-cost diversity among types.\footnote {Storage costs can be very diverse. According to  \cite{ralon2017electricity}, the  compressed-air energy storage (CAES) has cheap capital costs about 53-84\$/kWh, with the lifespan of  20-100 years. The Lithium battery's cost is high. Typically,  Tesla Powerwall's price is 6500\$ for 13.5 kWh, with the warranty of 10 years \cite{Teslap}.}  

\subsection{Social welfare loss due to an improperly designed ToU pricing scheme}
	
We show that a properly designed ToU pricing scheme can incentivize users' storage investment and reduce the social cost, while an improper one  may  fail to incentivize users' storage investment and  even  lead to a much higher social cost compared with no storage in the system.

	  \begin{figure}[t]
		\centering
		\hspace{-1ex}
		\includegraphics[width=2.5in]{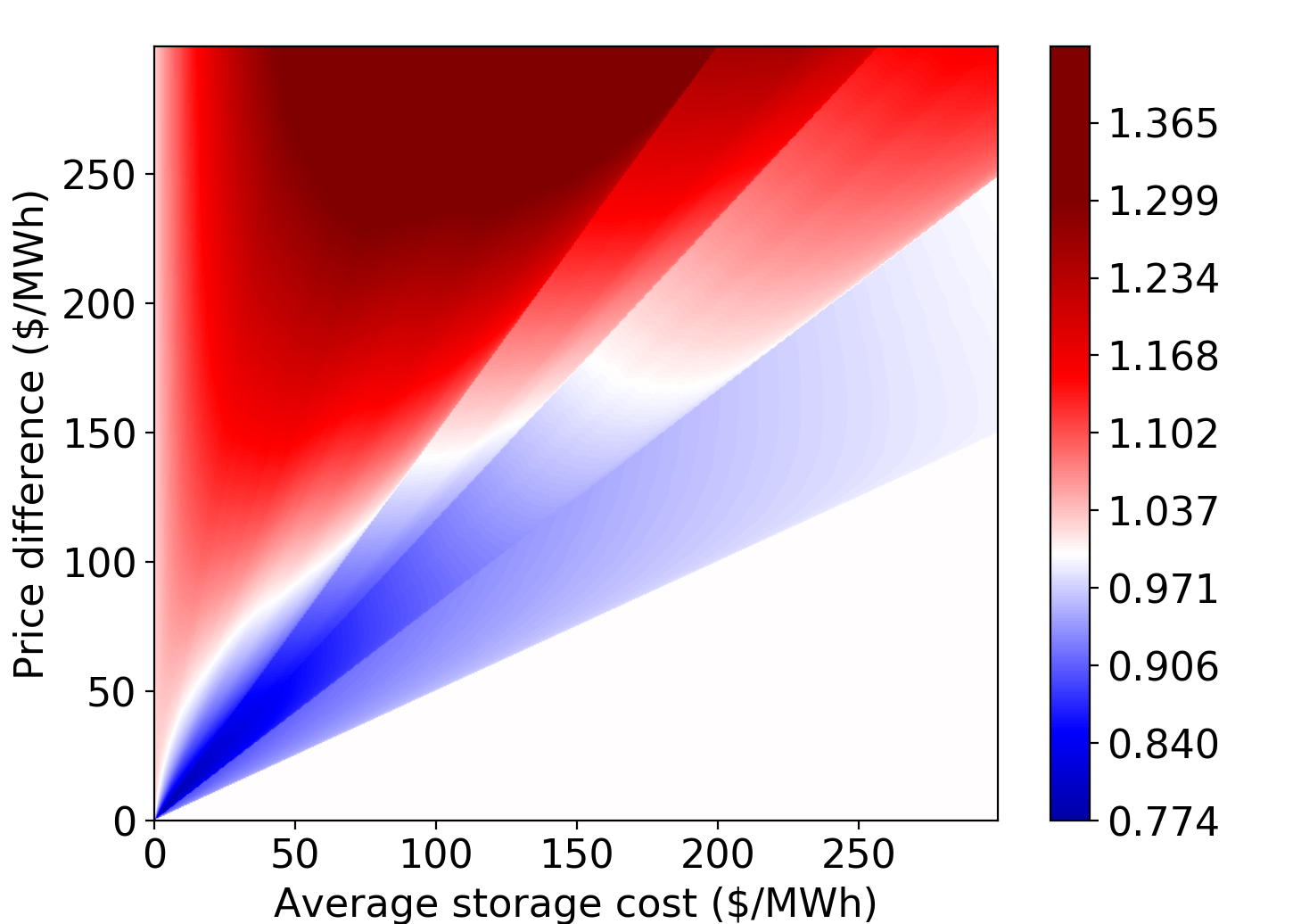}
		\vspace{-0mm}
		\caption{\small Ratio $\lambda$ with price difference and average storage costs.}
		\label{fig:motivation}
		\vspace{-4mm}
	\end{figure} 

We examine a ratio $\lambda$   between  the social cost under various ToU pricing (affecting users' storage investment decisions) and the social cost under no storage in the system. In Figure \ref{fig:motivation}, we show the ratio $\lambda$ in different colors under different price differences and different average  storage costs.

We can see that the figure can be divided into 3 parts: (a) white region in the bottom right; (b) red region in the upper left; (c) blue region in the middle. In the white region (a), the price difference is low compared with the storage cost. Thus, users will not be incentivized to invest in any storage in the ToU pricing and the ratio  $\lambda$ is 1. In the red region (b),  the price difference is high compared with the storage cost. This leads to the over-investment of storage in the system and the ratio  $\lambda$ is higher than 1 (sometimes even higher than 1.35). This shows that an improperly designed high price difference can lead to the over-investment of storage and a much higher social cost,  compared with no storage in the system. In the blue region (c), the price difference is not too high or too low, so it can incentivize proper storage investment to reduce the social cost, which drives the ratio  $\lambda$ below 1 (sometimes even lower than 0.78). Note that between the red region ($\lambda>1$) and blue region ($\lambda<1$), there is a transition of small white space ($\lambda=1$), where the social cost under the positive amount of  storage investment is equal to the no-storage case.

In summary, in the ToU pricing,  a price difference that is too low can not incentivize storage investment, while a price difference that is too high will incentivize too much investment. They can both lead to a high social cost compared with a properly designed price difference in the ToU pricing.

\subsection{Performance of the pricing scheme \textbf{PT}}

We will first show the optimal ToU pricing in the scheme \textbf{PT}. Then, 
we show that the pricing scheme \textbf{PT} can achieve a  good performance with an empirical  $\kappa^{\text{PT}}$  close to 1. Furthermore, we find that the performance is  robust across different data sets, different average  storage costs $\bar{\theta}$, and different storage cost diversities  $\delta^s$, where  $\kappa^{\text{PT}}$ is always less than 1.05.

\subsubsection{Optimal price difference}

\begin{figure}[t]
	\centering
	\subfigure[]{
	\hspace{-1ex}
		\raisebox{-2mm}{\includegraphics[width=1.74in]{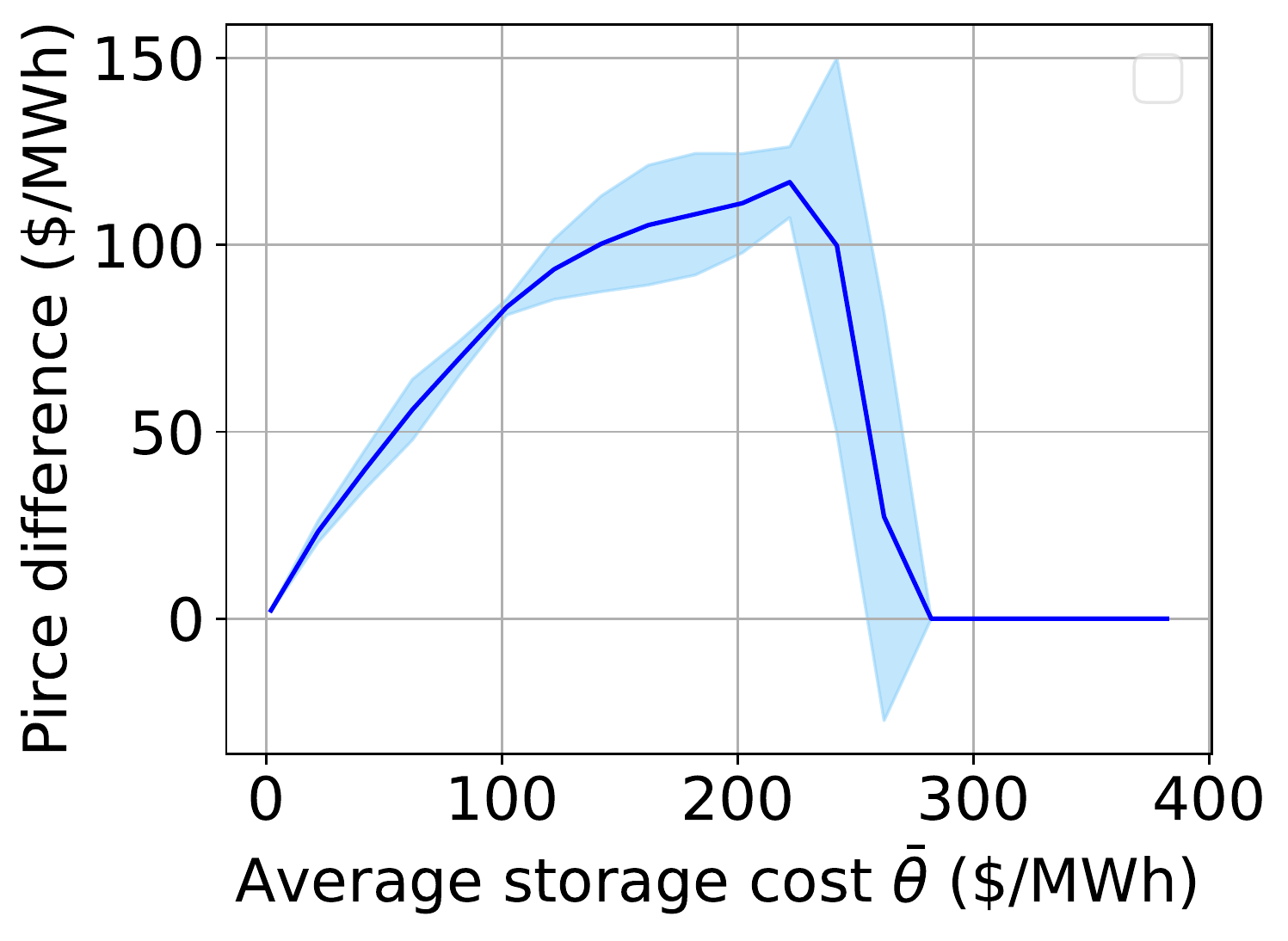}}}
	\hspace{-3ex}
	\subfigure[]{
		\raisebox{-2mm}{\includegraphics[width=1.74in]{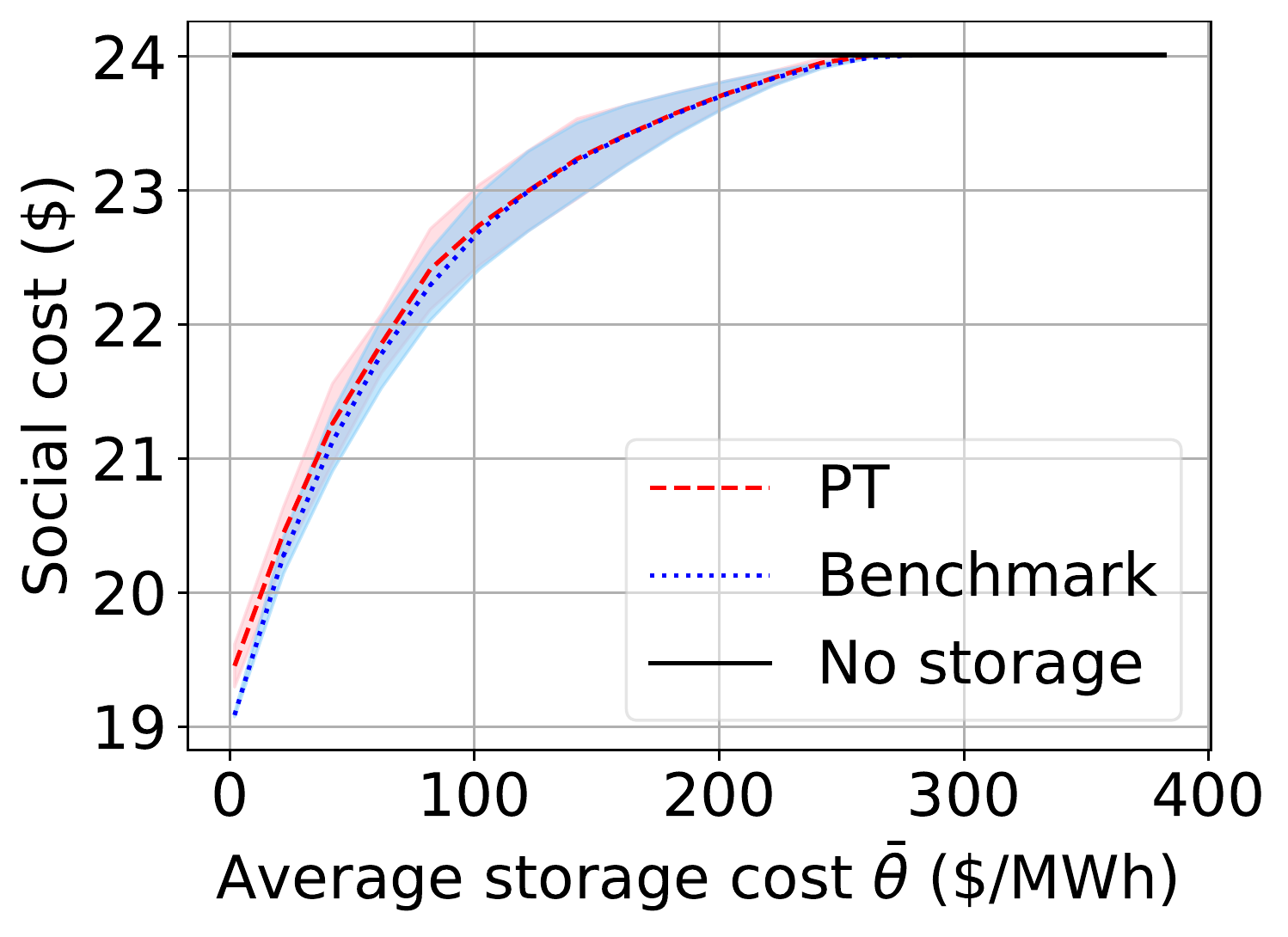}}}
	\caption{\small (a) The optimal price difference with $\bar{\theta}$ from Austin set for multiple type grouping. (b) The social cost with $\bar{\theta}$ from Austin set for multiple type grouping. }
	\label{fig:abs}
	\vspace{-4mm}
\end{figure}

In Figure \ref{fig:abs}(a),  we show  the optimal price difference $p^{\Delta*}$ as the average storage $\bar{\theta}$ increases. We report the overall results for 50 random groupings of 16 users into 4 types using Austin data. We show the mean  value of optimal price difference $p^{\Delta*}$  as well as the one-standard-deviation range  (in shades).

In Figure \ref{fig:abs}(a), we can see that the optimal price difference first increases (when $\bar{\theta}<220$\$/MWh) and then decreases (when $\bar{\theta}\geq 220$\$/MWh) as the average storage cost $\bar{\theta}$ increases. The reason is that, when the storage cost is close to zero, the optimal price difference should also be close to zero, otherwise it will cause the over-investment of storage.  As  the storage cost increases, the price difference will also increase to incentivize the storage investment. However, if the storage cost is too high, the storage investment is no longer beneficial to the social welfare, so the optimal price difference will decrease to prevent users' storage investment. 

\subsubsection{Good performance of \textbf{PT}}
We will first show the social costs in the pricing scheme \textbf{PT}, the social-optimum benchmark, and no-storage case, respectively. Then, in order to more clearly show  the good performance of \textbf{PT}, we examine the ratios between the social cost in  \textbf{PT} and the social cost in the social-optimum benchmark.

In Figure \ref{fig:abs}(b), considering 50 random groupings of 16 users into 4 types using Austin data, we show the mean values of social costs under the no-storage case (black curve), under  the pricing scheme \textbf{PT} (red curve),  and under the social-optimum benchmark (blue curve).  We also show the one-standard-deviation range  (in shades). We can see that the social costs under the pricing scheme \textbf{PT} (red curve) and under the social-optimum benchmark (blue curve) both increase with the  average storage cost $\bar{\theta}$. These two costs are very close. When the storage cost is too high ($\bar{\theta}>250$\$/MWh), no storage will be invested under both  \textbf{PT} and  the social-optimum benchmark, which will be the same as the no-storage case.

Then, to further show  the good performance of \textbf{PT}, we examine the ratios between the social costs. Similar to the definition of ratio $\kappa^{PT}$, we define  $\kappa^{\text{PI}}=	SC^{\text{PI}}/	SC^{\text{SO}}$ as the ratio between the social costs under  {\textbf{PI}} and under the benchmark {\textbf{SO}}, and  $\kappa^{\text{no}}=SC^{\text{no}}/	SC^{\text{SO}}$ as the ratio between the social costs under no storage in the system and under the benchmark  {\textbf{SO}}.

	\begin{figure}[t]
	\centering
	\subfigure[]{
		\label{fig:subfig:ra1} 
		\raisebox{-2mm}{\includegraphics[width=1.7in]{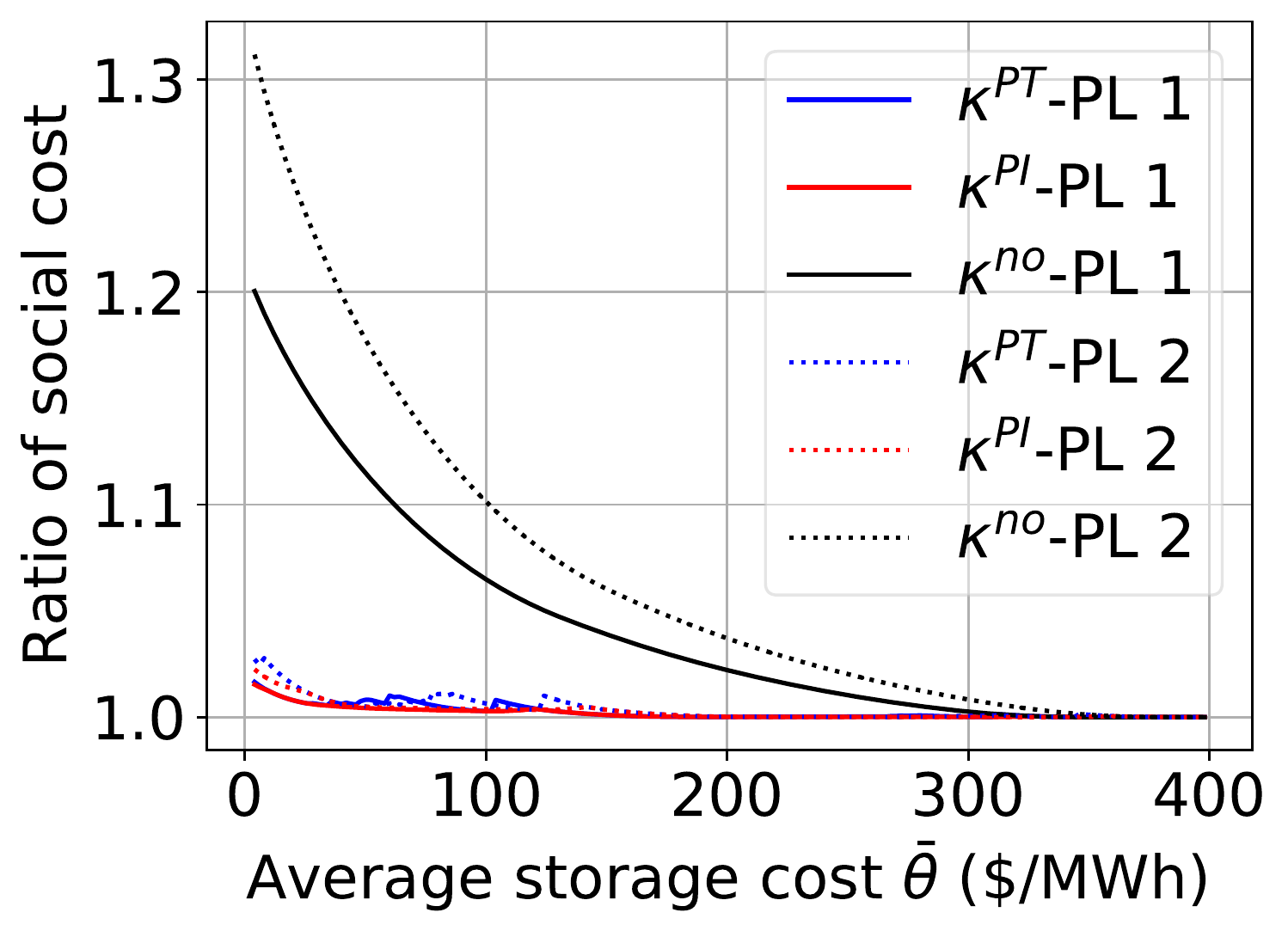}}}
	\hspace{-2ex}
	\subfigure[]{
		\label{fig:subfig:ra2} 
		\raisebox{-2mm}{\includegraphics[width=1.7in]{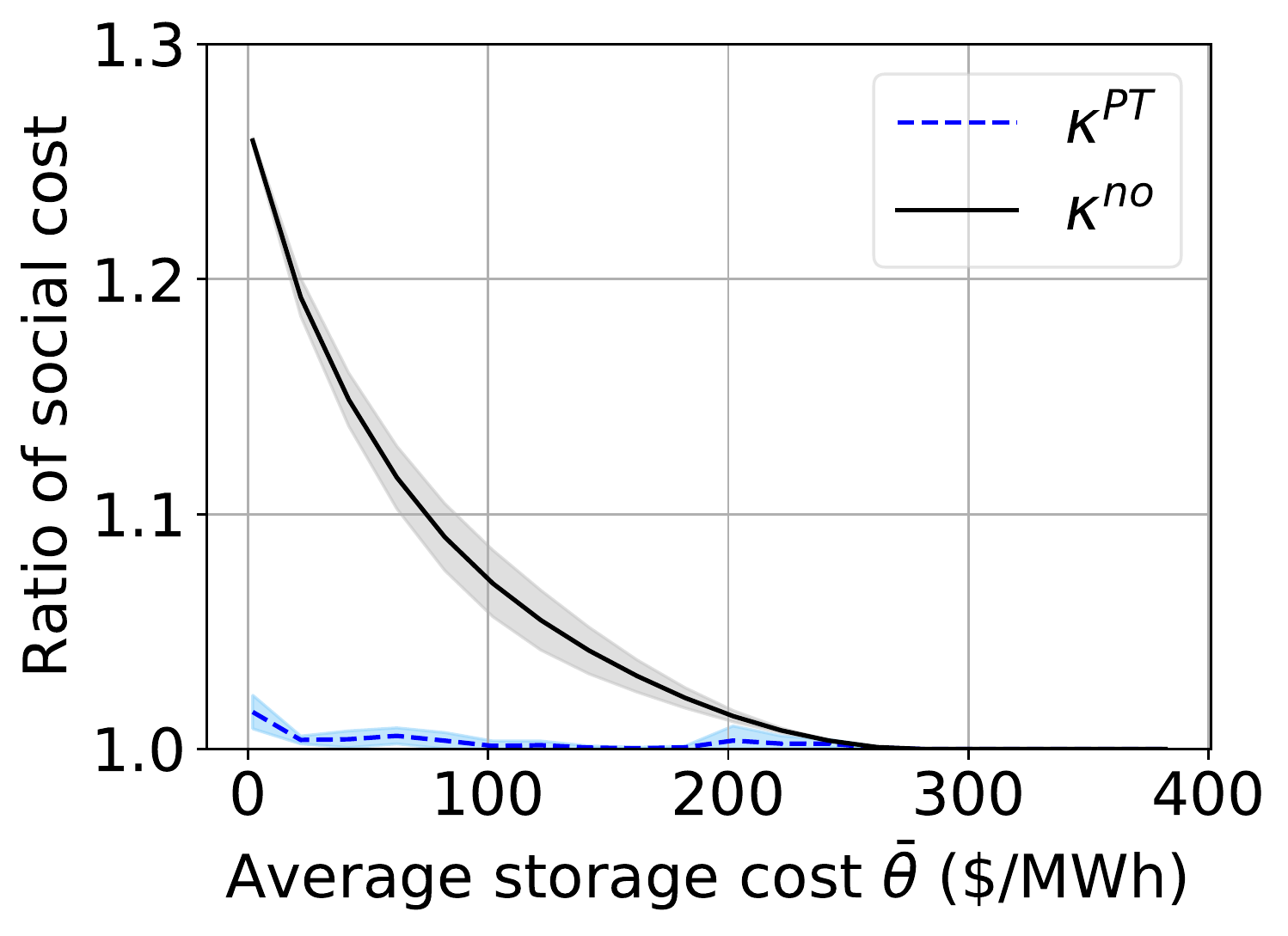}}}
	\hspace{-2ex}
	\subfigure[]{
		\label{fig:subfig:ra3} 
		\raisebox{-2mm}{\includegraphics[width=1.7in]{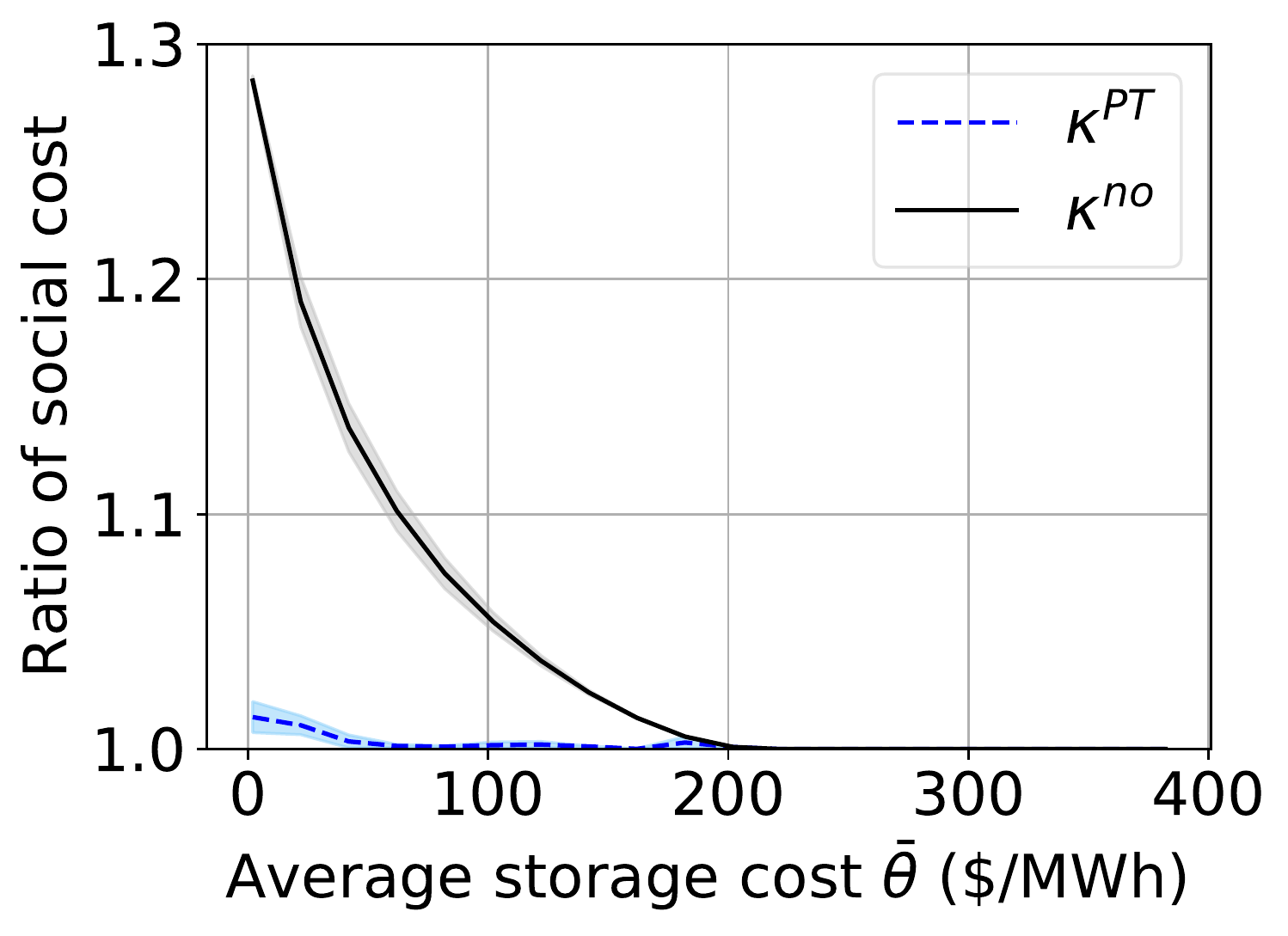}}}
	\hspace{-2ex}
	\subfigure[]{
		\label{fig:subfig:ra4} 
		\raisebox{-2mm}{\includegraphics[width=1.69in]{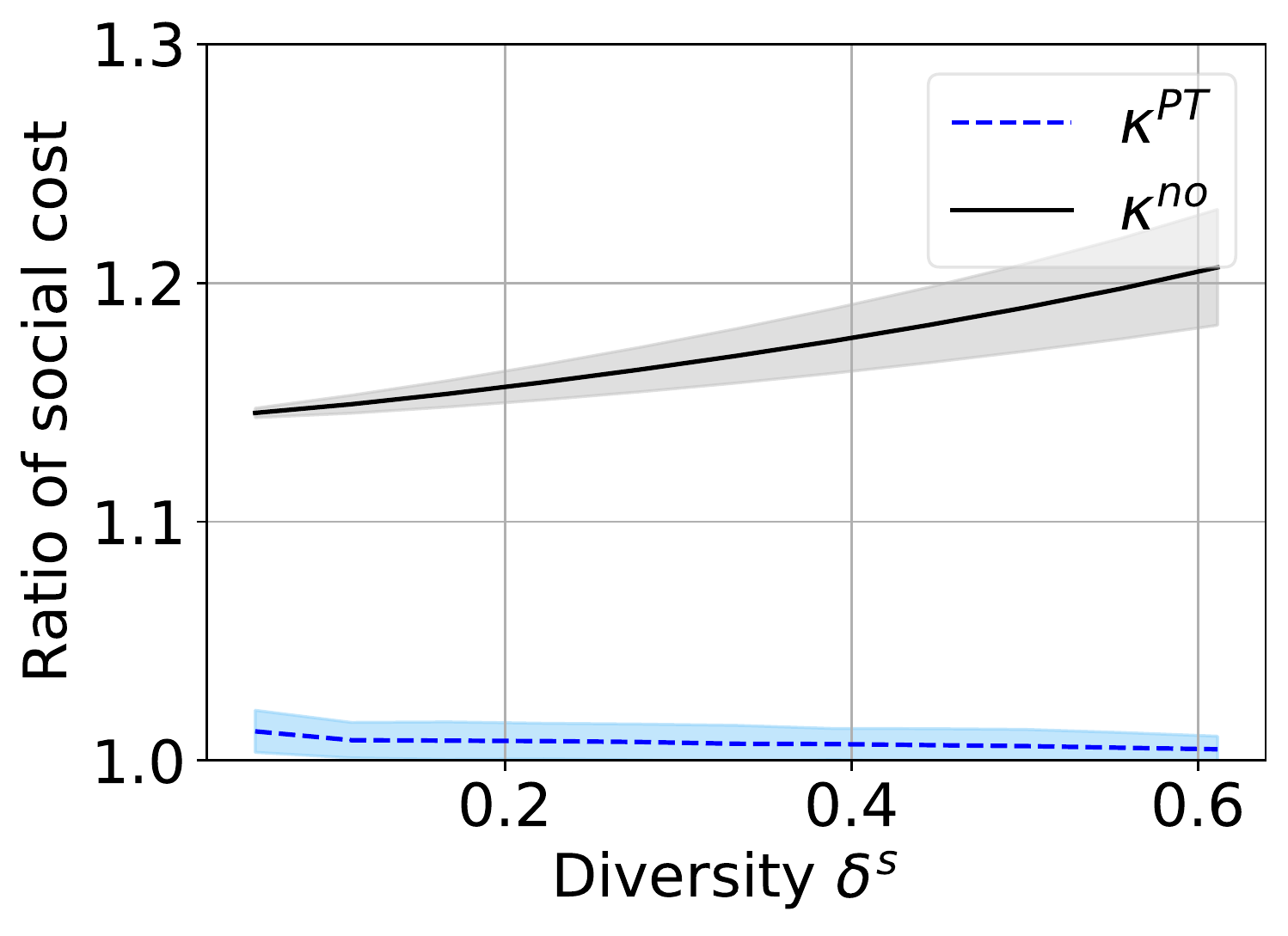}}}
	\vspace{-3mm}
	\caption{\small (a) Ratios with $\bar{\theta}$ from Austin set for one type grouping.  (b) Ratios with $\bar{\theta}$ from Austin data set for multiple type grouping. (c) Ratios with $\bar{\theta}$ from New York data set for multiple  type grouping. (d) Ratios with $\delta^s$ from Austin data set for multiple  type grouping.  }
	\label{fig:ratio}
	\vspace{-6mm}
\end{figure} 

In Figure \ref{fig:ratio}(a), we randomly group 16 users into 4 types. We show the ratios $\kappa^{\text{no}}$ (black curves), $\kappa^{\text{PT}}$ (blue curves), and  $\kappa^{\text{PI}}$ (red curves), which vary as the average storage cost $\bar{\theta}$ increases.  The solid curves correspond to the actual penetration level of solar energy as in the data set. The dotted curves correspond to the setting  when we double the solar energy amount comparing with the actual data, which can represent the future situation when the renewable energy penetration level is high.   
While Figure \ref{fig:ratio}(a) reports the results for one grouping, Figure  \ref{fig:ratio}(b) reports the overall results for 50 random groupings of 16 users into 4 types,\footnote{
To facilitate the computation with multiple random grouping results, we adopt a scenario-reduction method that reduces the original 361 outcomes (days)  to 100 outcomes \cite{heitsch2003scenario}.} where both  the mean ratios  $\kappa^{\text{no}}$ (black curve) and $\kappa^{\text{PT}}$ (blue curve) as well as the one-standard-deviation range  (in shades) are shown.\footnote{In Figures \ref{fig:ratio}(b)-(d), we focus on the case of double solar energy amount.}   Figure \ref{fig:ratio}(b) shows the results using the Austin data, while Figure \ref{fig:ratio}(c) shows similar results using the New York data. In Figures 5(a)-(c), we fix $\delta^s = 1/3$ and vary the average storage cost $\bar{\theta}$. Instead, in Figure 5(d), we vary the storage cost diversity  $\delta^s$ while fixing $\bar{\theta}=10\$$/(MWh). 
 
We have the following observations based on Figure \ref{fig:ratio}.

\vspace{0.7ex}
 \textit{Observation 1}: \textit{The pricing scheme \textbf{PT} can achieve a good performance with an empirical  ratio  $\kappa^{\text{PT}}$  lower than 1.05.}
\vspace{0.7ex}
    
    To see this,  note that the ratio $\kappa^{\text{PT}}$ in one-standard-deviation range  (blue curves with shades) is lower than 1.05 in all subfigures. Such good performance is also robust across different average  storage costs $\bar{\theta}$ (in Figure \ref{fig:ratio}(b)), different data sets (in Figure  \ref{fig:ratio}(b) and Figure  \ref{fig:ratio}(c)), and different storage cost diversities  $\delta^s$ (in Figure  \ref{fig:ratio}(d)). Furthermore, \textbf{PT} performs as well as  \textbf{PI} (comparing the blue  and red curves in Figure  \ref{fig:ratio}(a)).
  
  \vspace{0.7ex} 
  \textit{Observation 2}:   \textit{Compared with the case of no storage investment in the system, the  \textbf{PT} scheme can incentivize proper  storage investment and  significantly reduce the social cost.}
\vspace{0.7ex}
          
      Indeed, in all subfigures the ratio $\kappa^{\text{no}}$ (black curves) is much higher than 1, while the ratio $\kappa^{\text{PT}}$ (blue curves) is close to 1.  Furthermore, our pricing scheme  can reduce the social cost  more significantly compared with no storage, if the solar energy penetration level  is high (comparing the dotted and solid curves  in Figure \ref{fig:ratio}(a)). The reason is that more solar energy further reduces the load in daytime hours, which makes the system peak load  more significant. Thus, a larger storage capacity can shift load and reduce the social cost.

\subsection{Impact of demand variance on the performance of \textbf{PT}}
	   \begin{figure}[t]
		\centering
		\hspace{-1ex}
		\subfigure[]{
			\raisebox{-2mm}{\includegraphics[width=1.7in]{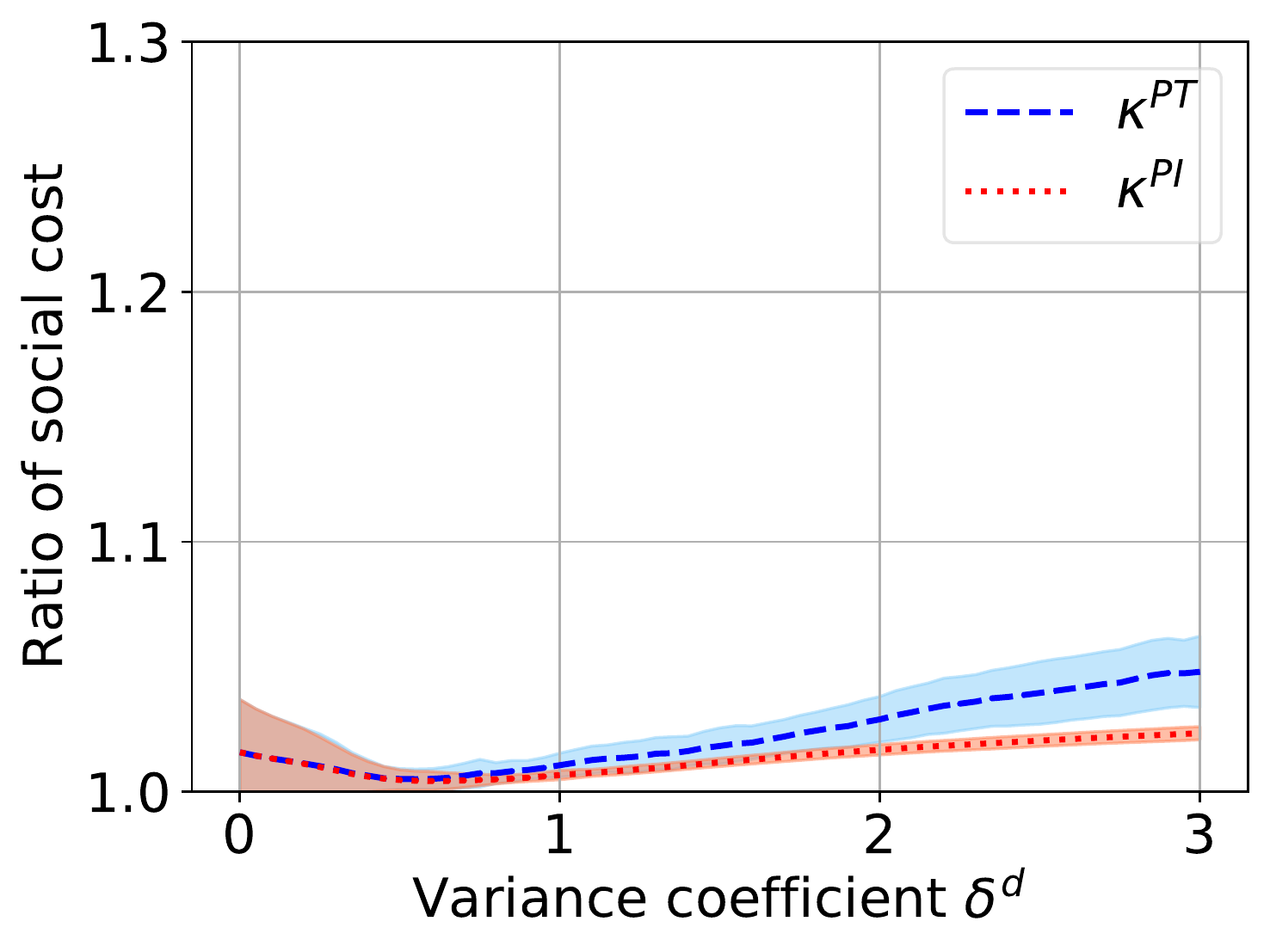}}}
		\hspace{-2ex}
		\subfigure[]{
			\raisebox{-2mm}{\includegraphics[width=1.7in]{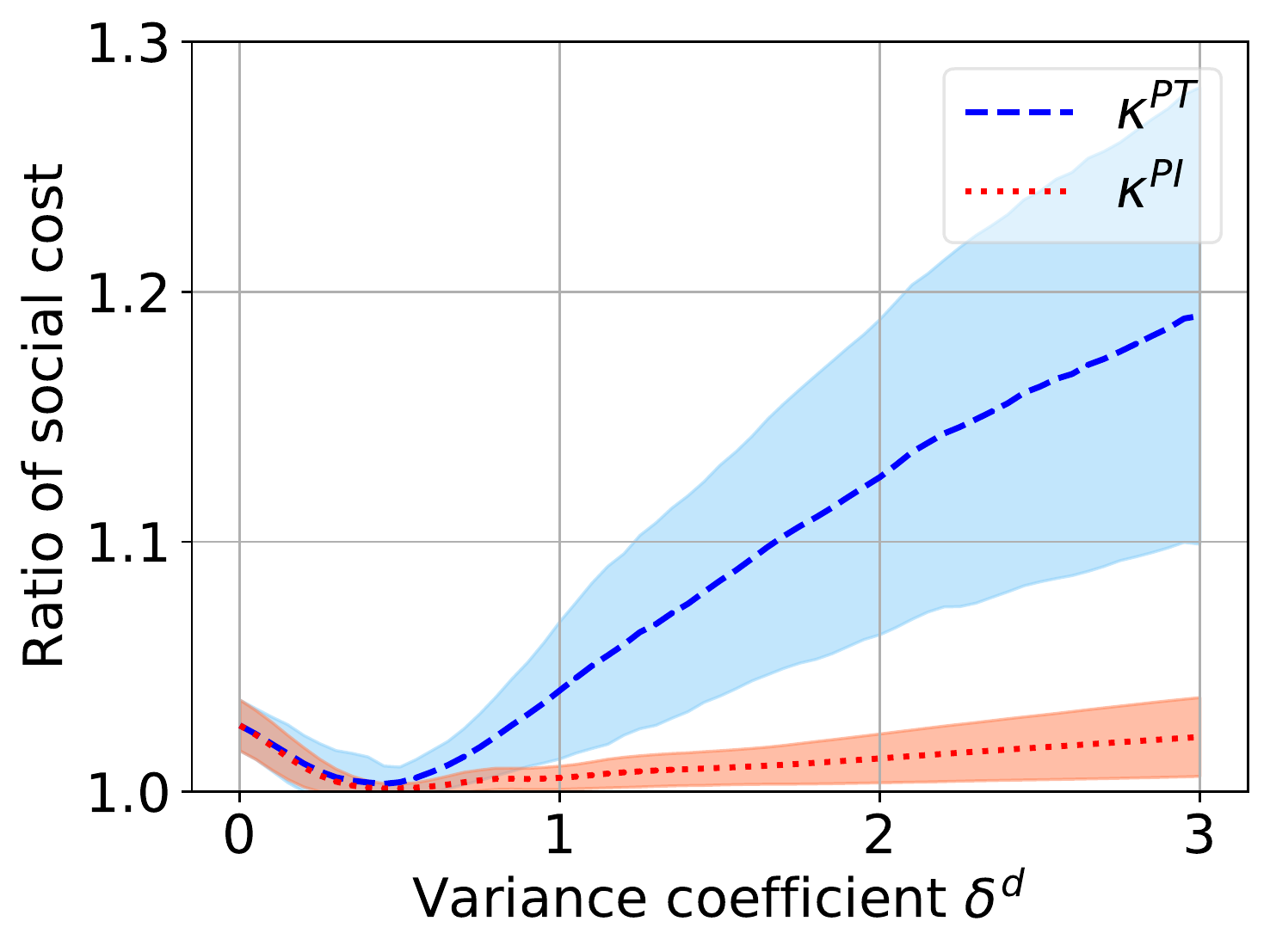}}}
		\vspace{-3mm}
		\caption{(a) \small Ratio based on realistic data; (b) Ratio based on synthetic data. Both with  $\bar{\theta}=10$ (\$/MWh) and $\delta^s=1/3$.}
		\label{fig:var}
		\vspace{-3mm}
	   \end{figure} 
	   
	   	   \begin{figure}[t]
		\centering
		\hspace{-1ex}
		\subfigure[]{
			\raisebox{-2mm}{\includegraphics[width=1.7in]{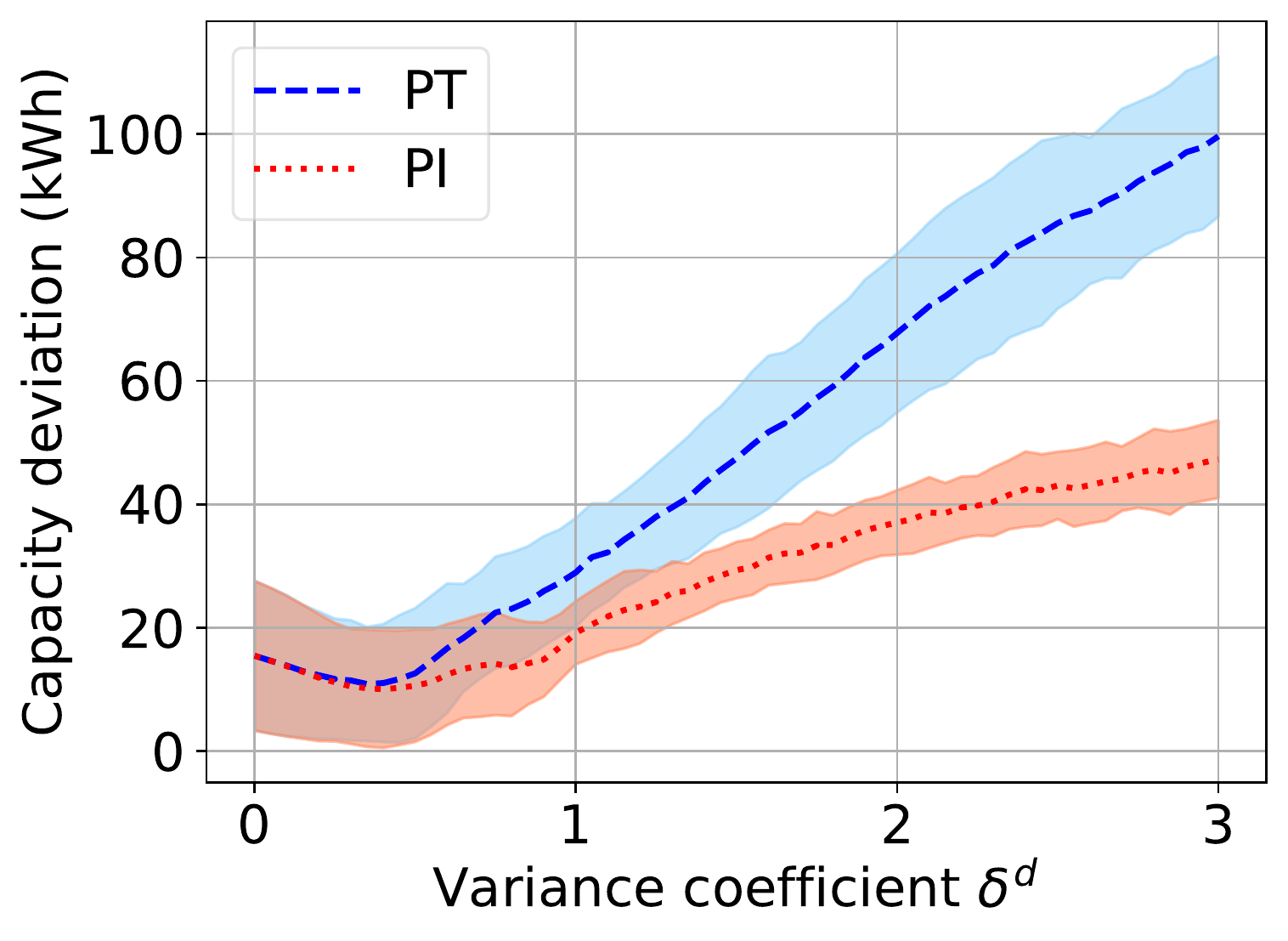}}}
		\hspace{-2ex}
		\subfigure[]{
			\raisebox{-2mm}{\includegraphics[width=1.7in]{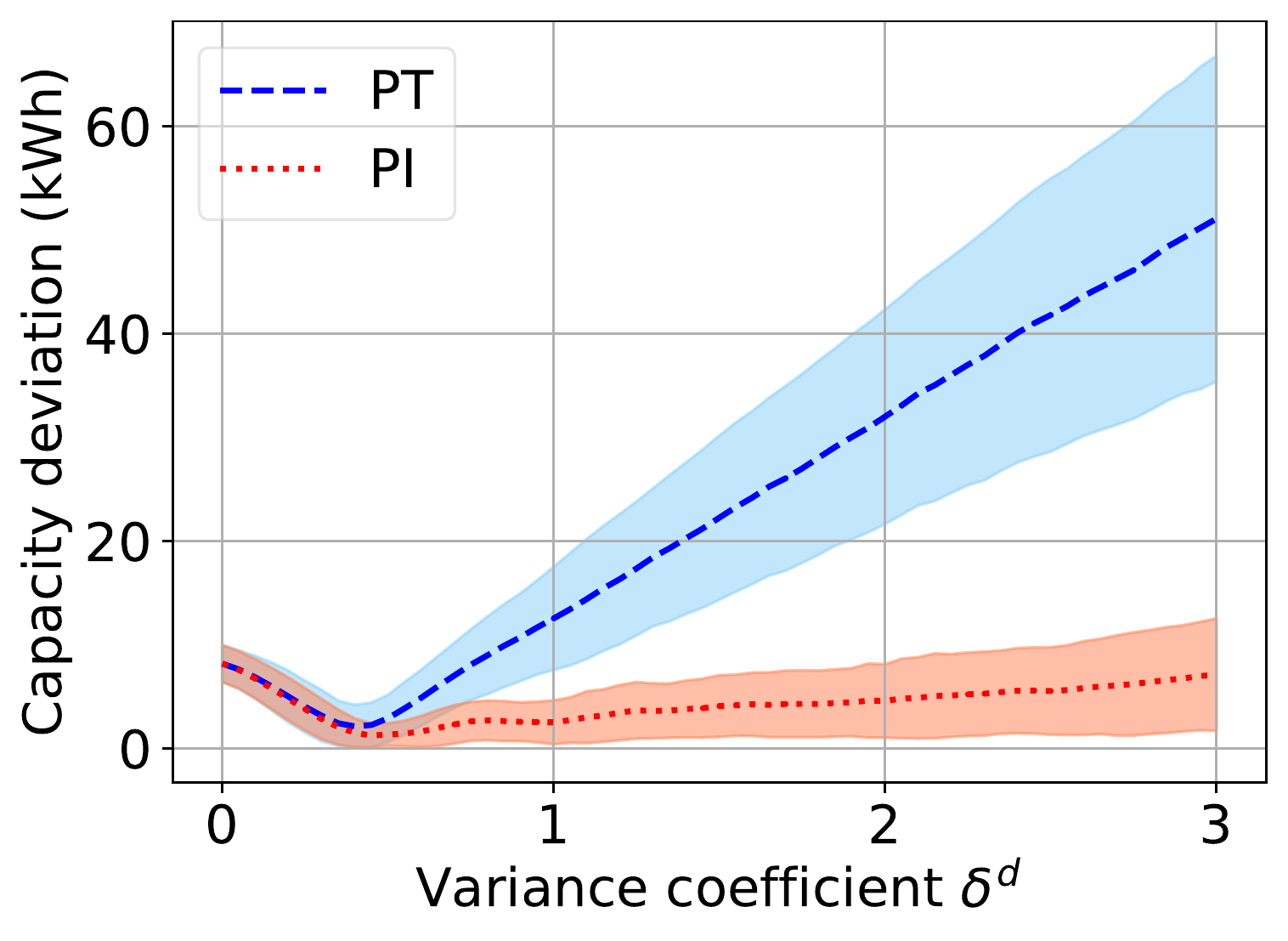}}}
		\vspace{-3mm}
		\caption{ \small Capacity deviation from benchmark based on (a) realistic data; (b) synthetic data. Both with  $\bar{\theta}=10$ (\$/MWh) and $\delta^s=1/3$.}
		\label{fig:varcap}
		\vspace{-6mm}
	   \end{figure} 
Intuitively a higher demand uncertainty may cause a larger gap of storage investment  between the ToU pricing and the social optimum, which can reduce ToU pricing's performance. However, counter-intuitively, we find that the  ratios $\kappa^{\text{PT}}$ and $\kappa^{\text{PI}}$ are not monotonic in the  demand variance, where both ratios first decrease and then increase in the demand variance. 

To see this, we will first describe how we adjust the variance of each user's peak demand  under one demand distribution. Then, we generate different demand distributions based on realistic data and synthetic data so as to examine the average results among different distributions. Finally,  we report the mean value of ratios $\kappa^{\text{PT}}$ and $\kappa^{\text{PI}}$ with respect to the demand variance among different distributions.
\subsubsection{Demand  variance adjustment}
Under a given distribution of demand, we adjust the  original peak load $D_i^{p,\omega}$ of each user $i$ and each outcome $\omega$  to  $D_i^{p,\omega'}=D_i^{p,\omega}-(1-\delta^d) (D_i^{p,\omega}-\mathbb{E}_\omega[D_i^{p,\omega}])$. Here, $\mathbb{E}_\omega[D_i^{p,\omega}]$  is the mean of user $i$'s peak  demand. We adjust the variance of the off-peak demand in the same way. Note that we control the variance of demand through the parameter $\delta^d$.  When $\delta^d=0$, the demand is deterministic at the mean.  When $\delta^d=1$, the load is just the same as the original one in the data set. The case $\delta^d>1$  means that we increase each outcome's demand variance comparing with the original data, while the case $0<\delta^d<1$ means that we reduce the variance.

We set up different distributions based on both realistic data and synthetic data.  Realistic data is the data used as  in Section \ref{section:simulation}.B and synthetic data for users' demand  is generated uniformly within a range so as to capture the generality.  We present the details in Appendix.E.

\subsubsection{Results}

In Figure \ref{fig:var}, we show the average  ratios $\kappa^{\text{PT}}$ (blue curves) and  $\kappa^{\text{PI}}$ (red curves) as well as the one-standard-deviation range (among different distributions) as the variance coefficient $\delta^d$ increases. Figure \ref{fig:var}(a) is based on realistic data,  while Figure \ref{fig:var}(b) is  based on synthetic data.  In Figure \ref{fig:varcap}, we show the absolute value of how much the total invested storage capacities under  \textbf{PT} (blue curve) and \textbf{PI} (red curve) deviate from the total capacity under the benchmark \textbf{SO}, respectively,  as  $\delta^d$ increases. Similarly, we show the mean value and one-standard-deviation range, where Figure \ref{fig:varcap}(a) is based on realistic data and  Figure \ref{fig:varcap}(b) is  based on synthetic data. We have the following observation.

\vspace{0.7ex}
 \textit{Observation 3: A larger demand uncertainty  may decrease the ratios  $\kappa^{\text{PT}}$ and $\kappa^{\text{PI}}$, i.e., improving the relative  performance of   \textbf{PT} and \textbf{PI}  comparing with the social optimum.}
\vspace{0.7ex}

As shown in Figure \ref{fig:var},  the  ratios $\kappa^{\text{PT}}$ and $\kappa^{\text{PI}}$  are non-monotonic in the demand variance. In both Figures \ref{fig:var}(a) and \ref{fig:var}(b), when  $\delta^d$ is high (e.g., $\delta^d >0.5$),  both the ratios $\kappa^{\text{PT}}$ and $\kappa^{\text{PI}}$  increase in $\delta^d$. This might seem intuitive as a higher uncertainty in demand leads to a higher gap of  the invested  storage capacity  between  ToU pricing  \textbf{PT} (\textbf{PI}) and the benchmark \textbf{SO}, as shown in Figure \ref{fig:varcap} (e.g., $\delta^d >0.5$).  This leads to a higher gap of social costs. 
However,  when $\delta^d$ is low (e.g., $\delta^d<0.5$), an increased $\delta^d$ can  reduce $\kappa^{\text{PT}}$ and $\kappa^{\text{PI}}$ as shown in Figure \ref{fig:var}. This is due to the smoothed boundary-user impact, which we explain below.
 
 Recall the discussions in  Propositions \ref{prop:benstorage} and \ref{prop:pricestorage} that there are boundary users in the benchmark \textbf{SO} but no such users in \textbf{PT} (or \textbf{PI}). Thus, in  \textbf{PT},  under a deterministic peak demand of $a$ ($\delta^d=0$), any user can only invest in either $a$ or 0 amount of energy storage (Proposition \ref{prop:pricestorage}). However, in the benchmark \textbf{SO}, the boundary users may be required to invest in a storage amount between $[0,a]$ (Proposition \ref{prop:benstorage}). This can lead to a large gap between \textbf{PT} and \textbf{SO}. In contrast, when the demand is random, the boundary-user effect may diminish. For example,  under a random demand over the support of $[0.7a,1.3a]$ with a mean value $a$, a user can invest in 0, or  any capacity in the range of $[0.7a,1.3a]$ depending on the \textbf{PT} pricing. This gives users more investment choices  compared with  all-or-nothing investment in the deterministic or near-deterministic case. As a result, the gap between the investment capacities under $\textbf{PT}$ (\textbf{PI}) and \textbf{SO} also decreases.  Figure \ref{fig:varcap} illustrates such a decreasing gap, where the capacity deviation of \textbf{PT} and \textbf{PI} from the benchmark decreases as  $\delta^d$ increases in $[0,0.5]$. 

\section{Conclusion}
This paper  designs an optimal ToU  pricing  explicitly considering the impact of  users' storage investment. We formulate a two-stage optimization problem between the utility and users to minimize the social cost.
Since the utility may not know individual users' private  information,  we  propose a pricing scheme for the utility  based on the storage type  information. We  design an efficient algorithm for the utility to determine the optimal price difference of ToU pricing, which only involves searching over  a finite set of threshold prices. Simulations based on realistic data demonstrate  the good performance of our proposed pricing scheme.  We also find that when the demand variance is low,  an increased variance  range may  improve the performance of the ToU pricing  by smoothing the all-or-nothing storage investment.

\bibliographystyle{IEEEtran}
\bibliography{storage}

%
\begin{IEEEbiography}[{\includegraphics[width=1in,height=1.25in,clip,keepaspectratio]{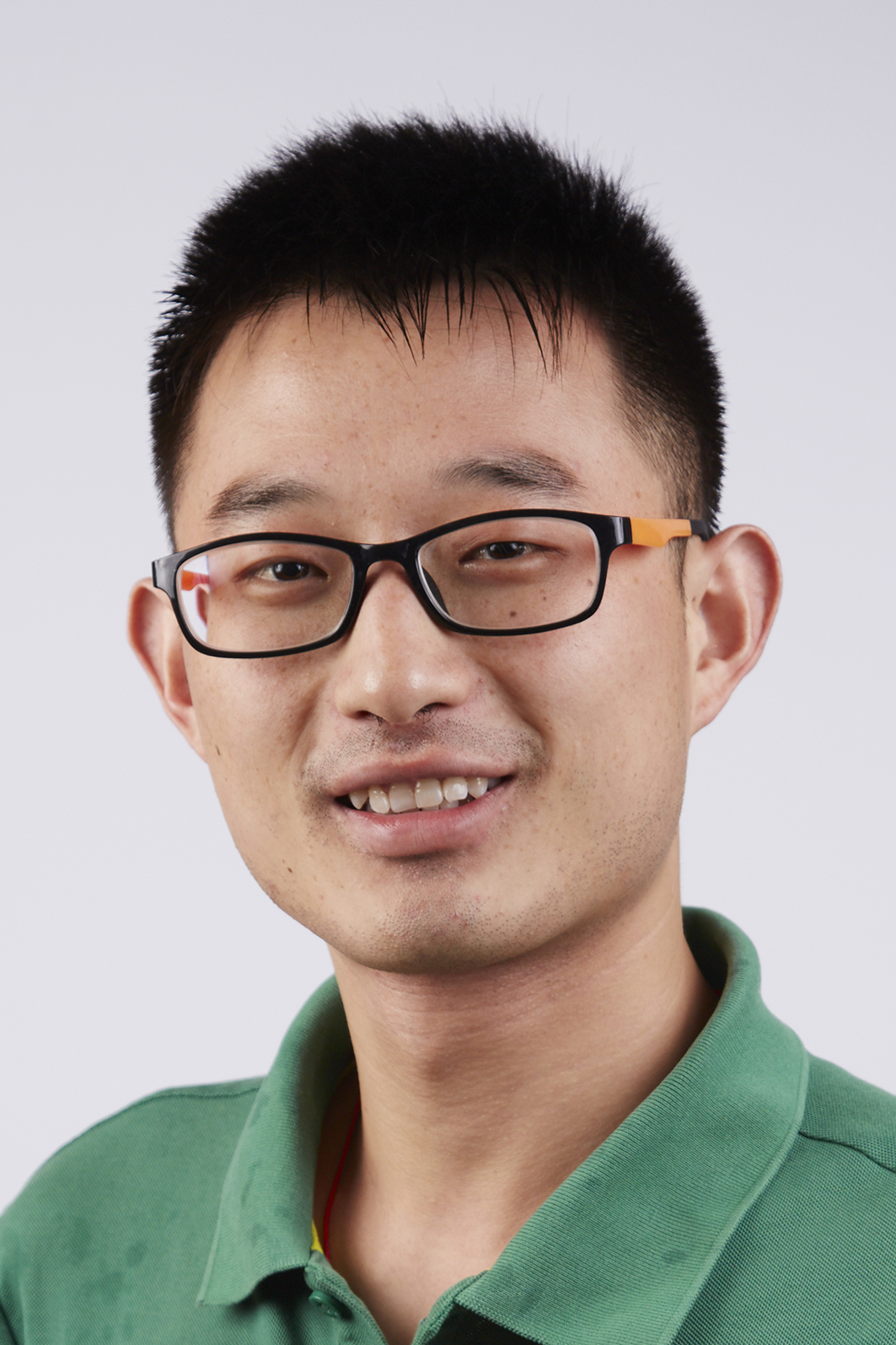}}]{Dongwei Zhao}(M'21) received the B.S. degree from Zhejiang University, Hangzhou,  in 2015, and  the  Ph.D.
degree from The Chinese University of Hong Kong, Hong Kong, in 2021. He is currently a postdoctoral associate in MIT Energy Initiative, Massachusetts Institute of Technology.
His main research interests are in the optimization and game theory of power and energy systems. More information at https://sites.google.com/view/joris-zhao
\end{IEEEbiography}
\begin{IEEEbiography}[{\includegraphics[width=1in,height=1.25in,clip,keepaspectratio]{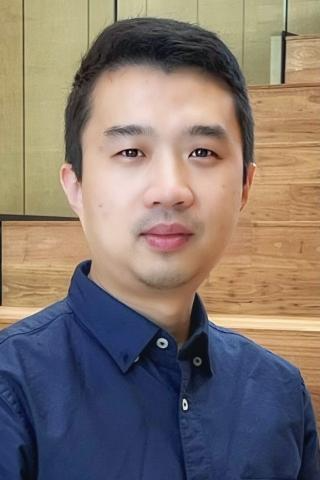}}]{Hao Wang}(M'16) received the Ph.D. degree from the Chinese University of Hong Kong, Hong Kong, in 2016. He was a Postdoctoral Research Fellow with Stanford University and a Washington Research Foundation Innovation Fellow with the University of Washington.
He is currently a Lecturer with the Department of Data Science and Artificial Intelligence, Faculty of Information Technology, Monash University, Australia. His research interests are in optimization, machine learning, and data analytics for power and energy systems. More information at https://research.monash.edu/en/persons/hao-wang.

\end{IEEEbiography}

\begin{IEEEbiography}[{\includegraphics[width=1in,height=1.25in,clip,keepaspectratio]{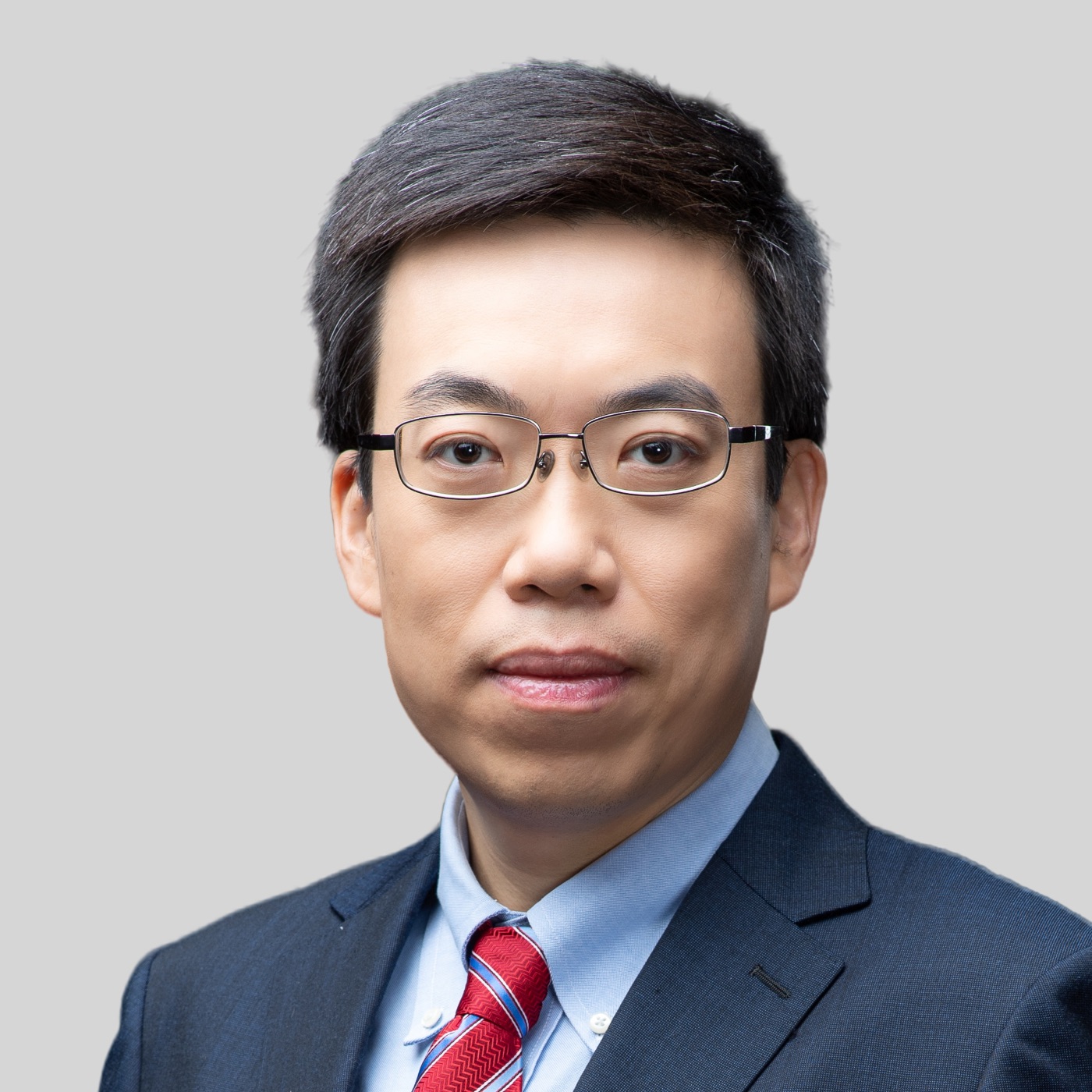}}]{Jianwei Huang} (F'16) received the Ph.D. degree in ECE from Northwestern University in 2005, and worked as a Postdoc Research Associate in Princeton University during 2005-2007. From 2007 until 2018, he was on the faculty of Department of Information Engineering, The Chinese University of Hong Kong. Since 2019, he has been on the faculty at The Chinese University of Hong Kong, Shenzhen, where he is currently a Presidential Chair Professor and an Associate Dean of the School of Science and Engineering. He also serves as a Vice President of Shenzhen Institute of Artificial Intelligence and Robotics for Society. His research interests are in the area of network optimization, network economics, and network science, with applications in communication networks, energy networks, data markets, crowd intelligence, and related fields. He has published more than 300 papers in leading venues, with a Google Scholar citation of 14000+ and an H-index of 59. He has co-authored 10 Best Paper Awards, including the 2011 IEEE Marconi Prize Paper Award in Wireless Communications. He has co-authored seven books, including the textbook on "Wireless Network Pricing." He is an IEEE Fellow, and was an IEEE ComSoc Distinguished Lecturer and a Clarivate Web of Science Highly Cited Researcher. He is the Editor-in-Chief of IEEE Transactions on Network Science and Engineering, and was the Associate Editor-in-Chief of IEEE Open Journal of the Communications Society.
\end{IEEEbiography}
\begin{IEEEbiography}[{\includegraphics[width=1in,height=1.25in,clip,keepaspectratio]{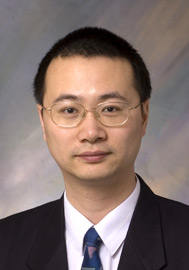}}]{Xiaojun
	Lin} (S'02 M'05 SM'12 F'17) received his B.S. from Zhongshan University, Guangzhou, China, in 1994, and his M.S. and Ph.D. degrees from Purdue University, West Lafayette, IN, in 2000 and 2005, respectively. He is currently a Professor of Electrical and Computer Engineering at Purdue University.

Dr. Lin's research interests are in the analysis, control and
	optimization of large and complex networked systems, including both communication networks and power grid.  He received the NSF CAREER award in 2007.  He received 2005 best paper of the year award from Journal of Communications and Networks, IEEE INFOCOM 2008 best paper award, and ACM MobiHoc 2021 best paper award. He was the Workshop co-chair for IEEE GLOBECOM 2007, the Panel
	co-chair for WICON 2008, the TPC co-chair for ACM MobiHoc 2009, the Mini-Conference co-chair for IEEE INFOCOM 2012, and the General co-chair for ACM e-Energy 2019.  He has served as an Area Editor for (Elsevier) Computer Networks Journal, an Associate Editor for IEEE/ACM Transactions on Networking, and a Guest Editor for (Elsevier) Ad Hoc Networks journal. 
\end{IEEEbiography}

\clearpage
\newpage

\section{Appendix}
\subsection{Proof of Proposition \ref{prop:usercontinuous} and Proposition \ref{prop:userdiscrete}}
We consider a general demand distribution for type $k$'s peak demand with the  CDF function  $F_k^p$.

We first analyze the optimal storage operation decision $s_k^*$ in Period-2 given the storage capacity $c_k$, and then characterize the optimal storage capacity $c_k^*$ in Period-1.
 
 First, in Period-2, given the storage capacity $c_k$, we have $s_k^*=\min\{c_k,D_k^p\}$  for any realization $D_k^p$ since $p^\Delta>0$ . 
 
 Then, we incorporate $s_k^*$ in Period-2 into Period-1, and  the objective is equivalent to 
\begin{align}
\min_{c_k}~ U(c_k)=&\mathbb{E}_{\mathcal{D}_k^p}\left[-p^\Delta \min\{c_k,\mathcal{D}_k^p\}\right]+\theta_k c_k\notag.
\end{align}
We will analyze the optimum of function $U(c_k)$. 
  We take the derivative of $U(c_k)$ with respect to $c_k$, and have
 \begin{align*}
U'(c_k)
&=-p^\Delta (1-F_k^p(c_k)) +\theta_k.
\end{align*}
Note that  $U''(c_k)\geq 0$, which shows the convexity of the function $U(c_k)$. Therefore, we will have the optimal solution $c_k^*$ as follows and prove Proposition  \ref{prop:usercontinuous}.
\begin{itemize}
	\item If $p_k^\Delta<\theta_k$, $c_k^*=0$.
	\item If $p_k^\Delta>\theta_k$, $c_k^*=F_k^{p^{-1}}(\frac{p^\Delta -\theta_k }{p^\Delta })$.
	\item If $p_k^\Delta=\theta_k$, $c_k^*$ can be any value in $ [0, \underline{\mathcal{D}}_k^p]$.
\end{itemize}

We can obtain Proposition \ref{prop:userdiscrete} based on Proposition  \ref{prop:usercontinuous}. When we consider the discrete distribution of the peak demand, the CDF function $F_k^p$ is step-wise. Note that if  $\sum_{\omega\geq m}\rho^\omega p^\Delta =\theta_k$, the solution $c_k^\star$ takes the value $D_k^{p,m}$ according to the definition of generalized diverse function $F_k^{p^{-1}}$ in Proposition  \ref{prop:usercontinuous}. In fact, the optimal investment  $c_k^\star$ can be any value within $[D_k^{p,m-1},D_k^{p,m}]$ due to the step-wise structure of $F_k^p$. \qed

\subsection{Proof of Theorem \ref{prop:utility}}
  This conclusion is due to that  the utility makes decision based on the optimal storage capacity $c_k^*$ and storage charge/discharge $s_k^*$ from Stage II.  The solution $c_k^*$ is step-wise in the price difference $p^\Delta$. The solution $s_k^*$ is determined by $c_k^*$ as in Proposition \ref{prop:userdiscrete}, which shows that  $s_k^*$ is also step-wise in  the price difference $p^\Delta$. 
  
  The utility's social cost includes the storage investment cost and energy supply cost. The storage investment cost is determined by all the users' storage capacity, and the supply cost is determined by the aggregated charge/discharge amount of all users. Therefore,  the social cost is also step-wise in the price difference $p^\Delta$, which has the threshold price set $\bigcup_k \mathcal{P}_k$. \qed

\subsection{Proof of Proposition \ref{prop:benstorage} and Proposition  \ref{prop:pricestorage}}

Proposition \ref{prop:pricestorage} can be directly proved by the solution structure in Proposition \ref{prop:userdiscrete} and Proposition \ref{prop:usercontinuous}. Note that if the price difference is higher than the storage cost of a user, i.e., $p^\Delta>\theta_i$, the invested capacity is between the lower support and upper support of the peak demand random variable. If the price difference is lower than the storage cost, i.e., $p^\Delta<\theta_i$, the invested capacity is zero. For the case $p^\Delta=\theta_i$, we assume that users will also not invest in storage.

We next prove Proposition \ref{prop:benstorage} by the following steps. In Step 1,  we show that  considering any two users with different storage costs, if the high-cost user invests in a  positive capacity, then the low-cost user  must also invest in positive capacity. Step 2:  We show that if  user $i$ invests in positive capacity, the capacity cannot be beyond the upper support of the peak demand variable, i.e., $0\leq c_i^{*} \leq \overline{\mathcal{D}}_i^{p}$. Step 3: Among all users who invest in storage with $M'$ storage types, for any user $i$ with  $\theta^{1}\leq \theta_i \leq \theta^{M'-1}$, we show that the optimal  capacity  is within the lower support and upper support of the peak demand , i.e., $\underline{\mathcal{D}}_i^{p}\leq  c_i^{*}  \leq \overline{\mathcal{D}}_i^{p}$. We show the steps in detail as follows.

Step 1:  We assume that user $i$ invests in capacity $c_i^{*}>0$ and user $j$ invests in capacity $c_j^{*}=0$, where $\theta_i>\theta_j$. In this case  we can always reduce $c_i^{*}$ by a sufficiently small  value $\epsilon>0$ and increase $c_j^{*}$ by corresponding $\epsilon$, such that $c_j^{*}+\epsilon \leq \underline{\mathcal{D}}_j^p$. In this case, the storage investment cost will be reduced while the aggregate charge/discharge amount can remain unchanged. This contradicts the social optimum. Therefore,  if users invest in positive capacity, they are with the lowest storage costs.

Step 2: We assume that   user $i$ invests in capacity $c_i^{*}>\overline{\mathcal{D}}_i^{p}$. Note that the charge/discharge decision $s_i^*\leq \min(c_i^{*},\mathcal{D}_i^{p})$. Thus, we can always reduce $c_i^{*}$ to $\overline{\mathcal{D}}_i^{p}$, which will reduce the investment cost without affecting charge/discharge decision. This contradicts the social optimum. Therefore, for any user $i$, we have $0\leq c_i^{*} \leq \overline{\mathcal{D}}_i^{p}$. 

Step 3:  Among all users who invest in storage with $M'$ storage types, for any user $i$ with  $\theta^{1}\leq \theta_i \leq \theta^{M'-1}$, we assume that $  c_i^{*} < \underline{\mathcal{D}}_i^{p}$. We can always increase $c_i^{*}$ by a sufficiently small value $\epsilon$ such that $c_i^{*}+\epsilon\leq \underline{\mathcal{D}}_i^{p}$, while reduce $c_j^{*}$ by $\epsilon$ such that $c_j^{*}-\epsilon\geq0$ for user $j$ with storage cost $\theta^{M'}$. In this case, the storage investment cost will be reduced while the aggregate charge/discharge amount can remain unchanged. This contradicts the social optimum. Therefore,  for any user $i$ with  $\theta^{1}\leq \theta_i \leq \theta^{M'-1}$, we have $\underline{\mathcal{D}}_i^{p}\leq  c_i^{*}  \leq \overline{\mathcal{D}}_i^{p}$.

We have Proposition \ref{prop:benstorage} proved based on the three steps above. \qed

\subsection{Proof of Proposition \ref{prop:zerocost}}

We will first characterize the upper bound in {subsection 1)} in this part and then show the upper bound is tight in {subsection 2)}.
\subsubsection{Characterize the upper bound}
 We construct the upper bound based on two sub-optimal solutions of the ToU pricing.
 \begin{itemize}
     \item Low price difference $p^\Delta$ that leads to no storage  invested in the system. We denote the social cost as
     $SC^{l}$ in this case.
     \item High price difference $p^\Delta$ that incentivizes all the users to invest in storage capacity at the maximum peak demand in the sample space (due to zero storage cost). Then, for each demand realization, the peak demand is totally shifted to the off-peak period. We denote the social cost as $SC^{h}$ in this case.
 \end{itemize}
 Thus, we have $\kappa^{\text{PT}}\leq \min(\frac{SC^{l}}{SC^{\text{SO}}},\frac{SC^{h}}{SC^{\text{SO}}} )$. We will first derive the social costs $SC^{l}$,  $SC^{h}$ and  $SC^{\text{SO}}$, respectively, and then we analyze the upper bound for  $\frac{SC^{l}}{SC^{\text{SO}}}$ and $\frac{SC^{h}}{SC^{\text{SO}}}$.  We denote the original aggregate peak demand and off-peak demand (with no storage in the system) as $\mathcal{D}_a^{p}$ and $\mathcal{D}_a^{o}$, which are random variables. 
  
 First, we have the social costs $SC^{l}$ and   $SC^{h}$ as follows.
 \begin{align*}
     SC^{l}&= \mathbb{E}\left[g^{p}(\mathcal{D}_a^{p})+g^{o}(\mathcal{D}_a^{o})\right]
 \end{align*}
 \begin{align*}
     SC^{h}&= \mathbb{E}\left[g^{p}(0)+g^{o}(\mathcal{D}_a^{p}+\mathcal{D}_a^{o})\right]
 \end{align*}

Then, we characterize the social cost  $SC^{\text{SO}}$. For the benchmark problem, since the storage cost approaches zero, the storage investment cost can be neglected. Also, the social planner can invest in enough storage capacity to shift the demand. The benchmark problem \textbf{SO} can be reformulated as follows.
\begin{align*}
	\min  ~&\mathbb{E}_{\bm{\mathcal{D}}}~  \left[ g^p\left((\mathcal{D}_a^{p}-\sum_i s_i)\right)+g^o\left( (\mathcal{D}_a^{o}+\sum_i s_i)\right)\right]\\
\text{s.t.~} 
& 0\leq s_i, ~\forall i\in \mathcal{I},\\
\text{var}: &\bm{s}
\end{align*}
We only need to derive the optimal aggregate charge/ discharge decision $\sum_i s_i$ for each realization of joint random demand. Such a problem is convex, and we can have the solution as follows.
\begin{align*}
    \sum_i s_i^*=\frac{ H^{o}\mathcal{D}_a^{p}-{H^{p}}\mathcal{D}_a^{o}   }{H^{o}+{H^{p}} }
\end{align*}
Recall that we assume $H^{o}\mathcal{D}_a^{p}-H^{p}\mathcal{D}_a^{o}  \geq 0$, which means that the average power in peak period is higher than  the average power in off-peak period.
We further calculate the social optimum $SC^{\text{SO}}$ as follows.
\begin{align}
   SC^{\text{SO}}=\mathbb{E}\Big[\alpha ({H^{p}}+H^o) (\frac{ \mathcal{D}_a^{p}+\mathcal{D}_a^{o} }{H^{o}+{H^{p}} })^2\notag\\+\beta (\mathcal{D}_a^{p}+ \mathcal{D}_a^{o})+ \gamma H^{o}+\gamma H^{p}\Big]. \label{eq:so}
\end{align}

Next, we characterize the upper bound for $\frac{ SC^{l}}{ SC^{\text{SO}}}$ and $\frac{ SC^{h}}{ SC^{\text{SO}}}$.
We first consider the ratio $\frac{ SC^{l}}{ SC^{\text{SO}}}$.
$$\frac{ SC^{l}}{ SC^{\text{SO}}}=\frac{\mathbb{E}\left[g^{p}(\mathcal{D}_a^{p})+g^{o}(\mathcal{D}_a^{o})\right]}{    \mathbb{E}\Big[\alpha ({H^{p}}+H^o) (\frac{ \mathcal{D}_a^{p}+\mathcal{D}_a^{o} }{H^{o}+{H^{p}} })^2+\beta (\mathcal{D}_a^{p}+ \mathcal{D}_a^{o})+ \gamma H^{o}+\gamma H^{p}\Big]  }.$$
We focus on each demand realization, and have
\begin{align}
    &\frac{g^{p}(\mathcal{D}_a^{p})+g^{o}(\mathcal{D}_a^{o})}{\alpha ({H^{p}}+H^o) (\frac{ \mathcal{D}_a^{p}+\mathcal{D}_a^{o} }{H^{o}+{H^{p}} })^2+\beta (\mathcal{D}_a^{p}+ \mathcal{D}_a^{o})+ \gamma H^{o}+\gamma H^{p}  }\\
    =&\frac{\frac{ \alpha }{H^{p}}(\mathcal{D}_a^{p})^2+\frac{ \alpha }{H^{o}}(\mathcal{D}_a^{o})^2+ \beta (\mathcal{D}_a^{p}+ \mathcal{D}_a^{o})+ \gamma H^{o}+\gamma H^{p}}{\alpha ({H^{p}}+H^o) (\frac{ \mathcal{D}_a^{p}+\mathcal{D}_a^{o} }{H^{o}+{H^{p}} })^2+\beta (\mathcal{D}_a^{p}+ \mathcal{D}_a^{o})+ \gamma H^{o}+\gamma H^{p}}\\
   \leq &\frac{\frac{ \alpha }{H^{p}}(\mathcal{D}_a^{p})^2+\frac{ \alpha }{H^{o}}(\mathcal{D}_a^{o})^2}{\alpha ({H^{p}}+H^o) (\frac{ \mathcal{D}_a^{p}+\mathcal{D}_a^{o} }{H^{o}+{H^{p}} })^2}\\
    = & \frac{\frac{ ({H^{p}}+H^o) }{H^{p}}(\frac{\mathcal{D}_a^{p}}{\mathcal{D}_a^{o}})^2+\frac{ ({H^{p}}+H^o)}{H^{o}}}{  ( \frac{\mathcal{D}_a^{p}}{\mathcal{D}_a^{o}}+1)^2} \label{x}
\end{align}
We define the function $f(x)=\frac{\frac{ ({H^{p}}+H^o) }{H^{p}}x^2+\frac{ ({H^{p}}+H^o)}{H^{o}}}{  ( x+1)^2}$, where $x\geq \frac{H^p}{H^o}$. We take the first order derivative and have
\begin{align*}
    f'(x)=\frac{ 2x\frac{ {H^{p}}+H^o }{H^{p}}-  2\frac{ {H^{p}}+H^o}{H^{o}}}{  ( x+1)^3} \geq 0, \forall x\geq \frac{H^p}{H^o},
\end{align*}
which shows that  $f(x)$ always increases over $[\frac{H^p}{H^o},+\infty)$. Note that when $x\rightarrow \infty$,  $f(x)\rightarrow \frac{ {H^{p}}+H^o}{H^{p}}$. Thus, we always have 
\begin{align}
\eqref{x} \leq  \frac{ {H^{p}}+H^o}{H^{p}}.
\end{align}
Considering the expectation overall the random variables, we  always have  
\begin{align*}
    \frac{ SC^{l}}{ SC^{\text{SO}}}\leq  \frac{ {H^{p}}+H^o}{H^{p}}.
\end{align*}

Similarly, for the ratio $\frac{ SC^{h}}{ SC^{\text{SO}}}$, we can have 

\begin{align*}
    \frac{ SC^{h}}{ SC^{\text{SO}}}\leq  \frac{H^{p}+H^o}{H^{o}}.
\end{align*}

Overall, we will have  $$\kappa^{\text{PT}}\leq \min(\frac{SC^{l}}{SC^{\text{SO}}},\frac{SC^{h}}{SC^{\text{SO}}} )\leq \min(\frac{H^{p}+H^o}{H^{p}},\frac{H^{p}+H^o}{H^{o}}).$$

\subsubsection{Tightness of the upper bound}We construct a special example to show the tightness of the upper bound.
We make the following assumptions.
 \begin{itemize}

		\item A1:  We consider zero off-peak demand.
		\item A2: Users' peak demands in each type have perfect positive correlations such that the pricing scheme \textbf{PT} is equivalent to the pricing scheme \textbf{PI}.
		\item A3: We assume  $K$ types for the users, whose joint peak demand distribution across types are constructed as follows.
		\begin{itemize}
			\item We assume $K$ outcomes of joint peak demand with equal probability $\frac{1}{K}$.
			\item  For Outcome $k$, type $k$ has peak demand $d$ and other types have peak demand $0$.\\
			Outcome 1: $(d,0,0,\ldots,0)$\\
			Outcome 2: $(0,d,0,\ldots,0)$\\
			Outcome 3: $(0,0,d,\ldots,0)$\\
			...\\
			Outcome $K$: $(0,0,0,\ldots,d)$

	\end{itemize}
	
		\item A4: We assume that the hourly supply cost only has the quadratic term, i.e., $g(x)=\alpha x^2$. Thus, the supply cost function for peak demand is  $g^{p}(x)= \frac{ \alpha }{H^{p}}x^2$ and the supply cost function for off-peak demand is  $g^{o}(x)= \frac{ \alpha }{H^{o}}x^2$. 
\end{itemize}

We next characterize the ratio $\kappa^{\text{PT}}$. First, we calculate the social cost   $SC^{\text{SO}}$ in the benchmark according to  \eqref{eq:so}.
\begin{align*}
   SC^{\text{SO}}=\alpha\frac{d^2 }{H^{o}+{H^{p}}}.
\end{align*}

\noindent Second, we characterize the optimal ToU pricing that minimizes the social cost.  According to Proposition \ref{prop:pricestorage}, any ToU pricing will always incentive some low-cost types to invest in storage and  other high-cost types not to invest in storage.  Furthermore, in the constructed example,  each user only has peak demand $0$ or $d$ in his sample space, so each user will either invest in 0 or $d$ capacity under the ToU pricing based on Proposition \ref{prop:userdiscrete}. 
Therefore, we assume that the optimal ToU pricing will incentivize $m$ types to invest in storage with capacity $d$, and  $K-m$ types not to invest in storage, i.e., $ K\theta^m<p^\Delta\leq K\theta^{m+1}$ based on Proposition \ref{prop:userdiscrete}. We will choose the optimal $m$ to get the optimal ToU pricing. In each outcomes, we have 
\begin{itemize}
	\item Type $1,2,\ldots,m$ will totally shift the peak demand to off-peak period in each outcome.
.	\item Type $m,m+1, \ldots,K$ will not shift any demand.

\end{itemize}

Thus, for any Outcome $k$, $1\leq k \leq m$, the aggregate peak demand is $0$, and the off-peak demand is $d$ in the system. For any Outcome $k$, $m+1\leq k \leq K$,  the aggregate peak demand is $d$, and the off-peak demand is $0$ in the system. Therefore, we can calculate the social cost under such conditions as follows.
$$ SC= \frac{m}{K} \cdot\frac{\alpha (d)^2}{H^o} +\frac{K-m}{K} \cdot \frac{\alpha (d)^2}{H^p}.$$

Then, we choose the optimal $m^*$ that minimizes $SC$ to get the optimal ToU pricing. If  $H^o\leq H^p$, we have  $m^*=0$ and $ SC^{\text{PT}}=  \frac{\alpha (d)^2}{H^p}$. If  $H^p < H^o$, we have  $m^*=K$ and $ SC^{\text{PT}}=  \frac{\alpha (d)^2}{H^o}$. Therefore, the ratio
$\kappa^{\text{PT}}=\min(\frac{H^{p}+H^o}{H^{p}},\frac{H^{p}+H^o}{H^{o}})$, which shows the upper bound is tight in the worst case. Overall, we have Proposition \ref{prop:zerocost} proved. \qed


\subsection{Setup of  different demand distributions in Section \ref{section:simulation}.C}

We set up different distributions based on realistic data and synthetic data, respectively.

\begin{itemize}
	\item \textit{Setting 1 with  realistic data}: We randomly classify 16 users from Austin data set  into 4 types. This simulates different demand distributions for types and different storage costs for users based on realistic data.
	\item \textit{Setting 2 with synthetic data}:  We consider 16 users of 4 types, and each type has 4 users. We set the off-peak demand to zero
	and vary the joint peak demands across users (with 7 outcomes).
	For the discrete joint distribution of peak demands across 16 users, we assume  7 outcomes with equal probability. 
	To construct one joint distribution, we
uniformly generating peak demand $D_i^{p,\omega}$ from  $[0,10] (\text{kWh})$ for each user  $i$ at each outcome $\omega$. Based on this, we randomly construct 500 joint peak demand   distributions across  users.
		\end{itemize}

\subsection{Impact of demand correlation within types on the performance of \textbf{PT}}

As we discussed in Section \ref{section:analysis}, the correlation of users' demand within each type will determine the difference between \textbf{PT} and \textbf{PI}. Next, we show that the positive correlation between users' demand in each type will cause a lower gap between  \textbf{PT} and \textbf{PI}.  In the realistic data, most users' demands are positively correlated, which helps the  pricing scheme \textbf{PT} to achieve good performance.

\subsubsection{Setup}

We consider 8 users of 4 types, and each type has 2 users. 
We set the off-peak demand to zero and vary the peak demand. 
We assume the following discrete joint distribution of peak demands across 8 users, which has 7 outcomes with equal probability. 
In each type,  we fix one user's peak demand distribution, and  adjust the other user's demand  by choosing different permutations across different outcomes. For example, in each type, we fix User 1's  peak demand over 7 joint outcomes as
\begin{align*}
 \bm{D}_1=(D_1^{p,\omega_1},D_1^{p,\omega_2}, D_1^{p,\omega_3},D_1^{p,\omega_4},D_1^{p,\omega_5},D_1^{p,\omega_6},D_1^{p,\omega_7})\\=(50,42,34,26,18,10,2)~\text{kWh}.
\end{align*}

The peak demand distributions of User 1 and User 2 are  positively correlated with  coefficient 1 if User 2's peak demand over 7  outcomes  is $\bm{D}_2=e \cdot \bm{D}_1, ~e>0.$
The peak demand distributions are  negatively correlated  with coefficient -1 if User 2's peak demand   is $\bm{D}_2=e \cdot (2,10,18,26,34,42,50)~\text{kWh}, ~e>0.$
We choose the same distributions of  User 1 and User 2 for all 4 types. 
We randomly generate 5000 permutations for User 2's peak demand with $e=0.8$. 
For each permutation, we calculate the ratios $\kappa^{\text{PT}}$ and $\kappa^{\text{PI}}$ as well as the   correlation coefficient of two users. We then report the mean value of the ratios
among different permutation results.

\begin{figure}[t]
	\centering
	\hspace{-1ex}
	\subfigure[]{
		\label{fig:simulation2} 
		\raisebox{-2mm}{\includegraphics[width=1.7in]{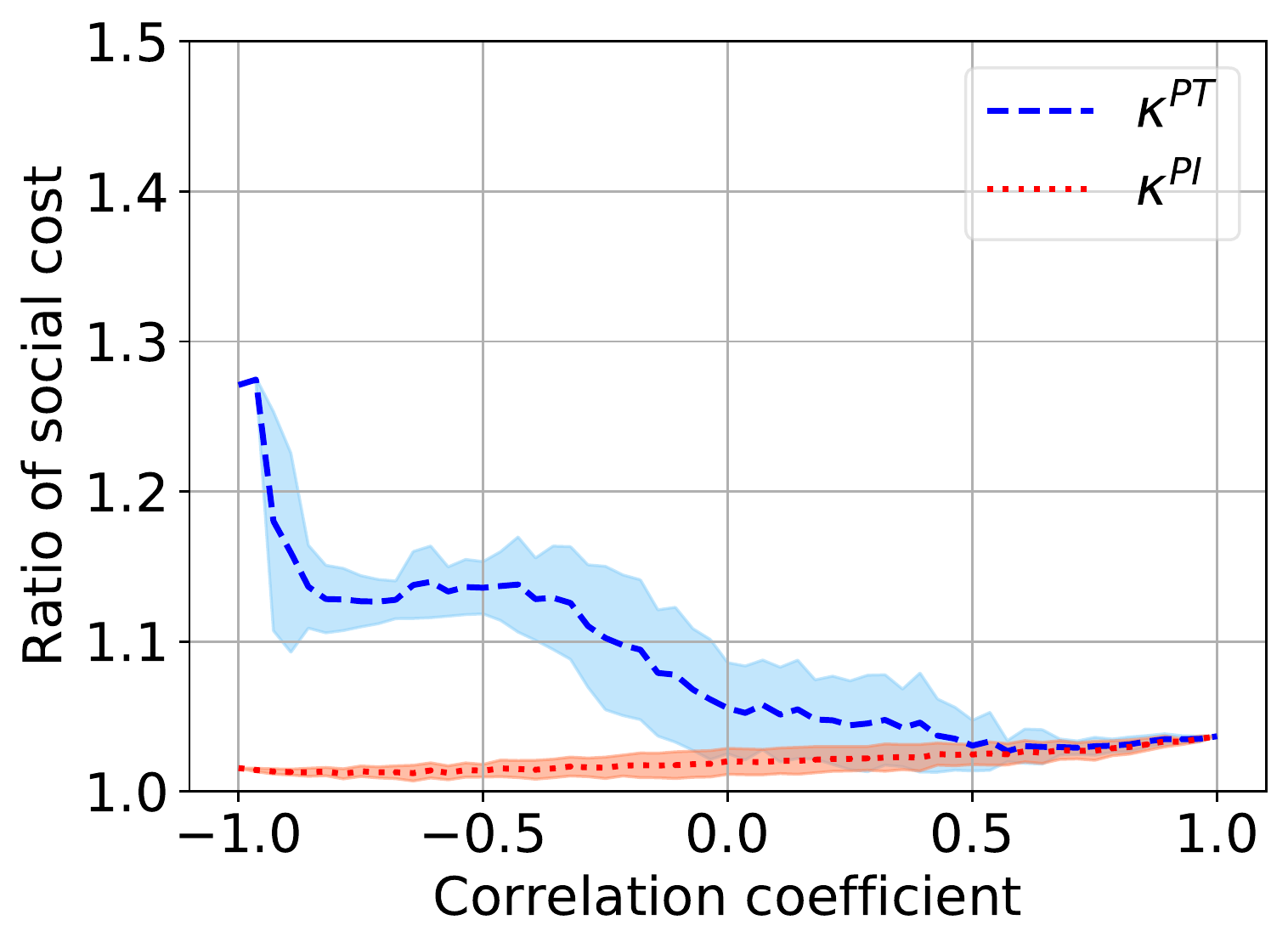}}}
	\hspace{-2.5ex}
	\subfigure[]{
		\label{fig:simulation3} 
		\raisebox{-2mm}{\includegraphics[width=1.7in]{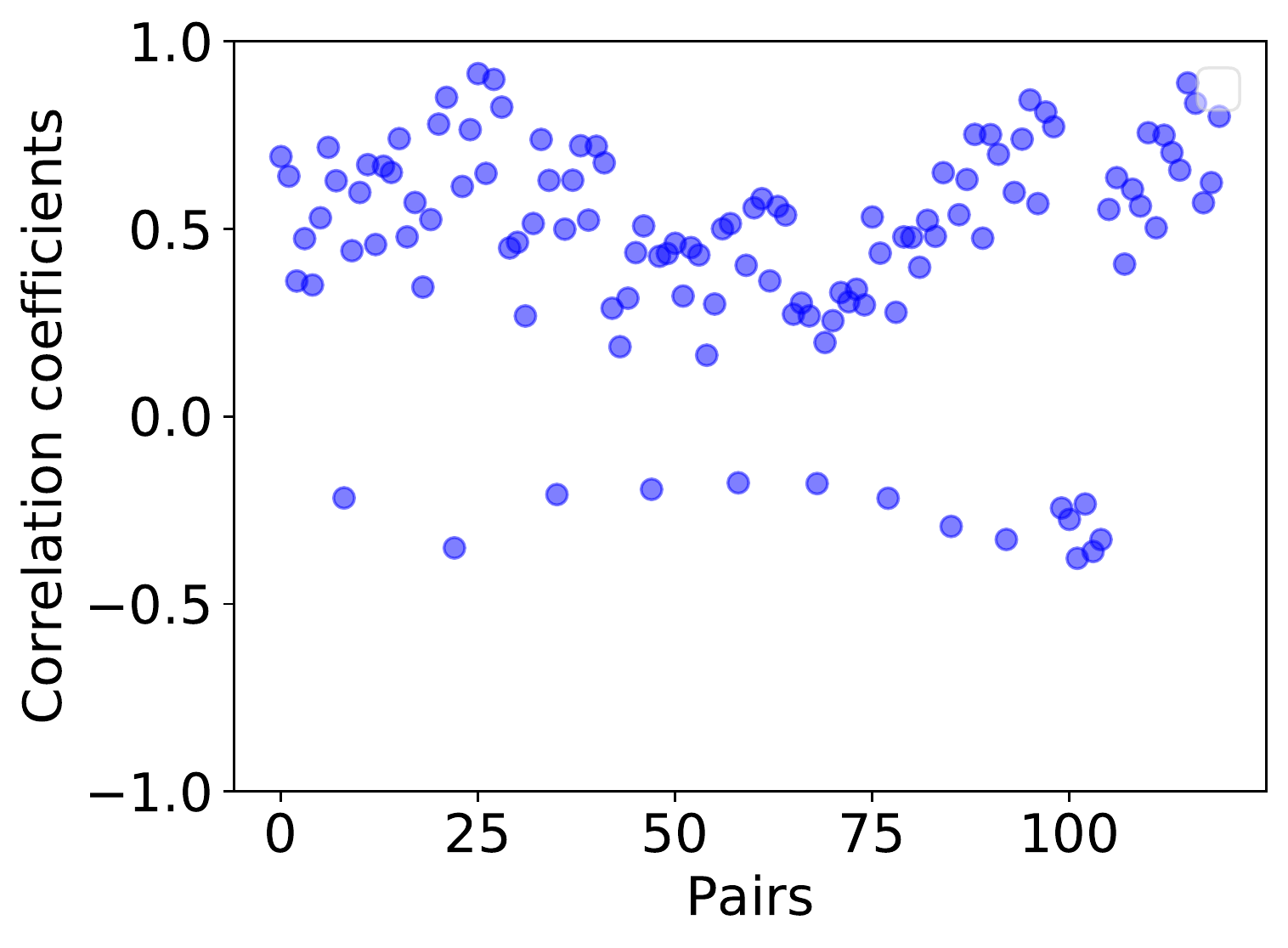}}}
	\vspace{-2mm}
	\caption{\small (a) Ratios $\kappa^{\text{PT}}$ and $\kappa^{\text{PI}}$   with the  correlation coefficient; (b) correlation coefficient of every two users among 16 users from Austin data set.}
	\label{fig:correlation}
	\vspace{-2mm}
\end{figure}

\subsubsection{Results} In Figure \ref{fig:correlation}(a), we show the average ratios $\kappa^{\text{PT}}$ (in blue curve) and $\kappa^{\text{PI}}$  (in red curve) as well as the  one-standard-deviation range as the  correlation coefficient increases. In Figure \ref{fig:correlation}(b),    we show the correlation coefficient of the peak demand distribution between every two users from the 16 users  of the  Austin data set. 

We have the following observations based on  Figure \ref{fig:correlation}. 

\vspace{0.8ex}
 \textit{Observation 4}:  \textit{A positive correlation  leads to a smaller  gap between the pricing	\textbf{PT} and 	\textbf{PI}.} 
\vspace{0.8ex}

 As shown in Figure \ref{fig:correlation}(a), a positive correlation  leads to a smaller  gap between the pricing schemes	\textbf{PT} and 	\textbf{PI}. When the correlation coefficient is 1,  the pricing schemes of 	\textbf{PT} and 	\textbf{PI} are equivalent. 

\vspace{0.8ex}
 \textit{Observation 5}:  \textit{Most users' demands are positively correlated in practice, which can improve the performance of the pricing scheme \textbf{PT}.} 
\vspace{0.8ex}

 As shown in Figure \ref{fig:correlation}(b) in the Austin data set,  most users' demands are positively correlated in practice, which can improve the performance of the pricing scheme \textbf{PT}.  The positive correlation can be because the users' demands are affected by the common  weather, climate, or social environment in one area.

\subsection{Demand Approximation in the ToU pricing}

We will first explain the reason of approximating the loads as constants  within the peak and off-peak periods, respectively. Then, we will show that the error due to such an approximation is relatively small, comparing with the benchmark considering hourly load variations. 
 
  Our model focuses on the two-period ToU pricing in practice, which charges users based on their total demands in the peak period and off-peak period, respectively. The two-period pricing does not directly regulate users' demand in each hour. To calculate the supply cost based on the total demand in the peak and off-peak periods, we adopt an approximation of constant load in each period. Specifically, we approximate the power of the peak period and off-peak
  period (with multiple hours) by the average power (in MWh
  per hour) in these periods, respectively.  The main purpose of such an approximation is to capture
 the load difference between the peak period and off-peak
 period for the two-period pricing structure.

 \begin{figure}[ht]
 	\centering
 	\includegraphics[width=2.7in]{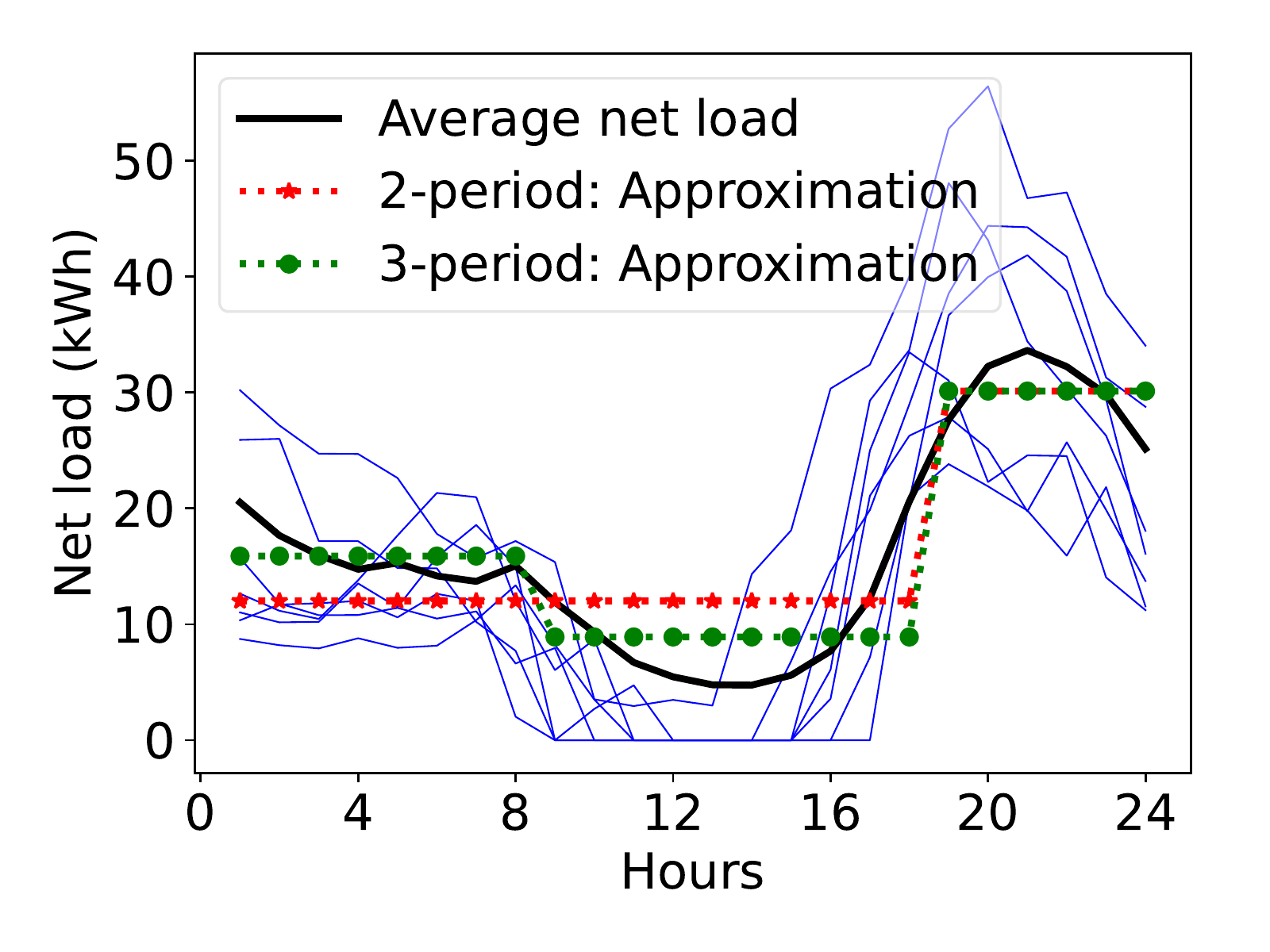}
 	\vspace{-2mm}
 	\caption{\small Two-period and three-period approximation.}
 	\label{fig:appr}
 	
 \end{figure}
 
 Based on  the realistic data of users’ aggregate load  (used in Section.VII of the main text), we can show that such an assumption of two-period constant power can still provide a good approximation for the more elaborate model of 24-hour variable load, in terms of  the supply cost.   We illustrate  the approximation in Figure \ref{fig:appr}. We show the  24-hour variable load (averaged over one-year data) in  black  curve. We also present the   2-period approximation with a red curve, together with the 3-period approximation with a green curve.  The supply cost under the 2-period constant-load approximation has a small gap of  6.2\% comparing with the supply cost
 computed based on the 24-hour variable load, while the 3-period approximation has a  gap of 3.5\%. This shows that the 2-period constant-power approximation is quite accurate in terms of predicting the total supply cost. In the future work, we may consider introducing an additional regularizer on the ToU  pricing so that the utility can regulate the users' demand at each hour, which will allow the utility to calculate the supply cost more accurately based on 24-period demand.

\subsection{Model generalization for elastic demand}

Users can have both inelastic demand and elastic demand in practice. If users do not install energy storage, under two-period ToU pricing, they can only shift the elastic demand from a high-price period to a low-price period. If they further invest in energy storage, then they can further shift the inelastic demand.  To incorporate the elastic demand, we generalize our model and provide additional simulation results about the impact of elastic demand. Our high-level finding is that having additional  elastic demand with a low shift cost  will reduce users' demand for storage but improve the social welfare.
 
 \subsection*{$\triangleright$ Model generalization}
 
Under the ToU pricing, for user $i$,  we model the  elastic demand  in the peak period as a random variable $\mathcal{D}_i^{e}$, which can be  shifted from the peak period to the off-peak period. We assume a linear inconvenience cost $e_i$ for user $i$ to shift  one unit  elastic demand. 

We note that the demand-shift cost can be higher or lower than the storage investment cost. Some demand can be easily shifted, such as the usage of washing machine, which will incur a low demand-shift cost. Some demand is more difficult  to shift, such as the need of using lights at night. Specially, the inelastic demand can be regarded as the elastic demand with an infinite demand-shift cost.  In the following model generalization of the elastic demand, we assume that the demand-shift cost is smaller than the storage cost,  i.e., $e_i<\theta_i$, since the storage investment cost is usually high for users. Thus, under the ToU pricing, each user will always first try to shift the elastic demand and then use storage to shift the remaining part. For the elastic demand with shift cost higher than the storage cost, we will just treat it as inelastic demand that has an infinite shift cost. In the future work, we will further study the case of elastic demand with bounded shift cost that is higher than the storage cost.

In our original model considering only inelastic demand, we group users into types based on the storage cost. With the elastic demand, we need to consider a set  $\mathcal{K}=\{1,2,\ldots,K\}$ of user types with two-dimensional private information, corresponding to different storage costs as well as  different demand-shift costs. The unit daily cost of storage capacity  and demand-shift cost for type $k$ is denoted as   $(\theta_k, e_k)$. Recall that we assume $e_k<\theta_k$. Similar to the original model, we denote the random daily aggregate peak and off-peak demands for a type $k$ as  $\mathcal{D}_k^{p}$ and $\mathcal{D}_k^{o}$, respectively. We denote the aggregate elastic demand for  type $k$ as  $\mathcal{D}_k^{e}$.

In the formulation of two-stage optimization  for the pricing schemes  \textbf{PT} and \textbf{PI}, we also need to include the demand-shift cost. We focus on  the  model for the pricing scheme  \textbf{PT}, where we highlight the new elements related to the  elastic demand  in purple. The model of  \textbf{PI} follows the same structure by considering each user as one  type.

 \begin{figure*}[t]
	\centering
	\hspace{-1ex}
	\subfigure[]{
		\raisebox{-2mm}{\includegraphics[width=1.9in]{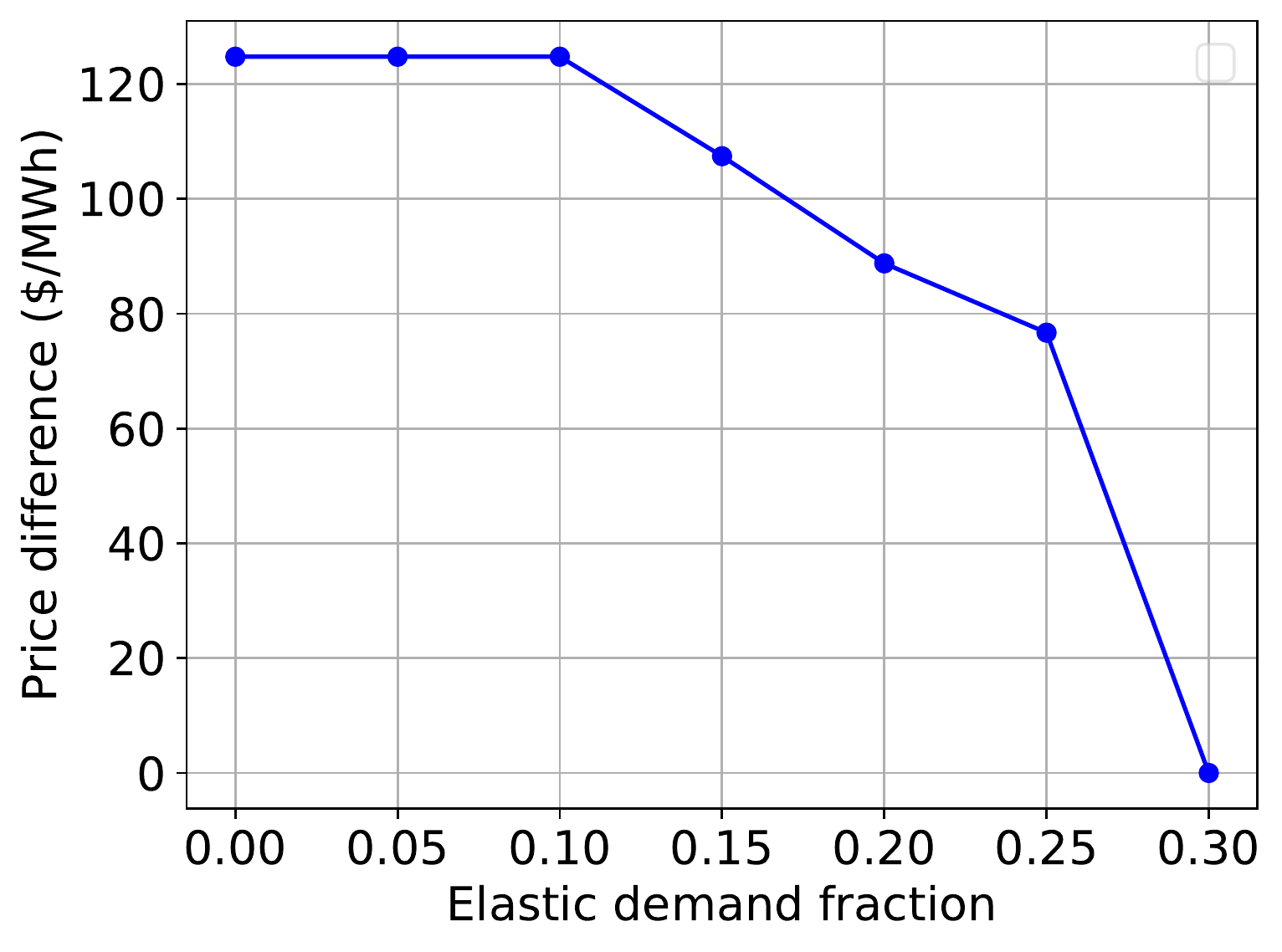}}}
	\hspace{-1.5ex}
	\subfigure[]{
		\raisebox{-2mm}{\includegraphics[width=1.9in]{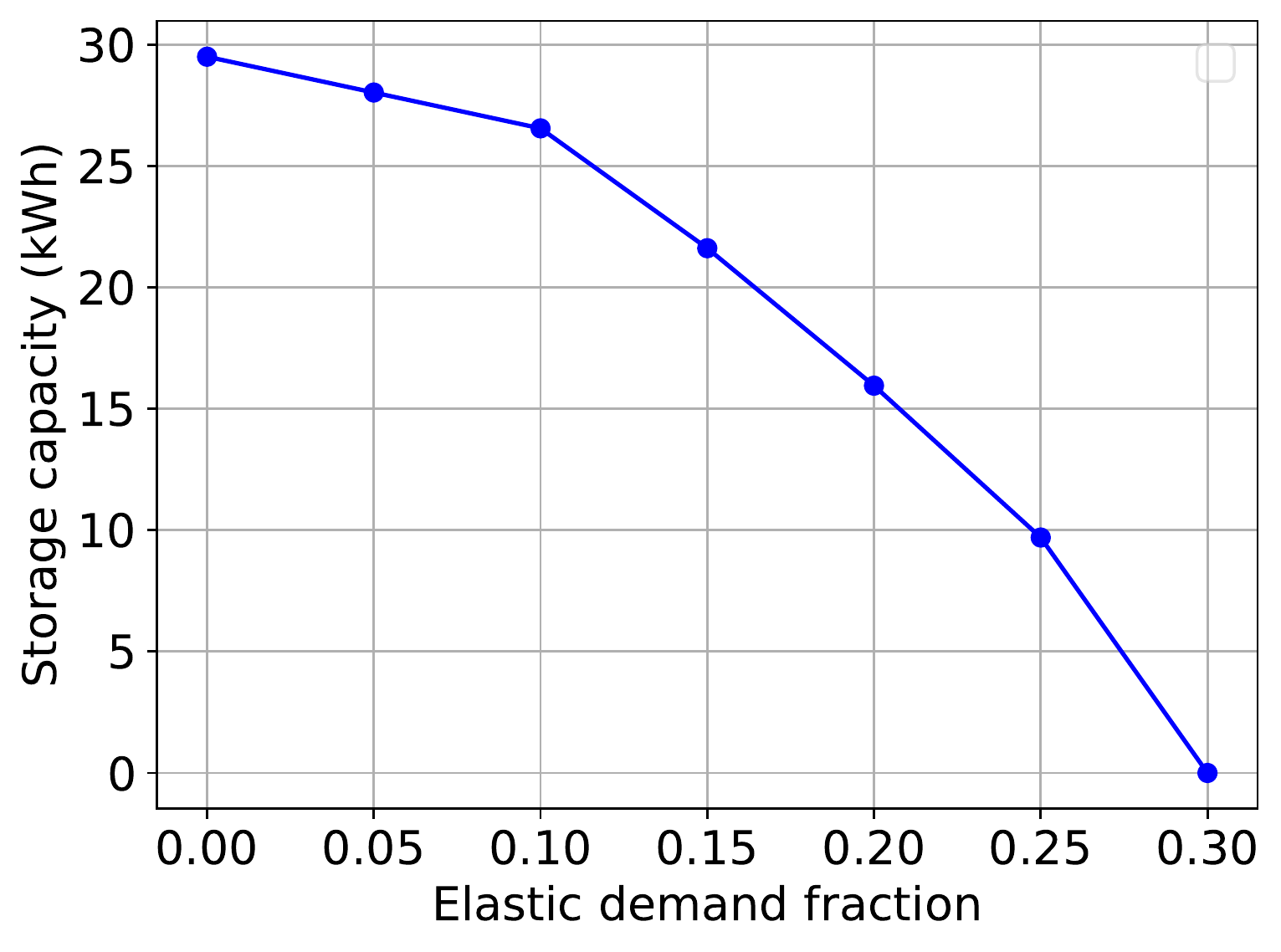}}}
		\hspace{-1.5ex}
	\subfigure[]{
		\raisebox{-2mm}{\includegraphics[width=1.9in]{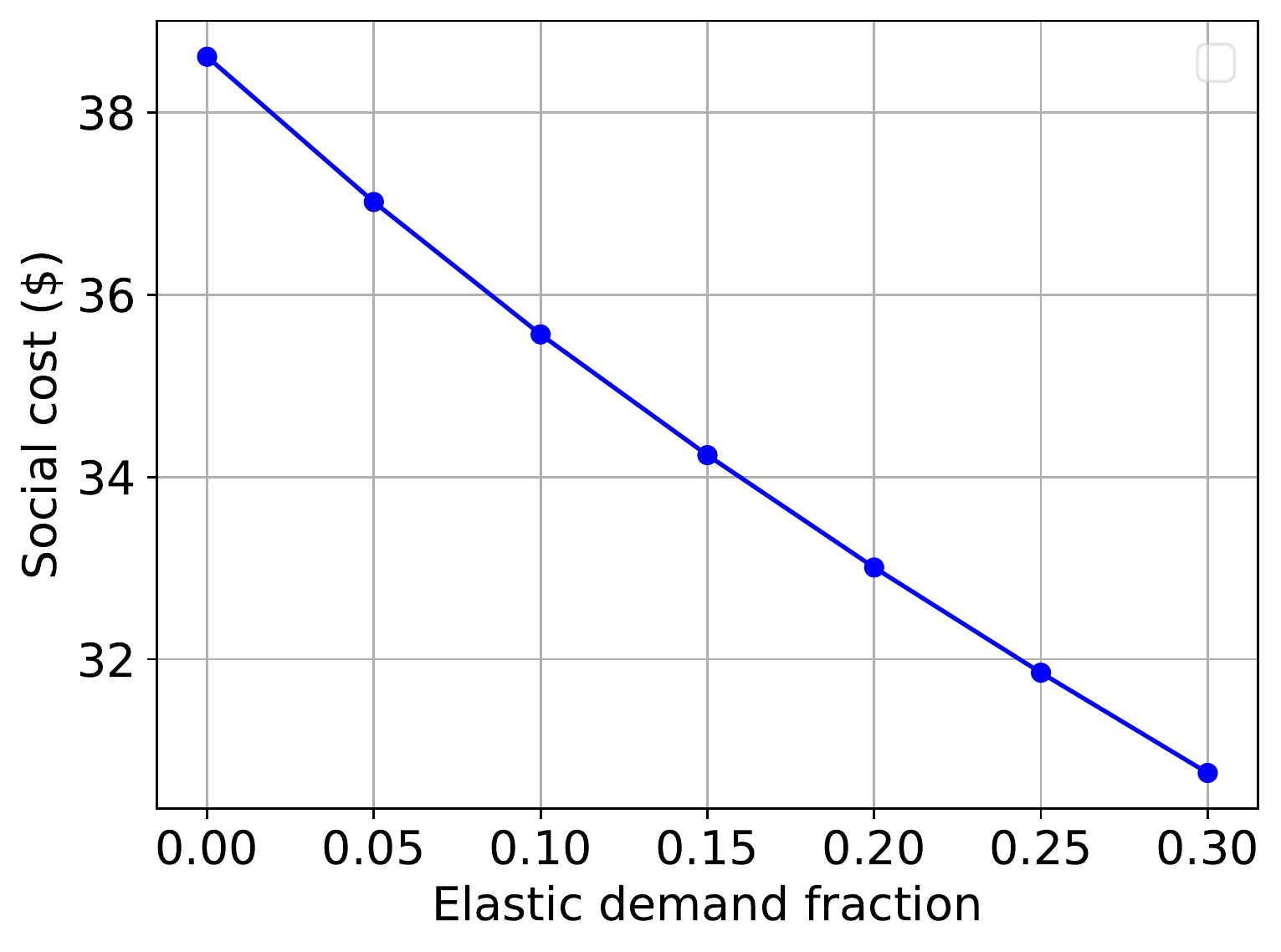}}}
	\vspace{-2mm}
	\caption{(a) \small Optimal price difference  $p^\Delta$; (b) Total invested storage capacity; (c) Social cost with the fraction of elastic demand. }
	\label{fig:elastic1}
	\vspace{-3mm}
\end{figure*} 

\subsubsection*{$\triangleright$ $\triangleright$ Stage II} In Stage II, compared with the original model, each type also needs to  decide how much elastic demand to be shifted under the ToU pricing.  The demand-shift cost is included in the total energy cost. We highlight those changes in purple.

\noindent \textbf{Problem {PT-Stage-II}:  Type $k$'s  Cost Minimization}
\vspace{-2mm}
\begin{align}
	\text{(Period-1)}~ \min ~&\theta_k c_k+\mathbb{E}_{\bm{\mathcal{D}}_k}[{Q}(c_k,\bm{\mathcal{D}}_k)]\\
	~\text{s.t.} ~&c_k\geq 0,\\
	\text{var:} ~&c_k.\notag
\end{align}
Given the storage capacity $c_k$, each type minimizes the energy cost in Period-2  for each demand realization $\bm{D}_k$, which decides the shifted amount of elastic demand $q_k$.
\vspace{-1mm}
\begin{align}
	\text{(Period-2)}~{Q}(c_k,\bm{D}_k):= \min~& \textcolor{purple}{e_k q_k}+ (p^p D_k^p\hspace{-0.8mm}-\textcolor{purple}{q_k}-\hspace{-0.8mm}s_k)\hspace{-0.8mm}\notag \\&+ p^o (D_k^o\hspace{-0.8mm}+\textcolor{purple}{q_k}+\hspace{-0.8mm}s_k)\\
	~\text{s.t.~} &0\leq s_k\leq c_k, \\
	&s_k\leq D_k^{p}, \\
	&\textcolor{purple}{0\leq q_k\leq D_k^e},\\
	\text{var}:&~ s_k,\textcolor{purple}{q_k}.\notag
\end{align}\par{\vspace{-1mm}}
\noindent Given the ToU pricing $\bm{p}=(p^p,p^o)$,  we denote each type $k$'s optimal charging decision as $s_k^*(\bm{p},\bm{D}_k)$ and optimal shift demand as $q_k^*(\bm{p},\bm{D}_k)$ for each realization demand $\bm{D}_k$.

\subsubsection*{$\triangleright$ $\triangleright$ Stage I}  

In Stage I, the social cost also needs to include the demand-shift cost of all types.

  \noindent \textbf{Problem PT-Stage-I: Type-based Pricing for  Social Cost Minimization }
  \vspace{-1mm}
  \begin{align}
  	\min~&  \sum_{k\in \mathcal{K}}\theta_k c_k(\bm{p})+\textcolor{purple}{\sum_{k\in \mathcal{K}} \mathbb{E}_{\bm{\mathcal{D}}}~  e_k q_k(\bm{p},\bm{\mathcal{D}})}+\mathbb{E}_{\bm{\mathcal{D}}}~  G(\bm{s}(\bm{p},\bm{\mathcal{D}}),\bm{\mathcal{D}})\\
  	\text{s.t.} ~&{p^p\geq p^o\geq 0}\\
  	\text{var:}~ &{p^p,p^o}\notag.
  \end{align}
The highlighted term $\textcolor{purple}{\sum_{k\in \mathcal{K}} \mathbb{E}_{\bm{\mathcal{D}}}~  e_k q_k(\bm{p},\bm{\mathcal{D}})}$ is the demand-shift cost of all types in Stage II.

\subsection*{$\triangleright$ Solution method}
We focus on  the  solution method for the pricing scheme  \textbf{PT}. The solution method of  \textbf{PI} follows the same structure by considering each user as one  type.

  \subsubsection*{$\triangleright$ $\triangleright$  Stage II}
  
If the price difference $p^\Delta\leq e_k<\theta_k$, type $k$ will not shift any demand in any demand realization, which is the same as the original model.
  
  If the price difference $p^\Delta>e_k$, type $k$ will shift all the elastic demand $\mathcal{D}_k^{e}$ from the peak period to the off-peak period. The new peak demand is changed to $\mathcal{D}_k^{p'}=\mathcal{D}_k^{p}-\mathcal{D}_k^{e}$ and the off-peak demand is changed to  $\mathcal{D}_k^{o'}=\mathcal{D}_k^{o}+\mathcal{D}_k^{e}$.  The storage investment and operation of type $k$  depend on the new demand $(\mathcal{D}_k^{p'}, \mathcal{D}_k^{o'})$, which can be solved in the same way as in Proposition 1 and  Proposition 2 of the main text.

  \subsubsection*{$\triangleright$ $\triangleright$  Stage I}
The utility searches a new threshold price set  $\mathcal{P}_k'$ of the price difference $p^\Delta$ as follows. This new set $\mathcal{P}_k'$  includes the original  threshold set 	$\mathcal{P}_k $  in (11) of the main text   and the additional elastic costs of all the users, i.e., 
\begin{align}
	\mathcal{P}_k'= 	\mathcal{P}_k \bigcup \{e_k, k \in \mathcal{K}\}.
\end{align}
 
The utility searches the threshold price set 	$\bigcup_k \mathcal{P}_k'$ as in Algorithm 1 of the main text to obtain the optimal solution.
 
 \subsection*{$\triangleright$ Numerical study}
 
We conduct a numerical study to show the  impact of the elastic demand. The simulation data is the same as the data in Section VII of the main text. We consider each user as one type. We investigate  the results as the fraction of elastic demand in the peak demand  increases from 0 to 30\%, where we show  the optimal price difference in Figure \ref{fig:elastic1}(a), the total invested storage capacity of all users in Figure \ref{fig:elastic1}(b), and the social cost in Figure \ref{fig:elastic1}(c). 

As shown in Figure \ref{fig:elastic1}, the optimal price difference, the invested storage  capacity, and the social cost  all decrease as the elastic demand increases. The reason is that more elastic demand decreases users' demand  for energy storage, which  also decreases the utility's demand for storage (in order to flatten the system load and reduce the social cost). Thus, the utility will also set a lower price difference and the social cost will also decrease. In summary, more elastic demand with low shift cost will reduce users' demand for storage but improve the social welfare.

\subsection{Model generalization for imperfect charge and discharge efficiency, and degradation cost}

Our original model only considers the investment cost as the storage cost. The degradation cost will further increase the storage cost. We can generalize our model to incorporate a linear  degradation cost that is  proportional to the  charge and discharge quantity. 	We can further incorporate  a generalization of  imperfect charge and discharge efficiency, which can also be viewed as an additional form of storage cost. We provide the details of these two generalizations below.

 \subsection*{$\triangleright$ Model generalization }
For the model of  user $i$, we denote the charge and discharge efficiency of storage as $\eta_i^c$ and $\eta_i^d$, respectively. For the degradation cost, we denote by $\tau_i$ the  cost of each unit of charge and discharge amount for user $i$. 
 
 Recall that we group users into types based on the storage cost. Since the storage cost depends on the storage technologies, we assume that users in each type also have the same charge and discharge efficiency and the storage degradation cost. We next present the optimization model  generalization of Stage I and Stage II, respectively.  We focus on  the  model for the pricing scheme  \textbf{PT}. The model of  \textbf{PI} follows the same structure by considering each user as one  type.
 
 \subsubsection*{$\triangleright$ $\triangleright$ Stage II}
 Compared with our original model, the imperfect charge and discharge efficiency will cause energy loss in the storage operation. If the purchased electricity amount $s_k$ is charged into storage in the off-peak period, only the amount $\eta_k^c s_k$ can be stored in the storage  and the amount $\eta_k^c\eta_k^d s_k$  can be discharged from the storage to serve the peak demand.
 Besides,  in each operation horizon, the charge and discharge will incur the degradation cost $\tau_ks_k$ and $\tau_ks_k \eta_k^d \eta_k^c$, respectively.
 
We show the generalized model in the following, where we highlight the changes in purple compared with the original model.

 \noindent \textbf{Problem {PT-Stage-II}:  Type $k$'s Cost Minimization}
 \begin{align}
 	(\text{Period-I})~~~\min~ &\theta_k c_k+\mathbb{E}_{\bm{\mathcal{ D}}_k}[{Q}(c_k,\bm{\mathcal{ D}}_k)]\\
 	~\text{s.t.~} 
 	&c_k\geq 0,\\
 	\text{var}: ~&c_k.\notag
 \end{align} 
 For each realization $\bm{D}_k$ of $\bm{\mathcal{D}}_k$, 
 \begin{align}
 	(\text{Period-II})~~~{Q}(c_k,\bm{D}_k)=&\min p^p (D_k^p-\textcolor{purple}{\eta_k^d \eta_k^c} s_k)+p^o (D_k^o+s_k)\notag \\&~~~~~+\textcolor{purple}{\tau_ks_k(1+\eta_k^d \eta_k^c)}\\
 	&~\text{s.t.~} 
 	0\leq \textcolor{purple}{\eta_k^c} s_k\leq   c_k, \\
 	&~~~~~~\textcolor{purple}{\eta_k^c \eta_k^d} s_k\leq D_k^{p}, \\
 	&~\text{var}: s_k. \notag
 \end{align} 
 
 \subsubsection*{$\triangleright$ $\triangleright$ Stage I}	
In Stage I, compared with the original model, the storage degradation cost is also included in the social cost highlighted in purple.
 
 \noindent \textbf{Problem PT-Stage-I: Type-based Pricing for  Social Cost Minimization }
 \vspace{-1mm}
 \begin{align}
 	\min~&  \sum_{k\in \mathcal{K}}\theta_k c_k(\bm{p})+	\textcolor{purple}{\sum_{k\in \mathcal{K}}  \mathbb{E}_{\bm{\mathcal{D}}} \left[\tau_k (1+\eta_k^d \eta_k^c) {s}_k(\bm{p},\bm{\mathcal{D}})\right]}\notag \\&+
 	\mathbb{E}_{\bm{\mathcal{D}}}~  G(\bm{s}(\bm{p},\bm{\mathcal{D}}),\bm{\mathcal{D}})\\
 	\text{s.t.} ~&{p^p\geq p^o\geq 0}\\
 	\text{var:}~ &{p^p,p^o}\notag,
 \end{align}
 where the invested capacity ${c}_k(\bm{p})$, and charging and discharging decision  $s_k(\bm{p},\bm{D}_k)$ are  type $k$'s decisions in Stage II.

 \subsection*{$\triangleright$ Solution method}
 
To solve the generalized model, the  key is to analyze the solution structure in Stage II.  In order to solve Stage-II problem with the method in the original model,  we will set  equivalent variables and parameters for the original model in the following equations. Specifically, (i) we set the equivalent variable  $p_k^{\Delta\dagger}$ for the original variable $p^\Delta$. Note that $p^\Delta$ is uniform for all the types in the original model. However, in the generalized model,  $p_k^{\Delta\dagger}$ is related to each type's charge/discharge efficiency and  degradation cost. (ii) We set the equivalent decision variable $c_k^{\dagger}$ to replace the original  variable $c_k$. (iii) We set the equivalent parameters $D_k^{p^\dagger}$ and $\theta_{k}^\dagger$ for the original parameters $D_k^{p}$ and $\theta_{k}$. (iv) For the other parameters and variables including $D_k^{o}$, $s_k$, $p^p$, and $p^o$, we keep them the same as those in our original model.
 
 \begin{itemize}
 	\item $p_k^{\Delta\dagger} =p^\Delta \eta_k^d \eta_k^c -p^o (1-\eta_k^d \eta_k^c)-\tau_k(1+\eta_k^d \eta_k^c)$
 	\item $c_k^{\dagger} =\frac{1}{\eta_k^c} c_k$
 	\item $D_k^{p^\dagger}= \frac{D_k^p}{\eta_k^c\eta_k^d}$
 	\item  $\theta_{k}^\dagger={\eta_k^c} \theta_k$
 \end{itemize}
 
 Based on the equivalent parameters and variables, we rebuild the optimization problems of Stage-II in the following. Note that we  do not need to modify Stage-I model.

 \noindent \textbf{Problem {PT-Stage-II}:  Type $k$'s Cost Minimization}
 \begin{align}
 	(\text{Period-I})~~~\min~ &\theta_k^\dagger c_k^\dagger+\mathbb{E}_{\bm{\mathcal{ D}}_k^\dagger}[{Q}(c_k^\dagger,\bm{\mathcal{ D}}_k^\dagger)]\\
 	~\text{s.t.~} 
 	&c_k^\dagger\geq 0,\\
 	\text{var}: ~&c_k^\dagger.\notag
 \end{align} 
 For each realization $\bm{D}_k^\dagger$ of $\bm{\mathcal{D}}_k^\dagger$, 
 \begin{align}
 	(\text{Period-II})~~~{Q}(c_k,\bm{D}_k^\dagger)=&\min \textcolor{purple}{\eta_k^d \eta_k^c} \cdot p^p (D_k^{p\dagger}- s_k)\notag\\&+p^o (D_k^{o}+s_k)+\textcolor{purple}{\tau_ks_k(1+\eta_k^d \eta_k^c)}\\
 	&~\text{s.t.~} 
 	0\leq  s_k\leq   c_k^\dagger, \\
 	&~~~~~~ s_k\leq D_k^{p\dagger}, \\
 	&~\text{var}: s_k. \notag
 \end{align}

 Next, we show how we solve the generalized two-stage optimization problem  using the equivalent variables and parameters.
 \subsubsection*{$\triangleright$ $\triangleright$ Stage II}
 
We can directly solve Stage-II problem by using the equivalent variables and parameters $p_k^{\Delta\dagger}, ~c_k^{\dagger},~D_k^{p^\dagger},~\theta_{k}^\dagger$ to replace $p^{\Delta}, ~c_k,~D_k^{p},~\theta_{k}$ in Proposition 1 and Proposition 2 of the main text.
 
 \subsubsection*{$\triangleright$ $\triangleright$ Stage I}
 
The generalized model increases the complexity for solving the utility's pricing problem in Stage I.   In our original model, each type makes decision by comparing its  cost $\theta_k$ to the price difference  $p^\Delta$. Thus, the utility only searches the threshold set for $p^\Delta$  to  decide the optimal $p^{\Delta*}$. The peak price $p^p$ and off-peak price $p^o$ can be freely determined without affecting the optimal results. That is not the case for the generalized model because in the generalized model, each type makes decision by comparing its equivalent cost $\theta_k^\dagger$ with the equivalent price difference $p_k^{\Delta\dagger} =p^\Delta \eta_k^d \eta_k^c -p^o (1-\eta_k^d \eta_k^c)-\tau_k(1+\eta_k^d \eta_k^c)$  that involves not only $p^\Delta$  but also off-peak price $p^o$.  
 
 To solve the Stage-I problem, we first exhaustively search  $p^o$ over a feasible range  $[\underline{p}^o,\overline{p}^o]$.  Then, for each given $p^o$, we search a threshold set of $p^\Delta$ to determine the optimal $p^{\Delta*}(p^o)$, which we explain in detail later. Finally, we  choose an optimal $p^{o*}$  in the range  $[\underline{p}^o,\overline{p}^o]$ that minimizes the social cost.  
 
We show how,  given $p^o$, we can construct and search in the threshold set of $p^\Delta$ to determine the optimal $p^{\Delta*}(p^o)$.  First, we construct a  threshold set $\mathcal{P}_{k}^\dagger$ of $p^\Delta$ for each type $k$, which is shown in the following. Then, we search the threshold price set 	$\bigcup_k \mathcal{P}_k^\dagger$ in the same way as in Algorithm 1 of the main text. 
 \begin{align}
 &	\mathcal{P}_k^\dagger=\{A_k\}\bigcup \left\{ \frac{\theta_k/\eta_k^d}{\sum_{\omega=m}^{\mid \Omega_k^p\mid} \rho^{\omega}}+A_k ,\forall m=1,2\ldots \mid \Omega_k^p\mid \right\},
 \end{align}
 where we let
$A_k:=\frac{\tau_k(1+\eta_k^d \eta_k^c)+p^o(1-\eta_k^c\eta_k^d)}{\eta_k^c \eta_k^d}$.


 
 Next, we conduct simulations to show the impact of storage charge and discharge efficiency as well as the degradation cost.

 \subsection*{$\triangleright$ Numerical study}

We conduct a numerical study to show the  impact of the charge and discharge efficiency and the degradation cost. The simulation data is the same as the Austin data in Section VII (Numerical Study of the main text).
 
 \begin{figure}[t]
 	\centering
 	\hspace{-1ex}
 	\subfigure[]{
 		\raisebox{-2mm}{\includegraphics[width=1.6in]{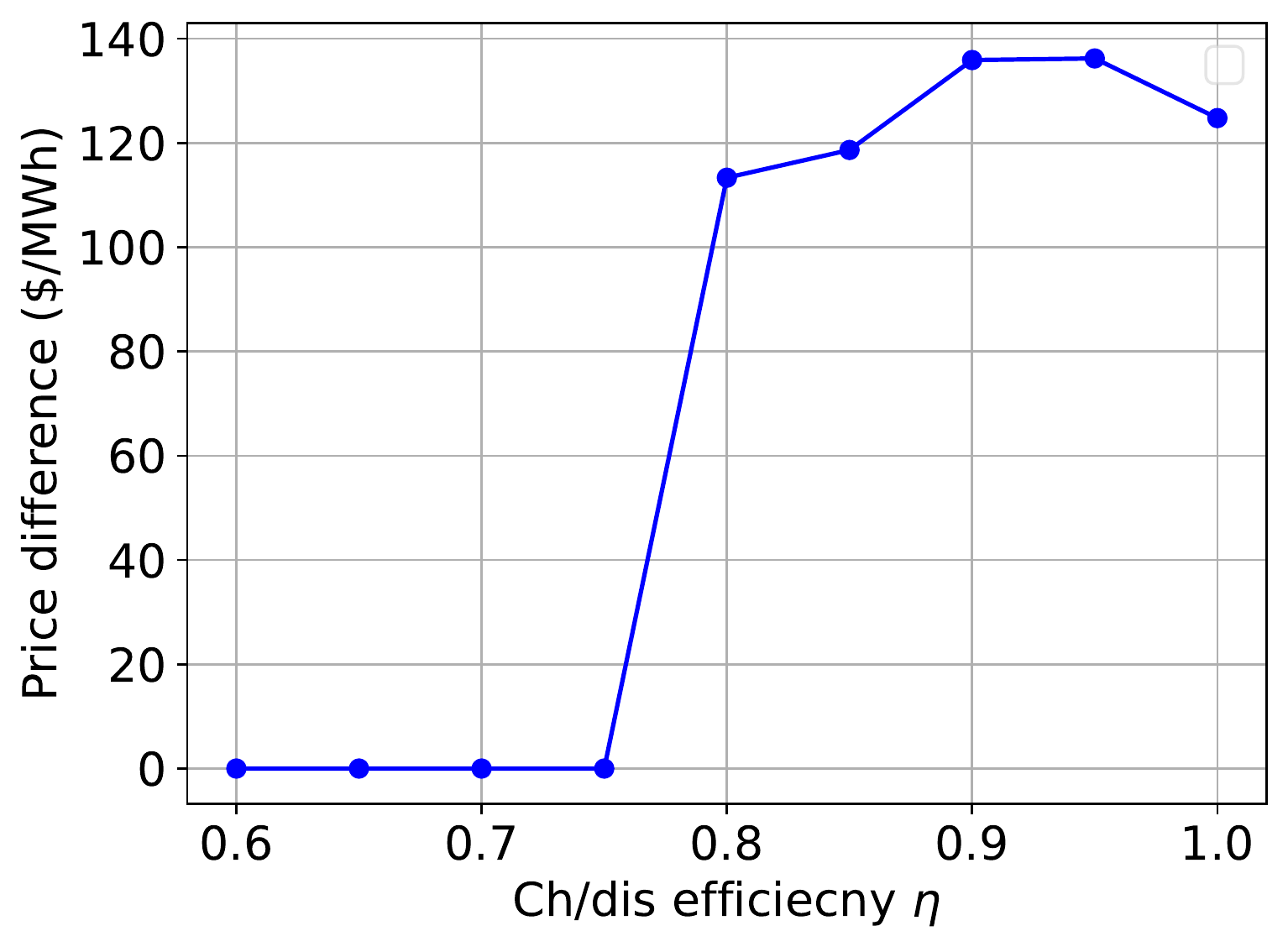}}}
 	\hspace{-1.5ex}
 	\subfigure[]{
 		\raisebox{-2mm}{\includegraphics[width=1.6in]{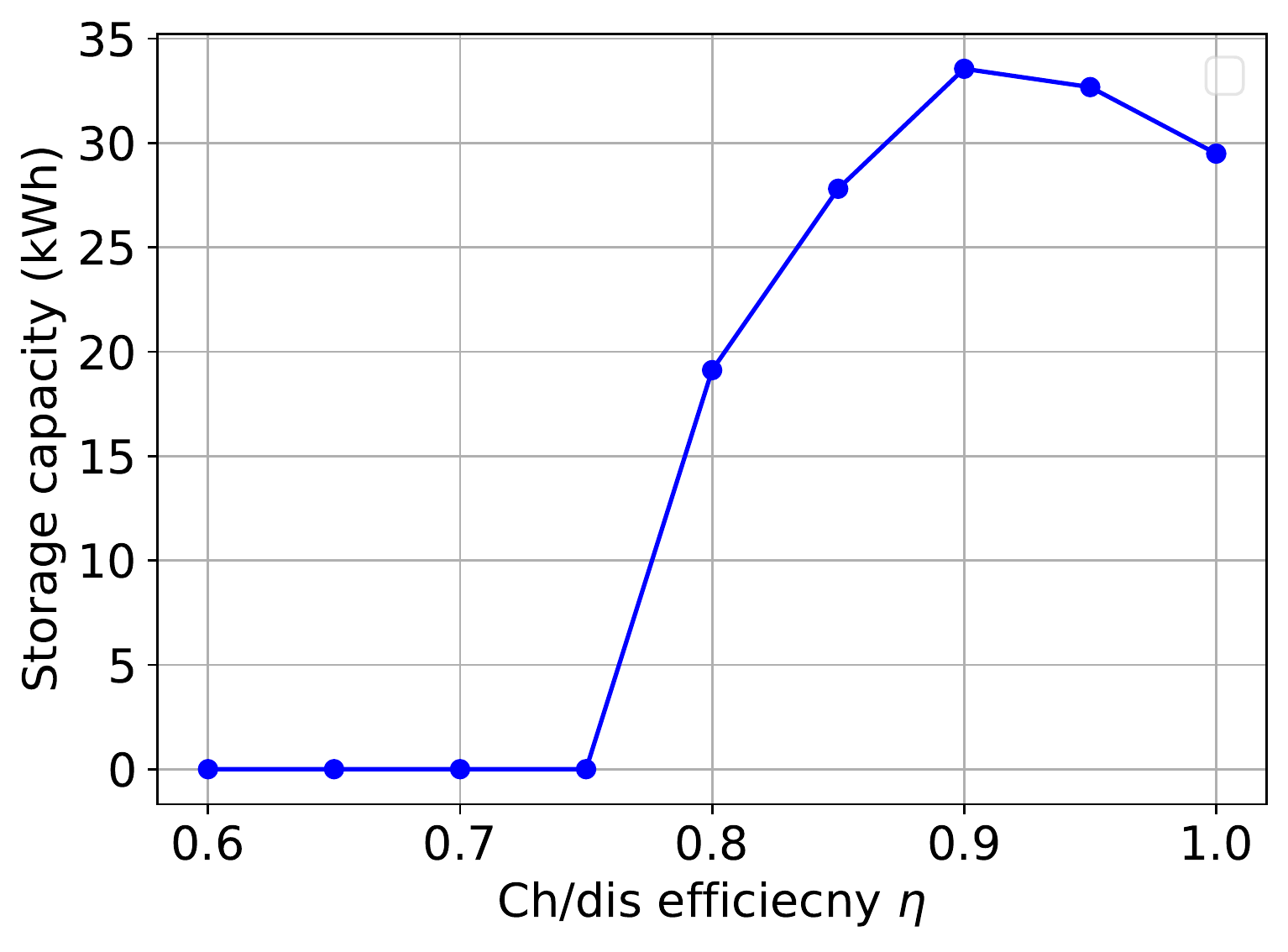}}}
 	\hspace{-1.5ex}
 	\subfigure[]{
 		\raisebox{-2mm}{\includegraphics[width=1.6in]{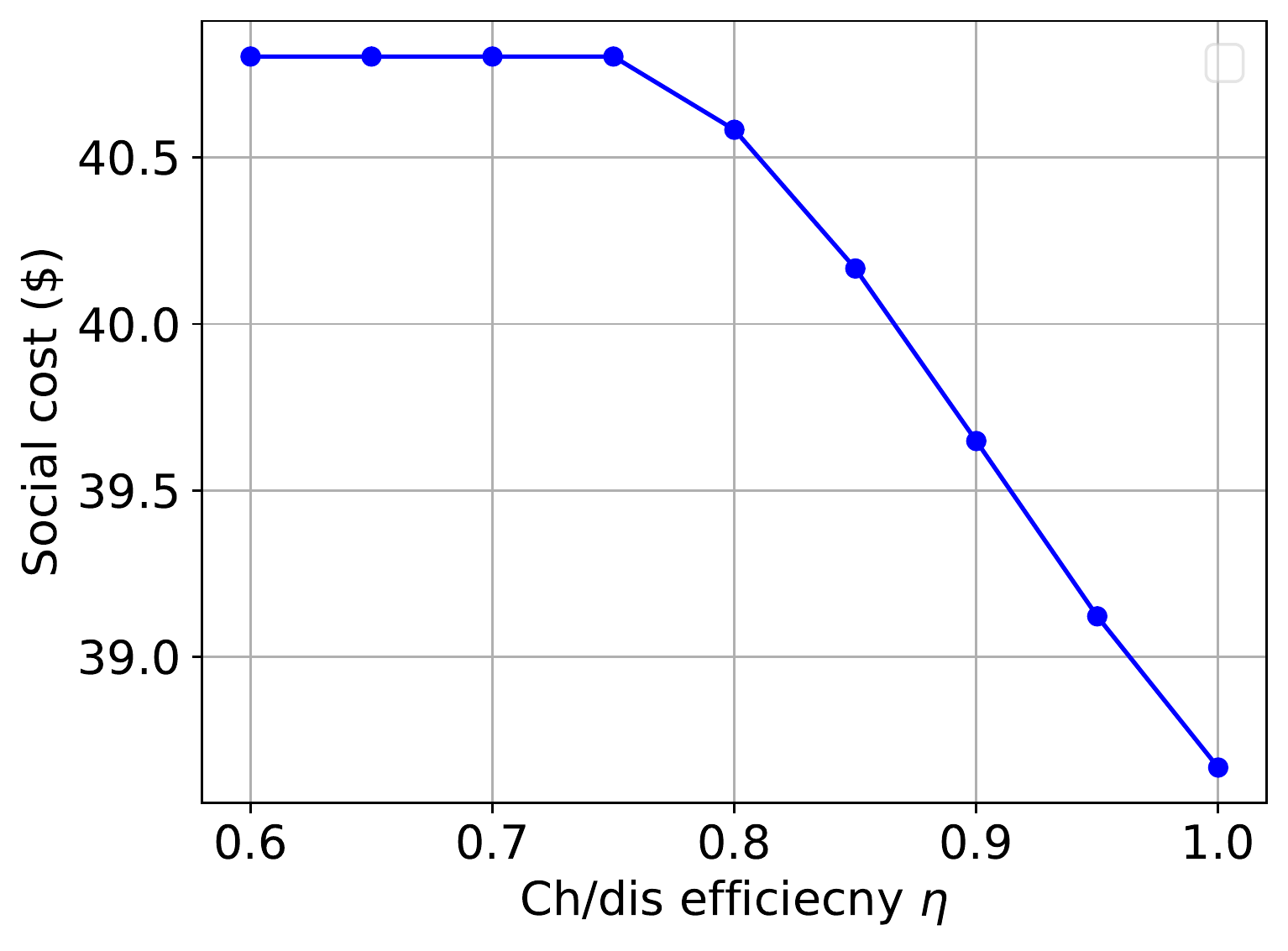}}}
 	\vspace{-2mm}
 	\caption{(a) \small Optimal price difference  $p^\Delta$. (b) Total invested storage capacity. (c) Social cost with the ch/dis efficiency. }
 	\label{fig:eff1}
 	\vspace{-3mm}
 \end{figure} 
 \vspace{1ex}
\noindent \textbf{Charge and discharge efficiency:} In Figure \ref{fig:eff1}, we consider each user as one type and set the same charge and discharge efficiency for all users, i.e., $\eta_i^c=\eta_i^d=\eta$ for all $i\in \mathcal{I}$.
 We investigate  the results as the charge and discharge efficiency $\eta$  increases from 0.6 to 1, where we show  the optimal price difference $p^{\Delta*}$ in Figure \ref{fig:eff1}(a), the total invested storage capacity of all users in Figure \ref{fig:eff1}(b), and the social cost in Figure \ref{fig:eff1}(c). 
 
 We have the following observations.  We see in Figure \ref{fig:eff1}(a) and (b)  that the optimal price difference and  the invested storage  capacity will first increase and then decrease with the efficiency. The reason is that when the efficiency is too low, it is not beneficial for users to invest in storage, which leads to zero storage capacity and zero price difference ($\eta\leq 0.75$). When the efficiency increases ($0.75<\eta\leq 0.9$), the price difference will increase and incentivize the increasing storage capacity. However, when the efficiency is too high ($\eta\geq 0.9$), to avoid the over-investment of storage, the price difference will decrease and incentivize less storage investment. In Figure \ref{fig:eff1}(c), the social cost always decreases as the efficiency increases.

 \begin{figure}[t]
 	\centering
 	\hspace{-1ex}
 	\subfigure[]{
 		\raisebox{-2mm}{\includegraphics[width=1.6in]{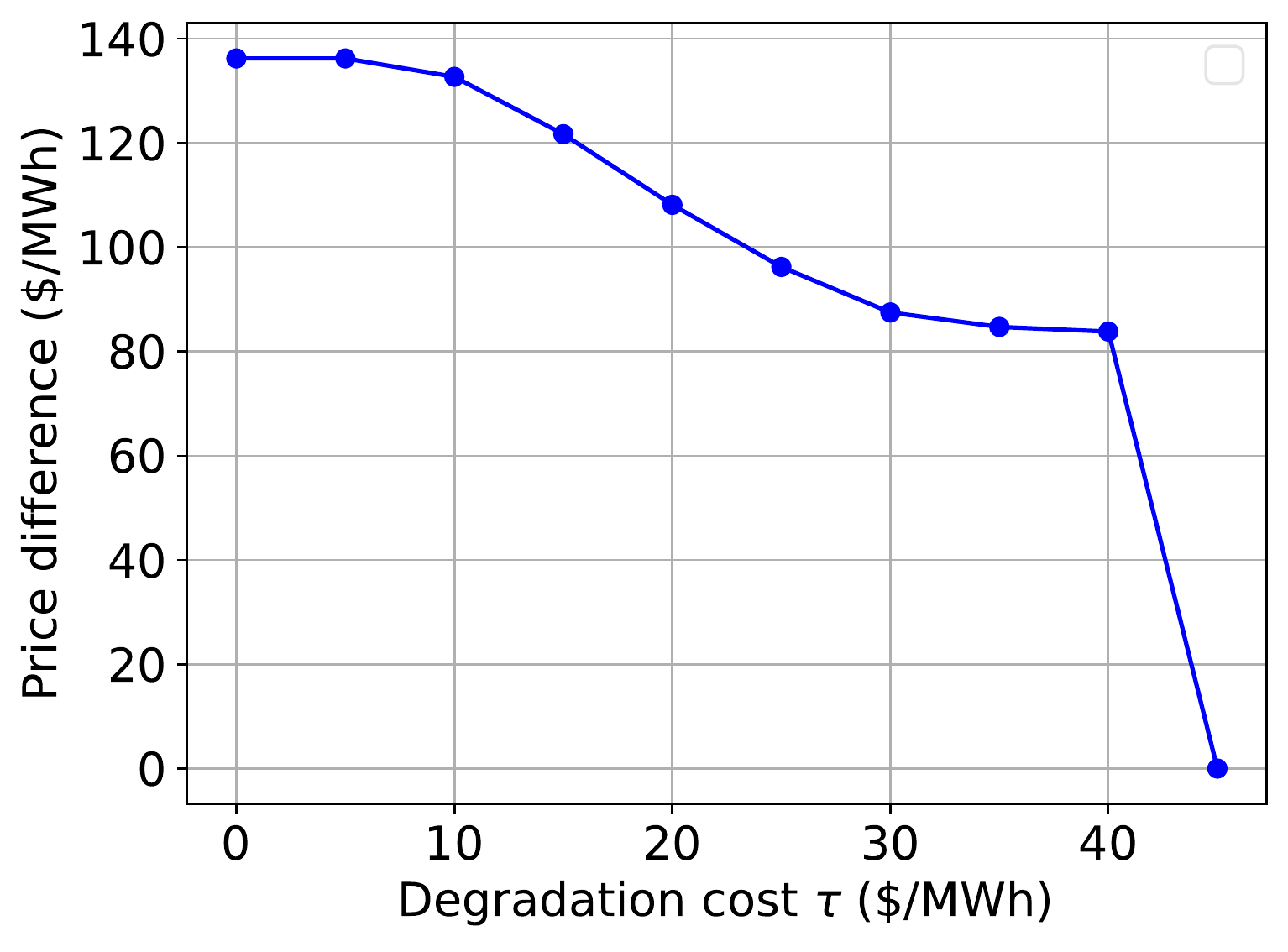}}}
 	\hspace{-1.5ex}
 	\subfigure[]{
 		\raisebox{-2mm}{\includegraphics[width=1.6in]{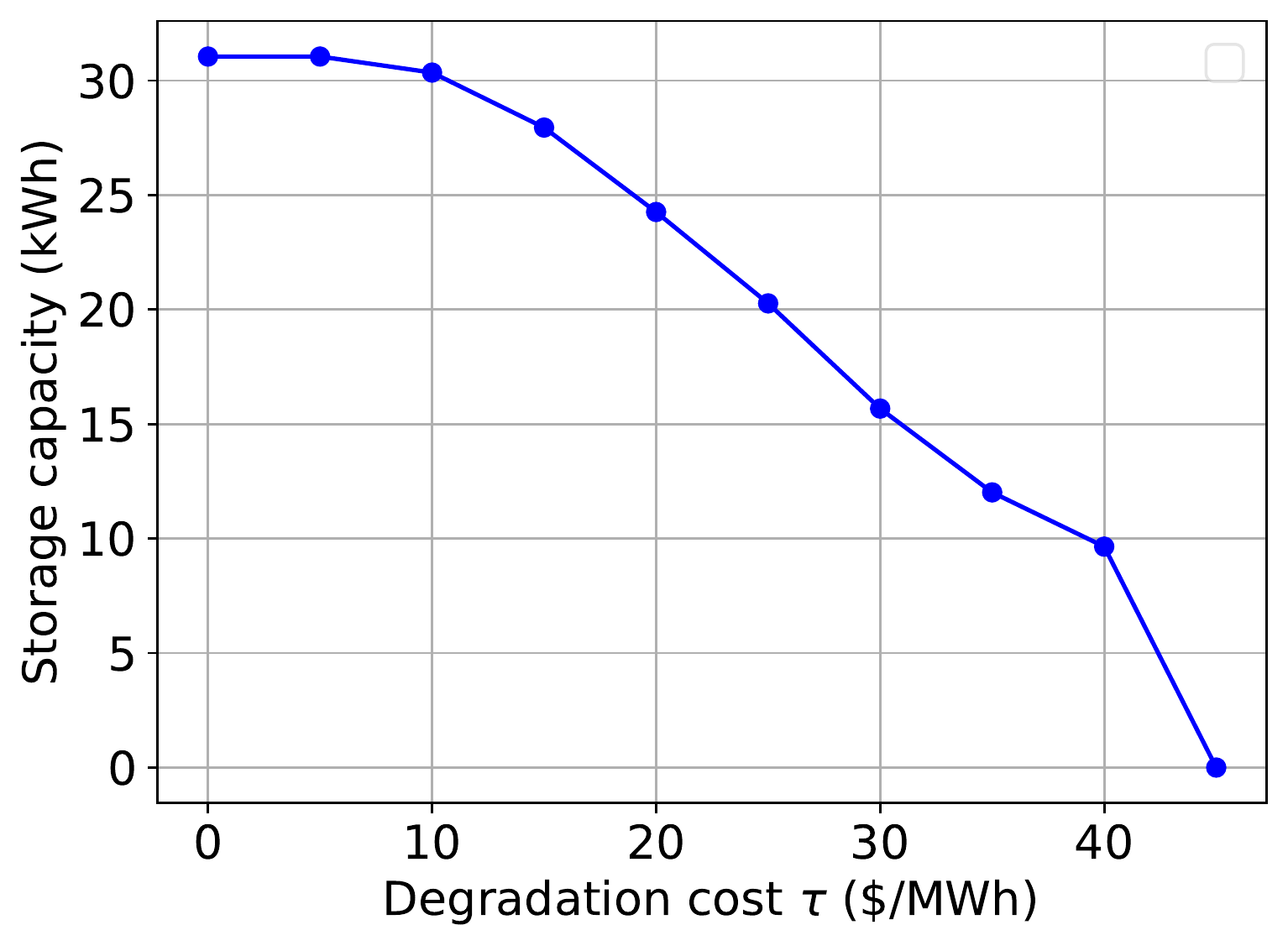}}}
 	\hspace{-1.5ex}
 	\subfigure[]{
 		\raisebox{-2mm}{\includegraphics[width=1.6in]{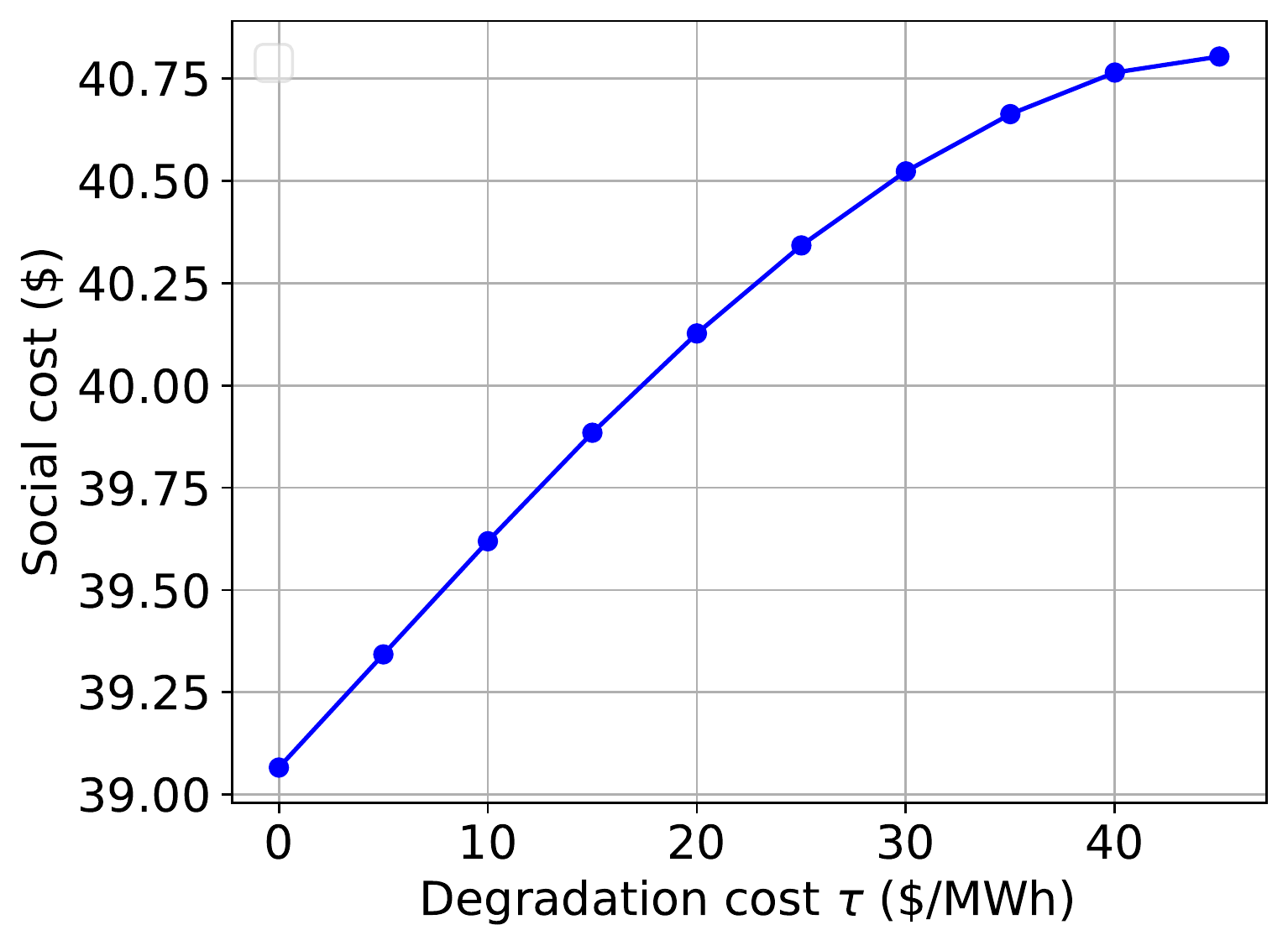}}}
 	\vspace{-2mm}
 	\caption{(a) \small Optimal price difference  $p^\Delta$. (b) Total invested storage capacity. (c) Social cost with the degradation cost. }
 	\label{fig:deg1}
 	\vspace{-3mm}
 \end{figure}

\vspace{1ex}
\noindent \textbf{Degradation cost:} In Figure \ref{fig:deg1}, we consider each user as one type and set the same degradation cost for all users, i.e.,$\tau_i=\tau$ for all $i\in \mathcal{I}$. we investigate  the results as the degradation cost  $\tau$  increases from 0 to 45 (\$/MWh), where we show  the optimal price difference $p^{\Delta*}$ in Figure \ref{fig:deg1}(a), the total invested storage capacity of all users in Figure \ref{fig:deg1}(b), and the social cost in Figure \ref{fig:deg1}(c). 

We have the following observations.  As shown in Figure \ref{fig:deg1}(a) and (b),  the optimal price difference and the invested storage  capacity will decrease with the degradation cost. The reason is that  the higher degradation cost increases the storage cost for the user. Thus, the price   difference will decrease  and incentivize less storage investment. In Figure \ref{fig:deg1}(c), the social cost always increases as the degradation  cost increases.

\end{document}